\documentclass{report}
\usepackage{amsfonts, amssymb, amsmath}
\usepackage[dvips]{graphicx}
\usepackage{t1enc}
\input xy
\xyoption{all}
\newcommand{\edth} {\mbox{\symbol{'360}}}
\newtheorem{theorem}{Theorem}

\newtheorem{lemma}[theorem]{Lemma}

\newtheorem{proposition}[theorem]{Proposition}

\input{tcilatex}

\catcode`\@=11 
\def\@evenhead{\underline{\makebox[\hsize]{{\small\sl\thepage\hfill 
\small\sl\leftmark}}}} 
\def\@oddhead{\underline{\makebox[\hsize]{{{\small\sl\rightmark}\hfill 
\small\sl\thepage}}}} 

\def\sectionmark#1{\markright {{\ifnum \c@secnumdepth >\z@ 
\thesection\ \ \fi #1}}} 

\def\mypart{
\thispagestyle{plain}%
\global\@topnum\z@ 
\@afterindentfalse 
\secdef\@mypart\@schapter} 

\def\@mypart[#1]#2{\ifnum \c@secnumdepth >\m@ne 
\refstepcounter{chapter}%
\typeout{\@chapapp\space\thechapter.}%
\addcontentsline{toc}{chapter}{#1}\fi 
\markboth{\ifnum \c@secnumdepth >\m@ne 
#1 \fi}{}%
\addtocontents{lof}%
{\protect\addvspace{10\p@}}
\addtocontents{lot}%
{\protect\addvspace{10\p@}}
\if@twocolumn 
\@topnewpage[\@makemyparthead{#2}]%
\else \@makemyparthead{#2}%
\@afterheading 
\fi} 

\def\@makemyparthead#1{%
\vspace*{50\p@}%
{\parindent \z@ \raggedright 
\ifnum \c@secnumdepth >\m@ne 
\huge\bf #1 
\par 
\vskip 20\p@ \fi 
\nobreak 
\vskip 40\p@ 
}}

\let\@ldthebibliography\thebibliography 
\renewcommand{\thebibliography}[1]{\newpage 
\addcontentsline{toc}{chapter}{\bibname} 
\@ldthebibliography{#1}}

\def\thanksname{Thank you} 
\def\listname{List of publications} 

\def\mythanks{\@restonecolfalse 
\addcontentsline{toc}{chapter}{\thanksname} 
\if@twocolumn\@restonecoltrue\onecolumn\fi 
\chapter*{\thanksname\@mkboth{\thanksname}%
{\thanksname}}\if@restonecol 
\twocolumn\fi} 

\def\mylist{\@restonecolfalse 
\addcontentsline{toc}{chapter}{\listname} 
\if@twocolumn\@restonecoltrue\onecolumn\fi 
\chapter*{\listname\@mkboth{\listname}%
{\listname}}\if@restonecol 
\twocolumn\fi} 

\def\myintro{\@restonecolfalse 
\addcontentsline{toc}{chapter}{\lowercase{\introname}} 
\if@twocolumn\@restonecoltrue\onecolumn\fi 
\chapter*{\introname\@mkboth{\lowercase{\introname}}%
{\uppercase{\introname}}}\if@restonecol 
\twocolumn\fi} 

\def\myoutro{\@restonecolfalse 
\addcontentsline{toc}{chapter}{\lowercase{\outroname}} 
\if@twocolumn\@restonecoltrue\onecolumn\fi 
\chapter*{\outroname\@mkboth{\uppercase{\outroname}}%
{\uppercase{\outroname}}}\if@restonecol 
\twocolumn\fi}

\begin{document}
\ifx\href\undefined\else\hypersetup{linktocpage=true}\fi
\thispagestyle{empty}
\begin{figure}[t!]
  \centering
  \includegraphics[width=2cm]{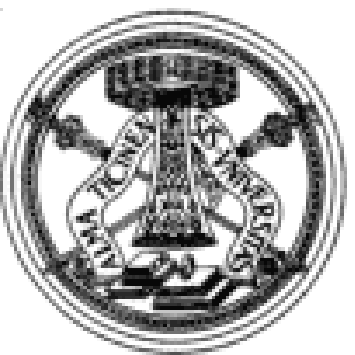}\\
\end{figure}
\begin{center}
{\LARGE \sc University of Pavia  } \\
\vspace{5pt} {\large \sc Department of Theoretical and Nuclear Physics}
\end{center}

\vspace{83pt}

{\LARGE

\begin{center}

{\bf SIMPLICIAL AND ASYMPTOTICAL ASPECTS OF THE HOLOGRAPHIC PRINCIPLE}

\end{center}}

\vspace{1 cm}

\nopagebreak[4] \vbox to 5cm {
\rm \large \noindent Supervisor:  \\
Prof. {\large\sc{Mauro Carfora}}\\
\\
\par
\noindent
\begin{flushright}
Doctoral thesis of \\
{ \large \sc Claudio Dappiaggi}
\end{flushright}
\begin{center}
\vspace{80pt} {\large Dottorato di Ricerca XVI ciclo}
\end{center}
}



\cleardoublepage

\thispagestyle{empty}

\hbox{} \vspace{200pt} \hfill{\large \em Se i fatti e la teoria non concordano, cambia i fatti

\begin{flushright}
(Albert Einstein)
\end{flushright}}

\clearpage{\pagestyle{empty}\cleardoublepage}%

\chapter*{Preface}
This thesis has been submitted on 12-02-2004 in partial satisfaction of the requirements for the degree \emph{Dottore in Ricerca in Fisica} in the Department of Nuclear and Theoretical Physics (Pavia University). 

\vskip 1cm

\noindent The evaluation commetee:

\vspace{.5cm}

\noindent Professor Alberto Rimini

\vspace{.5cm}

\noindent Professor Pietro Menotti

\vspace{.5cm}

\noindent Professor Michele Caselle

\tableofcontents

\newpage

\chapter{Introduction}
The history of physics has been characterized by the formulation of leading
principles that can describe the structure of different systems
and provide the skeleton of all theories. The equivalence principle and the
axioms of quantum mechanics are the main examples in this direction. The first
is at the basis of general relativity which describes gravity and the
macroscopic world whereas the latter gives us the rules to study microscopic
systems. 

The last fifty years have been characterized by numerous attempts to give a
unified description able to combine quantum mechanics with gravity. In this
spirit, in the seventies, Hawking studied the propagation of a scalar field in a
black hole geometry ultimately discovering that these objects emit a thermal 
radiation \cite{Hawking}. Furthermore Bekenstein \cite{Beck} was able to recognize that it is
possible to associate to a black hole an entropy 
\begin{equation}\label{BH}
S=\frac{A}{4G}+c,
\end{equation}
where $A$ is the area of the event horizon and $c$ is a constant which is assumed
to be bounded. Moreover if we consider a space-time geometry $M$ with a black
hole inside, the global system has to obey to a generalized second law of
thermodynamics which states 
$$\delta S_T>0,$$
where $S_T$ is the sum of (\ref{BH}) and of the entropy of the
matter evolving outside the event horizion. 

Together with this result, the most striking consequence of the evaporation of a
black hole has been derived still by Hawking who recognized that this effect
implies a very profound question about the consistency of the usual quantum
mechanical operations in a gravitational system \cite{Hawking2}. To understand this statement
let us consider an asymptotically flat space-time $M$ with a black hole $B$ and a scalar
field $\phi$ propagating from $\Im^-$ to $\Im^+$. The entire evoultion of $\phi$
can be thought as a scattering process where $B$ plays the role of an
intermediate state. Assuming that during this event no energy is lost and every
conservation law is satisfied then the ordinary rules of quantum mechanics grant
us that, if $|in>$ is the initial state of the system on $\Im^-$, than an
observer on $\Im^+$ will see a final state
$$|out>=S|in>,$$
where $S$ is a unitary scattering matrix satisfying $S^\dagger=S^{-1}$. A more
technical way to read this statement is to recognize that the final state of 
a system evolving from an initial pure quantum state, has to be pure
as well. Bearing this in mind and without entering into unnecessary
techniqualities, let us consider an initial scalar particle vacuum state $|0_->$ on
$\Im^-$  which can be expressed as a combination of states with different number
of particles entering into the black hole horizon or escaping to $\Im^+$ i.e.
$$|0_->=\sum\lambda_{AB}|A_+>|B_H>,$$
where $|A_+>$ is a state at the future infinity whereas $|B_H>$ is a state
on the horizon. Furthermore any measure of an observable $Q$ at $\Im^+$ will
lead to expectation values depending only on the states $|A_+>$ and thus in
ultimate instance on the creation and annihilation operators of the fields at
$\Im^+$. This implies that 
$$<0_-|Q|0_->=\sum\rho_{AC}Q_{CA},$$
where $Q_{CA}=<C_+|Q|A_+>$ and $\rho_{AC}$ is the density matrix describing all
observations made on $\Im^+$ and for this reason it depends only on the
expectation values of polynomials in the creation and annihilation operators on
$\Im^+$. In short, the analysis in \cite{Hawking2} shows that the above matrix is diagonal in a
basis of eigenstates of the number operator; thus after a black hole has
completely evaporated, the only possible final states for the radiation at
$\Im^+$ are those with total energy equal to the intial mass of the black hole
and all these configurations can be emitted with equal probability. Eventually
the whole process leads to a violation of the principles of quantum mechanics
where any consistent scattering process that can bring from a pure state to a mixed
one is forbidden.  \\
Furthermore the study of a physical system in presence of an extreme gravitational
field does not only lead to a contraddiction with quantum mechanics but even with
ordinary results of statistical mechanics. A canonical example comes from Susskind \cite{Susskind} who considers a three dimensional lattice 
of $\frac{1}{2}$ spin-like degrees of freedom whose fundamental spacing is of the order of the Planck length $l_p$. Ordinary statistical mechanics tells us that
the number of orthogonal quantum states of the system is $N(V)=2^n$ where $n$ is the number of sites occupying a certain region of volume $V$. The maximum
entropy available to the system is given by the logarithm of the number of degrees of freedom:
\begin{equation}\label{states}
S(V)=\log N(V)=\frac{V}{l^3_p}\log 2,
\end{equation}  
showing a proportionality with the volume. This is in sharp contrast with the
Beckenstein formula (\ref{BH}) since most of the states contributing to the
entropy have an energy far bigger than the one needed for the system to collapse
into a black hole and moreover the size of such a black hole could be even
greater than $V$ itself.

Going a step further with Susskind's argument we can consider a region $V$
filled with enough matter in order to have an entropy $S$ bigger than the one of 
a black hole but without enough energy to form it. Throwing inside other matter
increases both the entropy and the energy of the system until a gravitational
collapse begins leading to a final configuration with $S$ proportional to the
area. Thus the whole process violates the generalized second law of
thermodynamics since $\delta S<0$. 

\par

The above two examples are simply an indication of the apparent contraddiction
between general relativity on a side and quantum mechanics and statistical
mechanics on the other side. Both the arguments given by Hawking and Susskind
are convincing and difficult to circumvent unless a new radical interpretation
is given. The counterproposal was given at the beginning of the nineties by G.
't Hooft \cite{'thooft4} who considered a situation similar to the 
example discussed above. He argued to radically change the way we count degrees of
freedom in such systems: in presence of gravity, an observer can only excite the
states with an energy less than the one needed to form a black hole.
With this prescription in mind let us consider a gas of particles at a finite 
temperature $T$. The energy of the system is given by the Boltzmann law:
$$E\propto VT^4,$$
which implies
$$S\propto VT^3.$$
Imposing that $E<\frac{R}{2}$ where $R$ is the Schwartzschild radius of the
system (thus no gravitational collapse can occur), then the entropy satisfies the bound
$$S\propto A^{\frac{3}{4}}<A,$$
which could be read as the statement that a black hole is the physical system
with the highest reachable entropy. The whole hypotesis provides also an elegant
solution to Hawking paradox: the information "lost" inside the black hole which
is eventually emitted through a thermal radiation can be possibly recovered
bearing in mind that, since the entropy is a measure of the degrees
of freedom accessible to the system, then the Beckenstein bound is only another
way to say that all the datas inside the black hole can be encoded
on the event horizon with a density not exceeding $\rho_P$, i.e. on a lattice
with the Planck length $l_P$ as the characteristic one, one bit of information
per lattice site. In such a way there is no more need to break the laws of
quantum mechanics in presence of an extreme gravitational field. 

\par Moreover the above concept can be pushed a step further and it can be 
generalized to the
statement that all the information living in a D-dimensional manifold $M$ can
be encoded in an hypersurface of codimension $1$ (usually but not necessarly the
boundary $\partial M$) with a density of states not exceeding $\rho_P$. Thus the
main consequence of this idea is that the usual way of counting degrees of freedom in
a quantum field theory with a cut-off is highly redundant in a system where gravity is switched
on. In this situation if we try to excite more than $\frac{A}{4}$ degrees of
freedom, we are taking into account a number of states whose energy leads to the
formation of a black hole and thus in ultimate instance they are not accessible.
This is the key idea behind 't Hooft \emph{holographic principle}. 
\par 
Since its formulation in 1993 one the main task of modern physics has been to
associate to the holographic principle a consistent theory in
the same way as general relativity is the natural son of the equivalence
principle. Unfortunately up to now there is no such a candidate theory and the
main successes in this direction have been only limited to a certain class of
systems. We are going now to briefly review them.
\section{The covariant entropy conjecture}
In ultimate analysis the holographic principle represents a new interpretation
of the deep physical significance of the Beckenstein bound. Thus a natural step
in order to further motivate 't Hooft conjecture outside the black hole horizon
is to look for an entropic bound inside a more generic space-time region.

This question was addressed in a series of papers
by Bousso \cite{Bousso1} \cite{Bousso2} who considered a 3-dimensional space-like region $\Sigma$ in
arbitrary 4-dimensional manifold\footnote{Although we refer for simplicity to a
4 dimensional manifold, Bousso analysis can smoothly be extended to higher
dimensions} such that its boundary $\partial\Sigma$ is space-like as
well. From each point in $\partial\Sigma$ it is possible to construct four light rays, two
future directed and two past directed. The set of all these light-like
conguences forms the so called "light sheets" whose dynamics is governed by the
Raychauduri equation 
\begin{equation}\label{Ray}
\frac{d\theta}{d\lambda}=-\frac{\theta^2}{2}+\omega_{ab}\omega^{ab}+\sigma_{ab}\sigma^{ab}+8\pi R_{ab}k^a k^b,
\end{equation}
where $\theta=\frac{1}{A}\frac{dA}{d\lambda}$, $\omega_{ab}$ is the torsion tensor, and $\sigma_{ab}$ is the shear tensor.\\
Bousso prescription is to consider all the light-sheets $L$ with a negative
expansion and to follow them until the expansion changes sign or they end on a
caustic or at the boundary of the geometry. The covariant entropy conjecture states that the
entropy inside these regions satisfies the bound
\begin{equation}\label{Bousso}
S(L)\leq\frac{A}{4},
\end{equation}
where $A$ is the area of $\partial\Sigma$. \\
The conjecture can be clearly viewed as a generalization of Beckenstein work
and of all other entropy bounds derived in the framework of general relativity (see as an example
\cite{Fischler}). Moreover, by its own construction, the relation (\ref{Bousso}) is
manifestly invariant under time reversal which is a property that cannot be
understood at a level of thermodynamic only and since there is no
assumption about the microscopic physics of the underlying system, the most
natural way to interpret Bousso result is at a level of degrees of freedom
accessible to the system itself. 
Thus inspired by 't Hooft work, Bousso formulated a sort of
generalized holographic principle \cite{Bousso2}:

\vspace{0.3cm}

\noindent {\bf Bousso holographic principle}: {\it Consider a 2-dimensional region $B$ of area $A$ satisfying the covariant entropy conjecture, then the number $N$ of
orthonormal states in the Hilbert space $\mathcal{H}$ describing the physics in
the bulk region $L$ satisfies the bound
$$N\leq\exp\frac{A}{4},$$
or equivalently the number of degrees of freedom $N_{dof}$ accesible to the system is
$$N_{dof}\leq\frac{A}{4}.$$}  

\vspace{0.3cm} 

\noindent 
This recipe associate to any two dimensional surface two light-like
hypersurfaces with a bounded number of degrees of freedoms. Nonetheless this
result is clearly limited to a local a priori fixed region of space-time; 
on the other hand it would be more interesting to know if it is possible to apply
Bousso holographic conjecture to an hypersurface able to store
informations from the global space-time and not only from a limited portion. The solution works exaclty inverting the
previous prescription: starting from a null hypersurface, one should find the
geodesic generators with a non negative expansion until the expansion itself
changes sign. This procedure, called by Bousso "projection", identifies a
two dimensional spatial surface which is called \emph{screen}. Moreover inside
the class of screens, we can identify a certain subclass represented by the
surfaces with a vanishing expansion at any point. The latter are referred to as
preferred screens since it has been conjectured in \cite{Bousso2} that precisely
on these surfaces the holographic bound is saturated. In this thesis, we will
show that this concept plays a pivotal
role in the construction of an holographic correspondence at a level of quantum
field theory. In fact, since it is easy to identify a preferred screen in a wide
class of space-times, it will be natural to construct explicitly the dual field
theory on such submanifold. \par
Up to now we have given a background independent formulation of the holographic
principle and we have given as well the rules for finding those surface where
to encode the bulk information with a density not exceeding one bit of
information per Planck area. Eventually, in our hands we have a sort of
"geometric" (or classical) holography but up to now we have no clue of what sort
of theory we should expect on the (preferred) screens. Clearly we cannot
find, even a priori, a conventional quantum field theory since the holographic
principle is far more different and elusive than simply assigning some inital
datas on a
Cauchy surface. Moreover it has been pointed out in \cite{'thooft3} that the price for
protecting quantum mechanics and general relativity from a reciprocal contraddiction
could possibly be the loss of locality in the holographic theory. 
From one side this is not surprising since it is difficult to conceive a local quantum
field theory with cut-off without an entropy proportional to the volume.
Nonetheless we could still avert 
the loss of locality of the theory by introducing an unusual gauge simmetry. 
Stripping the details of such a formulation let us briefly comment that in this
approach, 
advocated by 't Hooft and mainly applied to the black hole scenario, we
distinguish between "ontological states" growing in number as the volume of the
system and "equivalence class" of states growing instead as the area and
reproducing the Beckenstein bound. This proposal, although extremely
interesting, is quite difficult to apply in a more general setting and it does
not reproduce yet Bousso results. For this reason we shall not discuss such an
approach any longer.\par
So far we have used the holographic principle only within the realm of general
relativity and the main result has been only to understand where the bound on
the number of degrees of freedom can effectively be implemented; the next natural step is
to enter the realm of quantum field theory and try to write a theory where such a
principle is manifest. The most natural and straightforward scenario would be to construct such an
"holographic theory" on a preferred screen where the bound of one bit of
information per Planck area is saturated. In this case we expect a
kind of dictionary relating the datas on the screen with those in the bulk and we
will refer from now on to this situation as the \emph{dual theory}. Although
it seems very difficult to write such a theory for a generic system, up to now
the only successful example of an holographic correspondence at a level of
quantum field theory, namely the AdS/CFT correspondence, is indeed a dual theory
living on the preferred screen of an asymptotically anti de Sitter
space-time.\\
Nonetheless as pointed out by Bousso in \cite{Bousso2}, the "dual"
approach is  in general doomed to failure since one of the peculiar aspect of asymptotically
AdS space-times is the space-like nature of its boundary. This implies that the
number of degrees of freedom is fixed whereas in presence of a time-like
boundary such number should evolve in Lorentzian time. In this case we should
talk only of a "holographic theory" where the information stored on the screen
does not necessarly saturate the Planck limit and a dictionary relating
bulk/boundary datas is elusive at best. In this thesis we will advocate this
point of view in a more radical way since we will deal with a class of manifolds, the
most notable ones being the asymptotically flat space-times, where the natural choice 
as a preferred screen, i.e. either the boundary $\Im^+$ or $\Im^-$, is a null 
submanifold. Thus by its own nature, in such a case, the screen
does not provide a good notion of time evolution and the recognition of the degrees of
freedom is difficult since we cannot even talk of an "activation" and a
"deactivation" of accessible states as advocated by Bousso for time-like screens. We will show that in this context it is
difficult as well to keep track of the concept of locality and it is likely that
the holographic theory will be most probably very different from a conventional
gauge theory. \\
The final message that we can read from Bousso conjecture is that it
can be very difficult to use the holographic principle to "describe nature"
since its realization can be extremely elusive and as far as we understand, it
is not manifest either in general relativity or in quantum field theory. 
\section{Holography and asymptotic symmetries}
At a quantum level the most notable example of the realization of the
holographic principle is the AdS/CFT correspondence \cite{ADS/CFT}. Although it is
not the purpose of this thesis to explain the details of such theory \cite{Petersen}
\cite{Klebanov}, let us nonetheless
point out some of the peculiar aspects characterizing this
correspondence as unique by its own nature.\par
The original Maldacena's statement deals with the existence of a complete equivalence
between type IIB superstring theory in the bulk of $AdS_5\otimes S_5$ and maximally
supersymmetric $3+1$ dimensional $SU(N)$ SYM theory (thus conformally invariant)
living on the boundary of the AdS space-time. This correspondence can be
geometrically read as a relation between a string theory invariant under
$SO(2,d-1)$, the symmetry group of the $AdS_d$ space time and a quantum field
theory living on $\partial AdS_d$ and invariant under the asymptotic symmetry
group of the bulk space-time. Moreover, although the gauge/gravity
correspondence has been originally introduced in the maximally symmetric
background, Maldacena ideas can be generalized to any asymptotically anti-de
Sitter background by means of an holographic renormalization group; this
technique allows to compute correlation functions of the dual fields only through
near boundary computations which rely only upon the asymptotic structure
\cite{skenderis}. In
this way the original AdS/CFT recipe that associates to every bulk field $\Phi$ a gauge invariant 
boundary operator $\mathcal{O}_\Phi$ is still valid and the boundary values of
the bulk fields are identified with sources that couple to dual operators;
eventually the on-shell bulk partition function is identified with the
generating function of the (bulk) qunatum field theory correlation functions:
$$Z_{sugra}[\Phi_{(0)}]=\int\limits_{\Phi\sim\Phi_{(0)}}[d\Phi]e^{-S[\Phi]}=<\exp(-\int\limits_{\partial
AAdS}\Phi_{(0)}\mathcal{O})>_{QFT},$$
where $<,>$ represents the Feynmann quantum field theory expectation value and
$\partial AAdS$ is the boundary of an asymptotically anti de Sitter manifold.
Following the
details of the foundational work of Henneaux
and Teitelbom \cite{Henneaux} it is clear that the case of negative cosmological
constant is quite peculiar; first of all the nature of the boundary itself is
extremely unique since, as mentioned before, it is a space-like submanifold which in the spirit of
Bousso conjecture represents a preferred screen. As an example let us
consider the AdS$_d$ space-time which is topologically equivalent to $\Re\times
S^{d-1}$ and can be described through the metric
$$ds^2=R^2\left[-\frac{1+r^2}{1-r^2}dt^2+\frac{4}{(1-r^2)^2}\left(dr^2+r^2d\Omega_{d-2}^2\right)\right],$$
where $R$ is the curvature radius. Slicing with hypersurfaces of constant time,
the boundary is a $2-$sphere (thus a space-like submanifold) at $r=1$ with a 
divergent area; in the spirit of Bousso's bound, we can consider
the past directed radial light-rays starting from $r=0$ at a certain fixed time.
They emanate from a caustic (i.e. $\theta=+\infty$ in (\ref{Ray})) and they form
a light cone with a spherical boundary which grows in time until it reaches the
boundary $r=1$ where it is easy to check that $\theta\to 0$ which,
in the language of \cite{Bousso2}, implies that the AdS boundary is a preferred 
screen. At a quantum level, instead, the dual theory is invariant under the conformal group ;
this enables us to read the
bulk/boundary correspondence as a relation between the infrared sectors in
$AdS_d$ and the ultraviolets sector in $\partial AdS_d$ which is a remarkable
feature since it has allowed \cite{Susskind2} to show explicitly 
that on the boundary the number of degrees of freedom is bounded by a bit of
information per Planck area which is a necessary characteristic of any
holographic theory. \par
A natural question raising from these considerations is
whether and to what extent it is possible to follow the road settled by Maldacena in a different
background. The work done in (asymptotically) dS space 
\cite{DS/CFT}, is often based upon analytic continuation
from AdS solutions and thus the results so far are of no use in asymptotically flat
space-times since the transition of the value of physical relevant quantities from non zero 
cosmological constant to $\Lambda=0$ is
non smooth. Thus, in the $\Lambda=0$ framework what happens is quite different from the AdS 
case; at a geometric level and in the spirit of finding an
holographic theory we notice that in the class of asymptotically flat space-times
the notion of preferred screen does not apparently differ from the AdS scenario.
Taking as leading example Minkowski space-time, the boundaries $\Im^\pm$ are
topologically equivalent to $\Re\times S^{d-2}$; in absence of a black hole it
is easy to show \cite{Bousso2} that the value of $\theta$ goes to $0$ as we approach
the boundary. Nonetheless it is imperative to remark that, as soon as a
gravitational collapse occur, $\Im^+$ does not represent any more a screen
since, as we can see drawing the Penrose diagram, part of the
information coming from $\Im^-$ falls into the black-hole which means, from an
holographic point of view, that the datas are stored on the apparent horizon.
Nonetheless the past boundary $\Im^-$ can always been seen as a preferred screen
but this should not mislead us in concluding that the situation is identical to the
one experienced in the Maldacena conjecture. In fact, as we shall see in greater
details in chapter 4, in asymptotically flat space-times there is
a finer relation than in asymptotically AdS manifolds between the asymptotic
symmetry group and the background metric.
In the $\Lambda=0$ scenario, the group of asymptotic isometries 
contains not only the Poincar\'e group which is the bulk symmetry group but also angle-dependent
translations which ultimately lead to an infinite dimensional Lie group (the BMS
group) \cite{Bondi}. Leaving the details of its derivation to the forthcoming
chapters, suffice to say that the BMS group is the semidirect product of the Lorentz
group with the set of class $C^\infty$ maps from the sphere $S^{d-2}$ to the
real numbers and, moreover, it admits a unique four dimensional normal subgroup which
is isomorphic to the translation subgroup. A tempting conclusion would be to mod
out the non Poincar\'e part of the BMS group by adding some suitable (but
unnatural) boundary
conditions but, as we will show, this is impossible whenever the background metric
is not stationary. The latter is an unwanted
restriction since there is no reason why it should exist a different description
of a potential holographic theory in presence of a time dependent or a non time
dependent manifold.\par 
From the above remarks, it follows that there is a further
significant difference between asymptotically AdS and asymptotically flat
space-times.
In the first case the preferred screen is a space-like
submanifold and the group of asymptotic isometries is a
universal structure which ultimately does not depend on the choice of the
background metric; on the opposite, in the $\Lambda=0$ scenario, applying Bousso conjecture, 
it comes natural to store
the holographic datas on a null hypersurface whose symmetry group, the
Bondi-Metzner-Sachs group, can be reduced in static backgrounds to the
Poincar\'e subgroup. In the spirit of an holographic correspondence, this
is a characteristic very difficult to take into account since at first sight it
leads us to conclude that in an asymptotically flat background, the
bulk/boundary correspondence depends partially upon the chosen metric and at a geometric level it
is not a universal feature. Thus, in the spirit of a "universal"
gauge/gravity correspondence, we suggest not to distinguish between time and non
time-dependent backgrounds and to try to construct the bulk/boundary relation 
starting from the full BMS group\footnote{We expect that, in the eventual holographic
theory the existence of a preferred
Poincar\'e subgroup in static space-times will be recovered in some suitable
limit.}. For this reason we will not start
constructing the dual theory as in AdS in the maximally symmetric background (i.e.
Minkowski space-time) but we will address the problem in the most general
framework.  \par 
A further striking difference between the asymptotically AdS and flat scenario
comes from the quantum theory.
As mentioned before in the AdS/CFT correspondence (and partially in the dS/CFT
scenario) the boundary theory is a conformal field theory living on a spacelike
manifold; thus we have been able to explicitly derive both the particle spectrum
and the correlators for the boundary theory in any asymptotically AdS
background and eventually these results led us to discover new insights on the
dynamic of the
string theory living in the bulk. On the opposite in asymptotically flat space-times the situation is
almost reversed since in an holographic scenario we face a boundary theory living on
the $\Im^+$ or on $\Im^-$ which are null submanifolds.
This implies that the underlying metric is degenerate and the usual techniques
fail since at present there is no direct way to write either a classical or a
quantum field theory living on
$\Im^\pm$. Moreover the spectrum of a BMS field
theory is unknown and candidate wave equations are not available. This prevents
us to compare, as it has been done in AdS, the Poincar\'e (bulk) with the BMS
(boundary) spectrum
in order to understand some feature of the holographic correspondence and write
a sort of bulk/boundary dictionary. 
Nonetheless although we face a great number of odds trying to construct the
dual theory, in this thesis we advocate that a line of research looking for an
holgraphic theory in asymptotically flat space-times through the BMS group is
far more reliable than most of the alternatives proposed until now since most of
them deal with the large $R$ limit of the AdS scenario without taking into
account the non smooth transition of physical observables between non vanishing
and vanishing cosmological constant. Moreover in the above approach the
peculiar nature of the asymptotic symmetry group of asymptotically flat
space-times is not taken into account and for this reason the geometric
difference between stationary and non stationary background, which is non
existent in the AdS scenario, is always hidden.\\
Thus, in this thesis, our first step will be to look at BMS representation
theory following the approach that led Wigner \cite{Wigner} to construct the
wave equations for the
Poincar\'e group. Our final purpose will be to derive the spectrum of the candidate
dual theory and to compare it with the bulk one; nonetheless we can
anticipate that, being the Poincar\'e group a subgroup of the
BMS, the particle spectrum of the asymptotic simmetry group of flat manifolds
will be larger and richer compared to the Poincar\'e spectrum thus leading to the natural
conjecture that the boundary theory will contain an unnecessary large number of datas. 
\section{"Intrinsic" holography}
The holographic principle has been formulated as a conjecture over the counting
of degrees of freedom in a physical system without any reference to a particular theory. Nonetheless
the successes of AdS/CFT correspondence led often to think to holography as a
gauge/gravity correspondence allowing to describe quantum gravity or a quantum
field theory in a curved background through a second field theory without
gravity living on the
boundary. Such a way of thinking, although not without credit, requires 't Hooft
conjecture only as a necessary condition and not as a sufficient one. In
particular an interesting line of research would be to explore whether it is
possible to strengthen the above remark looking for an holographic
correspondence outside the realm of Maldacena example.
\par Bearing in mind such idea, the most natural candidate in this direction is
a $2+1$ space-time; in this framework it has been known since the late eighties
\cite{Witten} that three dimensional gravity (euclidean and non) with non vanishing
cosmological constant can be formulated as a Chern-Simons theory. More in detail
on a manifold with empty boundary $M$ the action for this system is
\begin{equation}\label{CS}
S=k\int\limits_M d^3x [A\wedge dA+ \frac{2}{3}A\wedge A\wedge A],
\end{equation}
where $A$ is a connection for the gauge group $SL(2,\mathbb{C})$. An interesting
feature of (\ref{CS}) emerges if we consider a manifold $M$ such that $\partial
M\neq 0$ since we have to add a boundary term to the above action. Supposing,
as an example, that the three dimensional variety can be splitted as
$M=\Sigma_g\times\Re$ where $\Sigma_g$ is a Riemann surface, the variation of
(\ref{CS}) is
$$\delta S=k\int d^3x \epsilon^{\mu\nu\rho}Tr(\delta A_\mu F_{\nu\rho})+k\int
d^3x \partial_{\nu}[\epsilon^{\mu\nu\rho}Tr(A_\mu\delta A_\rho)].$$
This formula implies that we have to implement a boundary condition such that
$\int\limits_{\partial M}Tr(A\delta A)=0$; taking for simplicity $\Sigma_g$ as a
disk and interpreting the real axis as a time direction, the above boundary
constraint can be achieved imposing that
one of the component of the connection, namely the temporal one $A_0$, vanishes.
With this condition imposed, the gravitational action becomes:
$$S=-k\int_{D\times\Re}d^3x\epsilon^{ij}Tr(A_i\dot{A}_j)+\frac{k}{3}\int_{D\times\Re}d^3x\epsilon^{ij}Tr(A_0F_{ij}).$$
The component $A_0$ acts as a Lagrange multiplier leading to the constraint
$F_{ij}=0$ which admits as a solution pure gauge connections i.e.
$$A=g^{-1}dg.$$
Substituting this result in the Chern-Simons action, it becomes
$$
S=-k\int\limits_{D\times\Re}d\theta dt Tr(g^{-1}\partial_\theta
gg^{-1}\partial_0
g)+$$
\begin{equation}\label{WZW}
+\frac{k}{3}\int_{D\times\Re}d^3x\epsilon^{\mu\nu\rho}Tr(g^{-1}\partial_\mu
g g^{-1}\partial_\nu gg^{-1}\partial_\rho g).
\end{equation}
This is exactly the action of a chiral WZW model and more in general it is
possible to demonstrate that, dropping the hypotesis that
$\Sigma=D$, the Chern-Simons action still induces a WZW boundary term for any compact surface.
Moreover a deeper relation between three dimensional gauge theories and
CFT can be realized upon quantization of the Chern-Simons action. Without
entering into the technical details \cite{Witten3} \cite{Elitzur}, 
we are facing a system which, although at first sight the action (\ref{CS}) looks 
highly non linear, can be quantized through a canonical formalism. Working
always for simplicity with $M=\Sigma\times\Re$ and with the gauge choice $A_0=0$,
the procedure consists first in
quantizing (\ref{CS}) and then in imposing the constraint that the curvature
vanishes, i.e. $\frac{\delta\mathcal{L}}{\delta A_0}=\epsilon^{ij}F_{ij}^a=0$;
this implies that we consider as a phase space the moduli space of flat
connections over $\Sigma$, modulo gauge transformations. At a geometric level
the above procedure is approximately equivalent to the construction of a suitable
Hilbert space $\mathcal{H}_\Sigma$ over the Riemann surface $\Sigma$; this
operation requires the choice of a complex structure $J$ over $\Sigma$ and
$\mathcal{H}^J_\Sigma$ can be interpreted as a holomorphic vector bundle on
the moduli space of Riemann surfaces. Moreover since we also require that $\mathcal{H}^J_\Sigma$
is indipendent from $J$ and depends only on $\Sigma$, this implies the the vector
bundle given by $\mathcal{H}^J_\Sigma$ has flat connections as we expect in a
Chern-Simons gauge theory. The above remarks are important since flat bundles
appear also in conformal field theory; if one consider, as an example, a current
algebra on a Riemann surface with a generic symmetry group $G$ at level $k$, it
is a canonical result that, at genus 0, the Ward identities uniquely determine
the correlation functions for the identity operator and its descendants. On the
opposite, when the genus is greater than $1$, the space of solution of Ward
identities for descendants fields of the operator $\mathbb{I}$ is a vector space
often named "space of conformal blocks" which, following Segal construction
\cite{Segal}, is
exactly identical to the above Hilbert space $\mathcal{H}_\Sigma$.\\
Thus starting from these consideration, we are entitled to claim that there is a
one to one correspondence between the space of conformal blocks in 1+1
dimensions and the Hilbert space obtained upon quantization of a Chern-Simons
gauge theory in 2+1 dimensions. Furthermore, although these results rigorously apply only to
a space-time factorizable as $\Sigma\times\Re$, Moore and Seiberg
\cite{Moore} brought Witten's analysis a step further conjecturing that all
the chiral algebras of any rational conformal field theory arise from the
quantization of a 3D Chern-Simons gauge theory for some compact Lie group. This
hypotesis is clearly very suggestive; it connects in three dimensions the bulk
datas which, at quantum level are encoded in the Hilbert space of the theory, to
the space of conformal blocks of the boundary theory that are ultimately related
to the correlation functions which basically encode the dynamic of the theory.  \par  
Although at this stage it would be tempting at first sight to read the
above statement as a sort of hint of an holographic correspondence, it is
imperative to remark that this would be at best premature. First of all, even if
it would be natural to read the CS/WZW relation as a sort of rephrasing of the
$AdS_3/CFT_2$ correspondence, this point of view is unjustified since
there is no way to exclude a priori that the two relations are differents.
Moreover in the spirit of a gauge/gravity correspondence it is necessary to find
a sort of decoupling regime which allows us to separate the bulk from the
boundary dynamics but at present time, this is a feature which has not been
discovered yet. Nonetheless, taking into account the deep relation described
before between the
space of states of a Chern-Simons theory and the space of conformal blocks of a
WZW model, we still advocate the existence of a sort of "intrinsic" holographic 
description of the CS/WZW system. Furthermore we
claim that, instead of studying an AdS type correspondence in the continuum,
the most natural way to look at the bulk/boundary theory in 2+1 dimensions is
through a complete different approach; in fact is known that three dimensional
euclidean gravity on a manifold with (or without) boundary is 
equivalent to a discretized model proposed by Ponzano and Regge \cite{Ponzano}.\par
Without entering for now in unnecessary technical details \cite{ambjorn}, in a discretized scenario a
D-dimensional manifold is approximated through a polyhedron whose underlying
constituent pieces are a collection of D-simplices\footnote{By a p-simplex $\sigma_p$
with vertices $x_0,...x_p$ we mean the subspace of $\Re^d$ defined as
$\sigma_d=\sum\limits_{i=0}^p\lambda_i x_i$ where $\lambda_i$ are real positive
numbers satisfying the relation $\sum\limits_i\lambda_i=1$.} glued
together along a common D-1 dimensional subsimplex.  
In this scenario, known also as Regge calculus, the main features are to assign to
each edge of the triangulation a length and to consider the
curvature as concentrated over simplices of codimension 2 known also as hinges or
bones; starting from this simple assumption it is possible to give a discretized
analogue of Einstein action:
$$S_E=\sum\limits_{bones}\epsilon_i\mid\sigma_i\mid,$$
where $\mid\sigma_i\mid$ represents the volume of the i-th flat simplex of codimension 2
and $\epsilon_i$ is the so called deficit angle which is equal to the difference
between $2\pi$ and the dihedral angles between the faces of the simplices meeting at
the hinge.\par
In the approach proposed by Ponzano and Regge, instead, the recipe is a little bit
different: even if we deal with a three dimensional triangulated manifold with the curvature
concentrated over one dimensional bones, we label each edge of the
simplices with an element $j_i$ of the group $SU(2)$ which is ultimately related to
the length through the relation $l_i=j_i\hbar+\frac{1}{2}$. Moreover, starting from
the recombinatorial theory of angular momentum, this
construction also lead us to naturally associate to each face of a triangle a Wigner
$3jm$ symbol and to each tetrahedra a Wigner $6j$ symbol. Starting from these
assumptions, the main purpose in \cite{Ponzano} was to study the asymptotic
formula for the 6j symbol in a sort of semiclassical limit where the assignement
$j_i$ is diverging, $\hbar\to 0$ but the product $j_i\hbar$ (thus in ultimate instance
the edge length), is kept fixed.
Eventually it was found\footnote{This formula was rigorously demonstrated by
Roberts in \cite{roberts}}:
$$\left\{\begin{array}{ccc}
j_1 & j_2 & j_3\\
j_4 & j_5 & j_6\end{array}\right\}\sim\frac{1}{2\pi
V}\cos\left(\sum\limits_{i=1}^6 
j_i\theta_i+\frac{\pi}{4}\right),$$
where $V$ is the volume of the tetrahedron and $\theta_i$ is the external
dihedral angle at the $i-th$ edge. To connect this formula to quantum gravity
let us consider the following topological state sum for a generic three dimensional manifold
$M$ with non empty boundary:
\begin{equation}\notag
Z[(M^3, \partial M^3)]\,=
\,\lim_{L\rightarrow \infty}\:
\sum_
{\left \{\begin{array}{c}
(T^3, \partial T^3)\\ 
J,j,m \leq L
\end{array}\right\}}
Z[(T^3,\partial T^3) \rightarrow (M^3, \partial M^3); L],
\end{equation} 
\begin{eqnarray}
\lefteqn{Z[(T^3(J),\partial T^3(j;m)) \rightarrow (M^3, \partial M^3); L]\equiv
Z[(T^3,\partial T^3);L]
=}\hspace{.5in}\nonumber \\
& & \Lambda(L)^{-N_0}\,\prod_{1}^{N_1} (-1)^{2J} (2J+1)\,\prod_{1}^{N_3} 
(-1)^{\sum J}\:\ \left\{6j\right\}(J)  \nonumber \\
& & \Lambda(L)^{-n_0}\,\prod_{1}^{n_1} (-1)^{2j} (2j+1)\,\prod_{1}^{N^F_3} 
(-1)^{\sum (J+j)}\:\, \left\{6j\right\}(J,j)  \nonumber \\
& & \prod_{1}^{n_2} (-1)^{(\sum m)/2}\:\, \left[3jm\right], 
\label{tutta}
\end{eqnarray}
where $N_i$ represents the number of i-th simplices in the triangulation $T^3$,
$\left\{6j\right\}$ and $\left[3jm\right]$ are respectively the 6j and 3jm $SU(2)$ symbols and
$\Lambda(L)$ is a suitable cutoff. In the limit where $\partial M=0$ the above
formula reduces to:
$$Z[M^3]=\lim_{L\to\infty}\sum_
{\left\{ T^3(j), j\leq L\right\}}
Z[T^3(j) \rightarrow M^3; L]$$
\begin{equation}
Z[T^3(j)\rightarrow M^3;
L]=\Lambda(L)^{-N_0}\prod\limits_{A=1}^{N_1}(-)^{2j_A}(2j_A+1)\prod\limits_{B=1}^{N_3}
(-)^{\sum\limits_{p=1}^6}\left\{\begin{array}{ccc}
j_1 & j_2 & j_3\\
j_4 & j_5 & j_6\end{array}\right\}.
\end{equation}
In the semiclassical limit, the state sum becomes
\begin{equation}\label{sts}
Z[T^3(j);
L]=\Lambda(L)^{-N_0}\prod\limits_{A=1}^{N_1}(-)^{2j_A}(2j_A+1)\prod\limits_{B=1}^{N_3}
\frac{(-)^{\sum\limits_{p=1}^6}}{2\pi
V}\cos\left(\sum\limits_{i=1}^6 
j_i\theta_i+\frac{\pi}{4}\right),\end{equation}
and by exploiting the relation $\cos\theta=\frac{e^{i\theta}+e^{-i\theta}}{2}$, we can
recognize in (\ref{sts}) a term which looks like a Feynmann sum over histories i.e.
$$\Lambda(L)^{-N_0}\prod\limits_{A=1}^{N_1}(-)^{2j_A}(2j_A+1)\prod\limits_{B=1}^{N_3}\exp
(i\sum\limits_{edges\;l}j_l\epsilon_l),$$
where $\epsilon_l$ is the deficit angle at the edge $l$. Considering the product over
the edge lengths as a sort of discretized measure we can interpret the last term in
the above formula as a statistical weight a la Feynmann $e^{iS}$ with a Regge type action $S=\sum
j_l\epsilon_l$. \par
In an holographic setting the wish is to decouple in (\ref{tutta}) bulk datas from boundary
ones and in order to complete this task, we first need to recognize that, in a 
fixed triangulation, the bulk 
pieces are described by the fluctuations of the $SU(2)$ assignements of the thetraedra 
completely inside the manifold whereas the remaining pieces describe the interaction
bulk-boundary and are given by those thetraedra having some component on the boundary. In particular, given a triangulation $T^3$, we can
always refine it in such a way that each thetraedra sharing with the boundary a certain number of faces, has actually only one face on $\partial T^3$;
this construction provides us with the so called "standard triangulation". In this framework it is easy to distinguish two types of components
contributing to the boundary datas: the previously mentioned thetraedra $\{\sigma_F\}$ sharing a face with $\partial T^3$ whose number will be denoted $N_3^F$ and those
thetraedra sharing a single edge $\{\sigma_E\}$ with $\partial T^3$ whose number will be $N_3^E$. These elements represent the coupling thetraedra.
In reality, in the spirit of the holographic scenario, in order to decouple the bulk/boundary datas, we should take a proper limit on some metric 
variable and this role in the PR model is played by those thetraedra with all the edges in the interior of $T^3$ but with a single vertex on $\partial T^3$; we shall
indicate them as $\{\sigma_V\}$ and their number will be running from $1$ to $N_3^V$.

The decoupling procedure implies that all the boundary component are kept fixed and the assignements of "bulk" edges are all rescaled by the same 
factor $R$. This clearly provides for each separate component a different asymptotic expression (see \cite{Arcioni} and references therein for further details):
\begin{itemize}
\item for the thetaedra $\sigma_F$ we have:
\begin{gather}
\{6j\}(\sigma_F,R)\equiv
\left\{ \begin{array}{ccc}
j_1 & j_2 & j_3 \\
J_1+R & J_2+R & J_3+R
\end{array}\right\}\:\:\:
\begin{array}{c}
\longrightarrow \\
R \gg 1
\end{array}\:\\
\frac{(-1)^{\Phi}}{\sqrt {2R}}\:
\left[ \begin{array}{ccc}
j_1 & j_2 & j_3 \\
\mu_1 & \mu_2 & \mu_3
\end{array}\right],
\label{astetra}
\end{gather}
where from now on the capital letters are referred to bulk assignments, the small to boundary assignments and $\Phi=j_1+j_2+j_3+2(J_1+J_2+J_3)$,
\item for the tetrahedra $\sigma_E$ we have:
\begin{gather}
\{6j\}(\sigma_E,R)\equiv
\left\{ \begin{array}{ccc}
J_1+R & j & J_2+R \\
J_3+R & J_4+R & J_5+R
\end{array}\right\}\:\:\:\\
\begin{array}{c}
\longrightarrow \\
R \gg 1
\end{array}\:
\frac{(-1)^{\Psi}}{2R}\,\:
d^j_{\nu_2\nu_3}(\theta),
\label{astetra1}
\end{gather}

\noindent where $\Psi=3J_1+j+2(J_2+J_3+J_4+J_5)+\nu_1$ and $\theta$ 
is the angle between the
the edge labelled by $j$ and the quantization axis,
\item for the tetrahedra $\sigma_V$ we have:
\begin{eqnarray}
\{6j\}(\sigma_V,R)\equiv
\left\{ \begin{array}{ccc}
J_1+R & J_2+R & J_2+R \\
J_4+R & J_5+R & J_6+R
\end{array}\right\}\:\:\:
\begin{array}{c}
\longrightarrow \\
R \gg 1
\end{array}\nonumber\\
\left( 12\pi {\cal V}(\sigma_V)\right)^{-1/2}\;
\exp \left\{\; i\left( \; \sum_{\alpha=1}^{6}l_{\alpha}\theta_{\alpha} + 
\pi/4 \right) \right\}, 
\label{asPR}
\end{eqnarray}

where ${\cal V}(\sigma_V)$ is the Euclidean volume of the tetrahedron  
spanned by the six edges
$\{l_{\alpha}\}$, $l_{\alpha}=J_{\alpha}+1/2$ and $\theta_{\alpha}$ 
is the 
angle between the outward normals to the faces
which share $l_{\alpha}$ (these angles can be obviously expressed in 
terms of the $J$'s).
\end{itemize}

\noindent Eventually the partition function associated with this configuration is
\begin{eqnarray}
\lefteqn{Z[(T^3,\partial T^3)_{st}
;R \gg 1;L]\,
=}\nonumber \\
& & \Lambda(L)^{-N_0}\,\:\prod_{1}^{N_1-N_1^*} (-1)^{2J} (2J+1)\,
\prod_{1}^{N_3-N_3^E-N_3^V} 
(-1)^{\sum J}\:\ \left\{6j\right\}(J)\nonumber \\
& & \prod_{1}^{N_1^*} (-1)^{2R} (2R)\:\:\Lambda(L)^{-n_0}\:\:
\prod_{1}^{n_1} (-1)^{2j}\:(2j+1)\nonumber \\
& & \prod_{1}^{n_2} \;(2R)^{-1/2}\:
(-1)^{\sum (J+R)+ \sum j+ \Phi}(-1)^{(\sum m)/2}
\left[3j\mu\right]\left[3jm\right]\nonumber \\
& & \prod_{1}^{N_3^E} (2R)^{-1}(-1)^{\sum (J+R)+ j+ \Psi}\:\:
d^j_{\nu_2\,\nu_3}(\theta)\nonumber \\
& & \prod_{1}^{N_3^V} \;(2R)^{-3/2}
(-1)^{\sum (J+R)}\:\exp\left\{i(\sum_{\alpha=1}^{6}l_{\alpha}
\theta_{\alpha} + \pi/4)\right\},
\label{Rstate}
\end{eqnarray}
where $N_1^*$ is the number of edges of those bulk thetraedra that are not part of the coupling datas.

The last step in order to completely determine the holographic decoupling is to recognize that the topological union of $\{\sigma_V\}\cup\{\sigma_E\}\cup\{\sigma_F\}$
fills in a thick shell of the order of the decoupling parameter $R$ close to the boundary $\partial T^3$; we are thus entitled to introduce the
triangulation $\tilde{T}^3\doteq\left(Int T^3\right)- \{\sigma_V\}\cup\{\sigma_E\}\cup\{\sigma_F\}$ and the fixed 2-dimensional triangulation $\Sigma^{in}(R)$ closing up 
$\tilde{T}^3$. The couple $(\tilde{T}^3,\Sigma^{in}(R))$ is topologically equivalent to $(T^3,\partial T^3)$ and moreover $\Sigma^{in}(R)$ represents
a sort of "inner boundary" whose intersection with $\partial T^3$ is empty. If we now introduce the set $N_i(\Sigma^{in}(R))$ of i-simplices in $\Sigma^{in}(R)$, we can associate
to $(\tilde{T}^3,\Sigma^{in}(R)$, the partition function:    
\begin{eqnarray}
\lefteqn{Z[({\tilde T}^3,\Sigma^{in}(R));L]\,
=}\hspace{.5in}\nonumber \\
& & \Lambda(L)^{-(N_0-N_0(\Sigma^{in}))}\:\,\prod_{1}^{N_1 - N^*_1}\; 
(-1)^{2J} (2J+1)\,\prod_{1}^{N_3} 
(-1)^{\sum J}\:\ \left\{6j\right\}(J)  \nonumber \\
& & \prod_{1}^{N_1(\Sigma^{in})} (-1)^{R} (2R)^{1/2}.
\label{bulkst}
\end{eqnarray}
In the end, 
the partition function (\ref{Rstate}) can formally be splitted in three parts:
$$Z=Z_{bulk} \mathcal{P}_{bulk\to\; bound} Z_{hol},$$
where $Z_{bulk}=Z[({\tilde T}^3,\Sigma^{in}(R));L]$, $\mathcal{P}_{bulk\to\; bound}$ is a projection map connecting
the inner boundary with $\partial T^3$ and finally 
\begin{eqnarray}
\lefteqn{Z^{hol}\:
[\Sigma^{out}\,;L]\,
=}\hspace{.7in}\nonumber \\
& & \Lambda(L)^{-n_0-N_0(\Sigma^{in})}\:
\prod_{1}^{n_1} (-1)^{2j}\:(2j+1)\nonumber \\
& &   \prod_{1}^{n_2} \:
(-1)^{\sum J+ \sum j+ \Phi}\:\,(-1)^{(\sum m)/2}
\left[3j\mu\right]
\:\,\: \left[3jm\right]\nonumber\\
& & \prod_{1}^{N_3^E}(-1)^{\sum J+ j+ \Psi}\:\:
d^j_{\nu_2\,\nu_3}(\theta)\nonumber \\
& & \prod_{1}^{N_3^V} \;
(-1)^{\sum J}\:\exp\left\{i(\sum_{\alpha=1}^{6}l_{\alpha}
\theta_{\alpha} + \pi/4)\right\},
\label{holostate}
\end{eqnarray}
represents the functional that has to be associated with the holographic partition function.
\par The first result which comes from the above decoupling is the non
topological origin of (\ref{holostate}) since it is not invariant under Pachner
moves; from one side this does not allow us to easily recognize the nature of
the above functional but on the other side this remark is rather encouraging in
the spirit of an holographic correspondence. Moreover (\ref{holostate}) shows a
residual dependance from bulk datas and more precisely from the asymptotic
beahviour of bulk (gravitational) fluctuations. This is in some sense reminding
us the AdS/CFT correspondence where the source terms for the boundary theory
couple a boundary operator to the asymptotic beahviour of the bulk fields. A
deeper look at (\ref{holostate}) allow us to give prominence to the presence of
a pair of 3jm symbols for each vertex with a Wigner symbol connecting them. From
one side, this
suggests that the boundary functional could possibly be associated to fat
trivalent graph which are ultimately dual to 2-dimensional triangulations. 
From the other side the presence of 3jm symbols in the boundary partition function
encourages us to carry on in the direction of finding a relation between 
(\ref{holostate}) and a conformal field theory since it is a standard result
from Moore and Seiberg \cite{Moore2} that it exists a sort of dictionary between CFT and
group theory. In particular it is known that the Clebsh-Gordan coefficients
(thus the 3jm symbols) for a compact Lie group $G$ are in one to one
correspondence with the chiral vertex operator of the corresponding
$G$-invariant conformal theory. Following this line it is also possible to show
that the Racah coefficients are associated with the fusion matrices which in the
peculiar case of the $SU(2)$ group are exactly the 6j symbols. Thus, within this
scheme, it is natural to interpret the functions of the group elements as
physical fields in a corresponding CFT and the product of these functions as the
standard operator product expansion. Moreover, considering the deep relation
existing between Chern-Simons and WZW models, it is also natural to
suppose that the boundary theory associated to the system that we are studying is
exactly a sort of discretized version of an SU(2) WZW conformal field theory. 
It is thus intriguing to pursuit
this idea trying to relate (\ref{holostate}) to a WZW action but unfortunately
this line of research cannot be swiftly exploited since at present a coherent and
universally accepted definition of a conformal field theory over a triangulated
surface is unknown. The reason of such "deficency" can be tracked in the
underlying big difference between the two models since the discretized
rely upon a combinatorial approach which is difficult to apply in the
setting of a conformal field theory that is more analytic in spirit.
Nonetheless in this thesis we will show that it is possible to circumvent such
problem exploting the above mentioned correspondence between triangulated
surfaces and ribbon graphs. Roughly speaking, starting from a standard
2-simplicial complex, we will associate to it a dual graph which is ultimately
defined through a collection of vertices and edges connecting them. In this
simple geometrical construction, that we will outline more in detail in the next
chapter, the underlying surface will be "divided" in a collection of closed two
dimensional cells; the main feature of this operation will be the chance to
define in each cell a uniformizing complex coordinate and a unique quadratic
differential (and as a consequence a metric). This result will allow us to
switch from the purely combinatorial approach of the simplicial subsdivision of
the Riemann surface to a setting which has the twofold advantage from one side 
to retain the datas from the purely discretized approach and from the other side
to introduce analytic tools which are more suitable if we wish to work in
the realm of a conformal field theory. Starting from this basis and keeping in mind
the correspondence between group theory and CFT outlined mainly by Moore and Seiberg, 
in this thesis we will adress the problem of the
definition of a conformal field theory on a triangulated polyhedron using the
techniques coming from the "graph approach" to the discretized setting; nonetheless, bearing in
mind that our ultimate goal is to study the holographic correspondence in the
Ponzano-Regge calculus and to gain some insight on the true nature of the
boundary functional (\ref{holostate}), we will specialize our analysis on the
definition of a WZW model on a triangulated Riemann surface with the gauge group
$SU(2)$. In particular, following the works of Gaberdiel \cite{Gaberdiel}, we will be
able also to explicitly write the partition function for the subcase of a WZW model
with gauge group $SU(2)$ at level 1 and eventually to compare this result with
(\ref{holostate}) showing striking similiraties between both formulas.

\section{Outline of the thesis}
As we have mentioned several times in the previous sections, the main purpose of this
thesis is to study different physical systems where the holographic principle appears
to be realized in a way completely different from Maldacena approach where
"holography" is a synonym of a gauge/gravity correspondence that associate to a field
theory in a curved background on a bulk manifold $M$, a gauge theory without gravity living 
on the boundary of $M$. Bearing this in mind and following the
motivations of section 1.3 in the next chapter, we study a two dimensional
triangulated surface; the leading line will be to introduce several mathematical tools
that we will exploit in chapter 3 in order to introduce a WZW model. In particular as
we have outlined in the previous section, the natural approach in this scenario is to
switch from a triangulation to its dual graph; thus in section 2.1 we start
introducing
canonical concepts concerning Regge triangulations and in particular we outline
the beautiful Troyanov formulation of the Gauss-Bonnet theorem for triangulated
manifolds and its implications. The key concepts of this chapter will be outlined
in section 2.2 and 2.3 where we will introduce the barycentric subdivision of a
two dimensional triangulation $T$. With this geometric construction we will be able to
associate to $T$ a dual trivalent graph $P_T$ and we will show how it is possible to
parametrize through ribbon graphs the moduli space of Riemann surface; moreover we
apply Strebel theory of quadratic differential in the context of triangulated surfaces
emphasizing the role of uniformizing complex coordinates. In this section as
anticipated before we will move from a purely combinatorial language to a more
analytic approach endowing the dual graph with a complex structure that will play a
pivotal role in the construction of a conformal field theory. The full extent of this
change of "language" will be emphasized in chapter 3 which represents the main core
of this thesis together with chapter 4. In particular, starting from a Regge polytope
we outline in section 3.1 the construction of a WZW model. Since we
always bear in mind that our ultimate goal is to apply our results in the context of
Ponzano-Regge calculus, we specialize our results to the group $SU(2)$. Starting from
this point we will be able to achieve a twofold result: in section 3.2 we construct
the Hilbert space for the theory with $G=SU(2)$ at level 1. In particular, since to
each dual cell of the dual polytope, is associated a separate Hilbert space, we
explicitly describe how they are related introducing insertion operators associated to
the ribbon graph that allow to switch from states from one cell into another. In
section 3.3 instead we explicitly construct the partition function for the WZW model
with group $SU(2)_1$ with the hope to relate the result with (\ref{holostate}). In the
end we find striking similarities between the two formulas but we are not still able
to make a full correspondence.

In the fourth chapter, instead, we switch from the study of a discretized system in
three dimensions to asymptotically flat space-times. As in the previous chapters, our
ultimate goal is to provide new insights in the realization of the holographic
principle in this framework. As for the Ponzano-Regge scenario, we study mainly the
asymptotic dynamics of the theory whose informations in a flat background are encoded 
in the asymptotic symmetry group, the Bondi-Metzner-Sachs group. For this reason in
section 4.1 we review the derivation and the main features of this group. In
particular we emphasize its twofold nature as an asymptotic symmetry group from one
side and on the other side as the intrinsic symmetry group of the manifold $\Im$.
Starting from this remark and bearing in mind Bousso covariant entropy conjecture, we
address in section 4.2 the problem of a geometric reconstrucion of the bulk space-time starting from
datas encoded in $\Im$ ultimately finding that, due to high focusing of light rays,
a geometric holography is possible only within a restricted class of manifolds, namely
the stationary ones. Thus this result forces us to switch our attention to the quantum
level in order to implement the holographic principle and the main problem that we
face is the absence of a BMS field theory. For this reason our first priority will be
to determine BMS wave equations and we will follow a purely group theoretical way;
after extensively reviewing in section 4.4 McCarthy theory of representation, we
present in section 4.5 our main result which is the construction of BMS wave equations
with the a method similar to the one used by Wigner for the Poincar\'e group. Moreover
we will emphasize the asymptotic role of these fields, i.e. they evolve only on $\Im$
and they never propagate in the bulk; this remark will suggest us to pursuit a road
similar to 't Hooft ansatz of introducing an S-matrix in order to describe propagation
of informations from $\Im^+$ to $\Im^-$. Furthermore the structure of the fields
equations will allow us to conjecture that any holographic correspondence in asymptotically
flat space-times has to show an high degree of non locality.\\
Finally in the last chapter, after reviewing our main results, we discuss open
questions mainly referring to the holographic principle and we present some further
relation between our work and different approaches. In the end our aim will be also to
briefly discuss some possible future line of research which could have this thesis as
a starting point.  \par
The content and the figures available in this thesis have been taken in part
from the following papers \cite{carfora2}, \cite{carfora4}, \cite{Arcioni2}.

\chapter{Ribbon graphs and Random Regge Triangulations}
The aim of this chapter will be mainly to describe the deep relation existing between random Regge triangulations
and trivalent ribbon graphs. We will follow the lines of \cite{carfora} and \cite{carfora2} even though we will not concentrate on
the description of the modular aspects of two dimensional gravity; instead we will emphasize the role of the uniformizing coordinates and of the 
Strebel theorem as the main tools needed in order to define a conformal field theory over a triangulated Riemann surface. In essence this chapter should
be seen as preparatory for an application to the WZW model and for the study of the relation between CS/WZW in a discretized setting.
\section{Triangulated surfaces and Polytopes}

\begin{figure}[ht]
\begin{center}
\includegraphics[bb= 0 0 520 450,scale=.4]{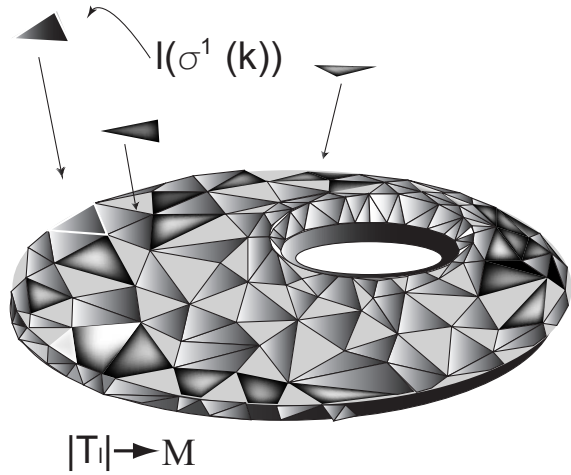}
\caption{A torus triangulated with triangles of variable edge-length.}\label{fig1}
\end{center}
\end{figure}

Let $T$ denote a $2$-dimensional simplicial complex with underlying
polyhedron $|T|$ and $f$- vector $(N_{0}(T),N_{1}(T),N_{2}(T))$, where $N_{i}(T)\in \mathbb{N}$ 
is the number of $i$-dimensional sub-simplices 
$\sigma ^{i}$ of $T$. Given a simplex $\sigma $ we denote by $st(\sigma )$,
(the star of $\sigma $), the union of all simplices of which $\sigma $ is a
face, and by $lk(\sigma )$, (the link of $\sigma $), is the union of all
faces $\sigma ^{f}$ of the simplices in $st(\sigma )$ such that $\sigma
^{f}\cap \sigma =\emptyset $. A Regge triangulation of a $2$
-dimensional PL manifold $M$, (without boundary), is a homeomorphism 
$|T_{l}|\rightarrow {M}$ where each face of $T$ is geometrically
realized by a rectilinear simplex of variable edge-lengths $l(\sigma
^{1}(k))$. A dynamical triangulation $|T_{l=a}|\rightarrow {M}$ is a
particular case of a Regge PL-manifold realized by rectilinear and
equilateral simplices of edge-length $l(\sigma ^{1}(k))=a$ (see figure \ref{fig1}). 

\begin{figure}[ht] 
\begin{center}
\includegraphics[bb= 0 0 480 440,scale=.4]{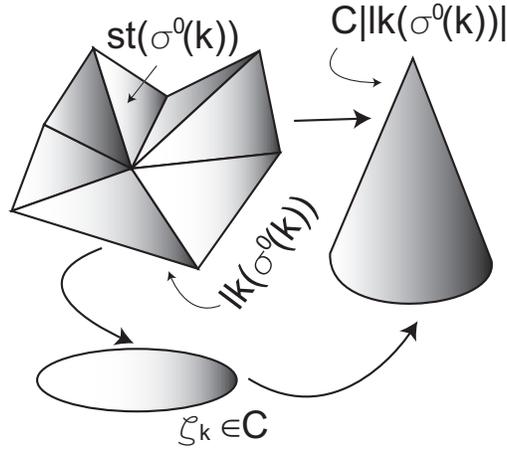}
\caption{The geometric structures around a vertex.}\label{fig2}
\end{center}
\end{figure}

The metric
structure of a Regge triangulation is locally Euclidean everywhere except
at the vertices $\sigma ^{0}$, (the \textit{bones}), where the sum of the
dihedral angles, $\theta (\sigma ^{2})$, of the incident triangles $\sigma
^{2}$'s is in excess (negative curvature) or in defect (positive curvature)
with respect to the $2\pi $ flatness constraint. The corresponding deficit
angle $\varepsilon $ is defined by $\varepsilon =2\pi -\sum_{\sigma
^{2}}\theta (\sigma ^{2})$, where the summation is extended to all $2$%
-dimensional simplices incident on the given bone $\sigma ^{0}$. If $K_{T}^{0}$ 
denotes the $(0)$-skeleton of $|T_{l}|\rightarrow {M}$, (\emph{i.e.}, the collection of vertices of the triangulation), 
then $M\backslash{K_{T}^{0}}$ is a flat Riemannian manifold, and any point in the interior of
an $r$- simplex $\sigma ^{r}$ has a neighborhood homeomorphic to $B^{r}\times {C}(lk(\sigma ^{r}))$, 
where $B^{r}$ denotes the ball in $\mathbb{R}^{n}$ and ${C}(lk(\sigma ^{r}))$ is the cone over the link 
$lk(\sigma ^{r})$, (the product $lk(\sigma ^{r})\times \lbrack 0,1]$ with 
$lk(\sigma ^{r})\times \{1\}$ identified to a point). In particular, let
us denote by $C|lk(\sigma ^{0}(k))|$ the cone over the link of the vertex 
$\sigma ^{0}(k)$. On any such a disk $C|lk(\sigma ^{0}(k))|$ we can introduce
a locally uniformizing complex coordinate $\zeta (k)\in \mathbb{C}$ in terms
of which we can explicitly write down a conformal conical metric locally
characterizing the singular structure of $|T_{l}|\rightarrow M$, \emph{viz.%
}, 
\begin{equation}
e^{2u}\left| \zeta (k)-\zeta _{k}(\sigma ^{0}(k))\right| ^{-2\left( \frac{%
\varepsilon (k)}{2\pi }\right) }\left| d\zeta (k)\right| ^{2},  \label{cmetr}
\end{equation}
where $\varepsilon (k)$ is the corresponding deficit angle, and
$u:B^{2}\rightarrow \mathbb{R}$ is a continuous function ($C^{2}$ on 
$B^{2}-\{\sigma ^{0}(k)\}$) such that, for $\zeta (k)\rightarrow \zeta
_{k}(\sigma ^{0}(k))$, we have $\left| \zeta (k)-\zeta _{k}(\sigma
^{0}(k))\right| \frac{\partial u}{\partial \zeta (k)}$, and $\left| \zeta
(k)-\zeta _{k}(\sigma ^{0}(k))\right| \frac{\partial u}{\partial \overline{%
\zeta }(k)}$ both $\rightarrow 0$, \cite{troyanov}. Up to the presence of $e^{2u}$, we
immediately recognize in such an expression the metric $g_{\theta (k)}$ of a
Euclidean cone of total angle $\theta (k)=2\pi -\varepsilon (k)$. The factor 
$e^{2u}$ allows to move within the conformal class of all metrics possessing
the same singular structure of the triangulated surface $|T_{l}|\rightarrow M$. 
We can profitably shift between the PL and the function theoretic point
of view by exploiting standard techniques of complex analysis, and making
contact with moduli space theory (see figure \ref{fig2}).

\subsection{Curvature assignments and divisors.} In the case of
dynamical triangulations, the picture simplifies considerably since the
deficit angles are generated by the numbers $\#\{\sigma ^{2}(h)\bot \sigma
^{0}(i)\}$\ of triangles incident on the $N_{0}(T)$ vertices, the \textit{%
curvature assignments}, $\{q(k)\}_{k=1}^{N_{0}(T)}\in \mathbb{N}^{N_{0}(T)}$,

\begin{equation}
q(i)=\frac{2\pi -\varepsilon (i)}{\arccos (1/2)}.  \label{curvat}
\end{equation}
For a regular triangulation we have $q(k)\geq 3$, and since each triangle
has $3$ vertices $\sigma ^{0}$, the set of integers $\{q(k)%
\}_{k=1}^{N_{0}(T)}$ is constrained by

\begin{equation}
\sum_{k=1}^{N_{0}}q(k)=3N_{2}=6\left[ 1-\frac{\chi (M)}{N_{0}(T)}\right]
N_{0}(T),  \label{vincolo}
\end{equation}
where $\chi (M)$ denotes the Euler-Poincar\'{e} characteristic of the
surface, and where $6\left[ 1-\frac{\chi (M)}{N_{0}(T)}\right] $, ($\simeq 6$
for $N_{0}(T)>>1$),\ is the average value of the curvature assignments $%
\{q(k)\}_{k=1}^{N_{0}}$. More generally we shall consider semi-simplicial
complexes for which the constraint $q(k)\geq 3$ is removed. Examples of such
configurations are afforded by triangulations with pockets, where two
triangles are incident on a vertex, or by triangulations where the star of a
vertex may contain just one triangle. We shall refer to such extended
configurations as generalized (Regge and dynamical) triangulations.

The singular structure of the metric defined by (\ref{cmetr}) can be
naturally summarized in a formal linear combination of the points $\{\sigma
^{0}(k)\}$ with coefficients given by the corresponding deficit angles
(normalized to $2\pi $), in the \emph{real divisor }\cite{troyanov} 
\begin{equation}
Div(T)\doteq \sum_{k=1}^{N_{0}(T)}\left( -\frac{\varepsilon (k)}{2\pi }%
\right) \sigma ^{0}(k)=\sum_{k=1}^{N_{0}(T)}\left( \frac{\theta (k)}{2\pi }%
-1\right) \sigma ^{0}(k)
\end{equation}
supported on the set of bones $\{\sigma ^{0}(i)\}_{i=1}^{N_{0}(T)}$. Note
that the degree of such a divisor, defined by 
\begin{equation}
\left| Div(T)\right| \doteq \sum_{k=1}^{N_{0}(T)}\left( \frac{\theta (k)}{2\pi }-1\right) =-\chi (M)  \label{rediv}
\end{equation}
is, for dynamical triangulations, a rewriting of the combinatorial
constraint (\ref{vincolo}). In such a sense, the pair $(|T_{l=a}|\rightarrow
M,Div(T))$, or shortly, $(T,Div(T))$, encodes the datum of the triangulation 
$|T_{l=a}|\rightarrow M$ and of a corresponding set of curvature assignments 
$\{q(k)\}$ on the vertices $\{\sigma ^{0}(i)\}_{i=1}^{N_{0}(T)}$. The real
divisor $\left| Div(T)\right| $ characterizes the Euler class of the pair 
$(T,Div(T))$ and yields for a corresponding Gauss-Bonnet formula.
Explicitly, the Euler number associated with $(T,Div(T))$ is defined, \cite{troyanov},
by

\begin{equation}
e(T,Div(T))\doteq \chi (M)+|Div(T)\mathbf{|.}  \label{euler}
\end{equation}
and the Gauss-Bonnet formula reads \cite{troyanov}:

\begin{lemma}
(\textbf{Gauss-Bonnet for triangulated surfaces}) Let $(T,Div(T))$ be a
triangulated surface with divisor 
\begin{equation}
Div(T)\doteq \sum_{k=1}^{N_{0}(T)}\left( \frac{\theta (k)}{2\pi }-1\right)
\sigma ^{0}(k),
\end{equation}
associated with the vertices $\{\sigma ^{0}(k)\}_{k=1}^{N_{0}(T)}$. Let $ds^{2}$ 
be the conformal metric (\ref{cmetr}) representing the divisor $Div(T)$ . Then 
\begin{equation}\label{Euclass}
\frac{1}{2\pi }\int_{M}KdA=e(T,Div(T)),  
\end{equation}
where $K$ and $dA$ respectively are the curvature and the area element
corresponding to the local metric $ds_{(k)}^{2}.$
\end{lemma}

Note that such a theorem holds for any singular Riemann surface $\Sigma $
described by a divisor $Div(\Sigma )$ and not just for triangulated surfaces
\cite{troyanov}. Since for a Regge (dynamical) triangulation, we have 
$e(T_{a},Div(T))=0$, the Gauss-Bonnet formula implies

\begin{equation}
\frac{1}{2\pi }\int_{M}KdA=0.  \label{GaussB}
\end{equation}
Thus, a triangulation $|T_{l}|\rightarrow M$ naturally carries a conformally
flat structure. Clearly this is a rather obvious result, (since the metric
in $M-\{\sigma ^{0}(i)\}_{i=1}^{N_{0}(T)}$ is flat). However, it admits a
not-trivial converse (recently proved by M. Troyanov, but, in a sense, going
back to E. Picard) \cite{troyanov}, \cite{picard}:

\begin{theorem}
(\textbf{Troyanov-Picard}) Let $(\left( M,\mathcal{C}_{sg}\right) ,Div)$ be
a singular Riemann surface with a divisor such that $e(M,Div)=0$. Then there
exists on $M$ a unique (up to homothety) conformally flat metric
representing the divisor $Div$.
\end{theorem} 

\subsection{Conical Regge polytopes.}
\begin{figure}
\begin{center}
\includegraphics[bb= 0 0 280 530,scale=.3]{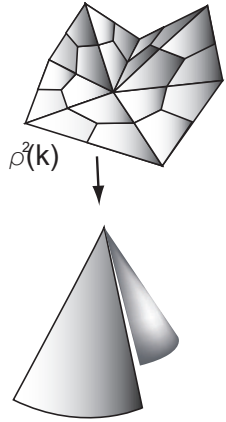}
\caption{The conical geometry of the baricentrically dual polytope.}\label{fig3}
\end{center}
\end{figure}

Let us consider the (first) barycentric subdivision of $T$. The closed stars, in such a subdivision,
of the vertices of the original triangulation $T_{l}$ form a collection of 
$2$-cells $\{\rho ^{2}(i)\}_{i=1}^{N_{0}(T)}$ characterizing the \emph{%
conical} Regge polytope $|P_{T_{l}}|\rightarrow {M}$ (and its rigid
equilateral specialization $|P_{T_{a}}|\rightarrow {M}$) barycentrically
dual to $|T_{l}|\rightarrow {M}$. If $(\lambda (k),\chi (k))$ denote polar
coordinates (based at $\sigma ^{0}(k)$) of a point $p\in \rho ^{2}(k)$, then 
$\rho ^{2}(k)$ is geometrically realized as the space 
\begin{equation}
\left. \left\{ (\lambda (k),\chi (k))\ :\lambda (k)\geq 0;\chi (k)\in 
\mathbb{R}/(2\pi -\varepsilon (k))\mathbb{Z}\right\} \right/ (0,\chi
(k))\sim (0,\chi ^{\prime }(k))
\end{equation}
endowed with the metric 
\begin{equation}
d\lambda (k)^{2}+\lambda (k)^{2}d\chi (k)^{2},
\end{equation}
as it can be seen in figure \ref{fig3}.

In other words, here we are not considering a rectilinear presentation of
the dual cell complex $P$ (where the PL-polytope is realized by flat
polygonal $2$-cells $\{\rho ^{2}(i)\}_{i=1}^{N_{0}(T)}$) but rather a
geometrical presentation $|P_{T_{l}}|\rightarrow {M}$ of $P$ where the $2$
-cells $\{\rho ^{2}(i)\}_{i=1}^{N_{0}(T)}$ retain the conical geometry
induced on the barycentric subdivision by the original metric (\ref{cmetr})
structure of $|T_{l}|\rightarrow {M}$.

\subsection{Hyperbolic cusps and cylindrical ends.}\label{degenerations}
\begin{figure}[ht]
\begin{center}
\includegraphics[bb= 0 0 450 480,scale=.3]{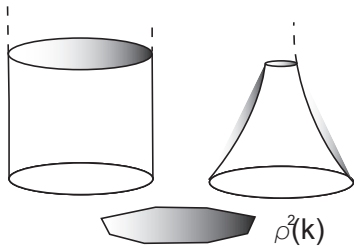}
\caption{The cylindrical and hyperbolic metric over a $\theta\to 0$ degenerating polytopal cell.}\label{fig4}
\end{center}
\end{figure}

It is important to stress that whereas a Regge triangulation characterizes
a unique (up to automorphisms) singular Euclidean structure, this latter
actually allows for a more general type of metric triangulation. The point
is that some of the vertices associated with a singular Euclidean structure
can be characterized by deficit angles $\varepsilon (k)\rightarrow 2\pi$\emph{i.e.}, 
$\sum_{\sigma ^{2}(k)}\theta (\sigma ^{2}(k))=0$.
Such a situation
corresponds to having the cone $C|lk(\sigma ^{0}(k))|$ over the link $%
lk(\sigma ^{0}(k))$ realized by a Euclidean cone of angle $0$. This is a
natural limiting case in a Regge triangulation, (think of a vertex where
many long and thin triangles are incident), and it is usually discarded as
an unwanted pathology. However, there is really nothing pathological about
that, since the corresponding $2$-cell $\rho ^{2}(k)\in
|P_{T_{l}}|\rightarrow {M}$ can be naturally endowed with the conformal
Euclidean structure obtained from (\ref{cmetr}) by setting $\frac{%
\varepsilon (k)}{2\pi }=1$, \emph{i.e. } 
\begin{equation}
e^{2u}\left| \zeta (k)-\zeta _{k}(\sigma ^{0}(k))\right| ^{-2}\left| d\zeta
(k)\right| ^{2},
\end{equation}
which (up to the conformal factor $e^{2u}$) is the flat metric on the
half-infinite cylinder $\mathbb{S}^{1}\times \mathbb{R}^{+}$ (a cylindrical
end, see figure \ref{fig4}). 

Alternatively, one may consider $\rho ^{2}(k)$ endowed with the
geometry of a hyperbolic cusp, \emph{i.e.}, that of a half-infinite
cylinder $\mathbb{S}^{1}\times \mathbb{R}^{+}$ equipped with the hyperbolic
metric $\lambda (k)^{-2}(d\lambda (k)^{2}+d\chi (k)^{2})$. The triangles
incident on $\sigma ^{0}(k)$ are then realized as hyperbolic triangles with
the vertex $\sigma ^{0}(k)$ located at $\lambda (k)=\infty $ and
corresponding angle $\theta _{k}=0$\cite{judge}. Since the Poincar\'{e} metric on the
punctured disk $\{\zeta (k)\in C|\;0<\left| \zeta (k)-\zeta _{k}(\sigma
^{0}(k))\right| <1\}$ is 
\begin{equation}
\left( \left| \zeta (k)-\zeta _{k}(\sigma ^{0}(k))\right| \ln \frac{1}{%
\left| \zeta (k)-\zeta _{k}(\sigma ^{0}(k))\right| }\right) ^{-2}\left|
d\zeta (k)\right| ^{2},
\end{equation}
one can shift from the Euclidean to the hyperbolic metric by setting 
\begin{equation}
e^{2u}=\left( \ln \frac{1}{\left| \zeta (k)-\zeta _{k}(\sigma
^{0}(k))\right| }\right) ^{-2},
\end{equation}
and the two points of view are strictly related. 
At any rate the presence of
hyperbolic cusps or cylindrical ends is consistent with a singular
Euclidean structure as long as the associated divisor satisfies the
topological constraint (\ref{rediv}),\emph{\ }which we can rewrite as 
\begin{equation}
\sum_{\{\frac{\varepsilon (k)}{2\pi }\neq 1\}}\left( -\frac{\varepsilon (k)}{
2\pi }\right) =2g-2+\#\left\{ \sigma ^{0}(h)|\;\frac{\varepsilon (h)}{2\pi }
=1\right\} .
\end{equation}
In particular, we can have the limiting case of the singular Euclidean
structure associated with a genus $g$ surface triangulated with $N_{0}-1$
hyperbolic vertices $\{\sigma ^{0}(k)\}_{k=1}^{N_{0}-1}$ (or, equivalently,
with $N_{0}-1$ cylindrical ends) and just one standard conical vertex, 
$\sigma ^{0}(N_{0})$, supporting the deficit angle 
\begin{equation}\label{lastmarked}
-\frac{\varepsilon (N_{0})}{2\pi }=2g-2+(N_{0}-1).  
\end{equation}

\section{Ribbon graphs on Regge Polytopes}

\begin{figure}[ht]
\begin{center}
\includegraphics[bb= 0 0 450 580,scale=.3]{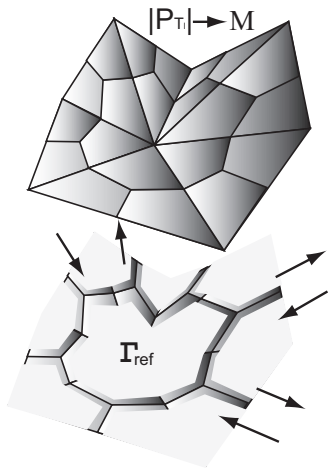}
\caption{The dual polytope around a vertex and it's edge refinement.}\label{fig7}
\end{center}
\end{figure}

The geometrical realization of the $1$-skeleton of the conical Regge
polytope $|P_{T_{l}}|\rightarrow {M}$ is a $3$-valent graph 
\begin{equation}
\Gamma =(\{\rho ^{0}(k)\},\{\rho ^{1}(j)\})
\end{equation}
where the vertex set $\{\rho ^{0}(k)\}_{k=1}^{N_{2}(T)}$ is identified with
the barycenters of the triangles $\{\sigma ^{o}(k)\}_{k=1}^{N_{2}(T)}\in
|T_{l}|\rightarrow M$, whereas each edge $\rho ^{1}(j)\in \{\rho
^{1}(j)\}_{j=1}^{N_{1}(T)}$ is generated by two half-edges $\rho ^{1}(j)^{+}$
and $\rho ^{1}(j)^{-}$ joined through the barycenters $\{W(h)%
\}_{h=1}^{N_{1}(T)}$ of the edges $\{\sigma ^{1}(h)\}$ belonging to the
original triangulation $|T_{l}|\rightarrow M$. If we formally introduce a
ghost-vertex of a degree $2$ at each middle point $\{W(h)\}_{h=1}^{N_{1}(T)}$%
, then the actual graph naturally associated to the $1$-skeleton of 
$|P_{T_{l}}|\rightarrow {M}$ is the edge-refinement \cite{mulase} of $\Gamma =(\{\rho^{0}(k)\},\{\rho ^{1}(j)\})$, \emph{i.e.} 
\begin{equation}\label{Gamma}
\Gamma _{ref}=\left( \{\rho^{0}(k)\}\bigsqcup_{h=1}^{N_{1}(T)}\{W(h)\},\{\rho^{1}(j)^{+}\}\bigsqcup_{j=1}^{N_{1}(T)}\{\rho ^{1}(j)^{-}\}\right),
\end{equation}
as it can be seen in figure \ref{fig7}.\\

The natural automorphism group $Aut(P_{l})$ of \ $|P_{T_{l}}|\rightarrow {M}$,
(\emph{i.e.}, the set of bijective maps $\Gamma =(\{\rho ^{0}(k)\},\{\rho
^{1}(j)\})\rightarrow \widetilde{\Gamma }=(\widetilde{\{\rho ^{0}(k)\}},
\widetilde{\{\rho ^{1}(j)\}}$ preserving the incidence relations defining
the graph structure), is the automorphism group of its edge refinement \cite{mulase}, 
$Aut(P_{l})\doteq Aut(\Gamma _{ref})$. The locally uniformizing complex
coordinate $\zeta (k)\in \mathbb{C}$ in terms of which we can explicitly
write down the singular Euclidean metric (\ref{cmetr}) around each vertex $%
\sigma ^{0}(k)\in $ $|T_{l}|\rightarrow M$, provides a (counterclockwise)
orientation in the $2$-cells of $|P_{T_{l}}|\rightarrow {M}$. Such an
orientation gives rise to a cyclic ordering on the set of half-edges $\{\rho
^{1}(j)^{\pm }\}_{j=1}^{N_{1}(T)}$ incident on the vertices $\{\rho
^{0}(k)\}_{k=1}^{N_{2}(T)}$. According to these remarks, the $1$-skeleton
of $|P_{T_{l}}|\rightarrow {M}$ is a ribbon (or fat) graph \cite{ambjorn}, 
a graph $\Gamma $ together with a cyclic ordering on the set of half-edges
incident to each vertex of $\Gamma$. Conversely, any ribbon graph $\Gamma 
$ characterizes an oriented surface $M(\Gamma )$ with boundary possessing $%
\Gamma $ as a spine, (\emph{i.e.}, the inclusion $\Gamma \hookrightarrow
M(\Gamma )$ is a homotopy equivalence). In this way (the edge-refinement of)
the $1$-skeleton of a generalized conical Regge polytope $%
|P_{T_{l}}|\rightarrow {M}$ is in a one-to-one correspondence with trivalent
metric ribbon graphs. The set of all such trivalent ribbon graph $\Gamma $
with given edge-set $e(\Gamma )$ can be characterized \cite{mulase}, \cite{looijenga} as a space
homeomorphic to $\mathbb{R}_{+}^{|e(\Gamma )|}$, ($|e(\Gamma )|$ denoting
the number of edges in $e(\Gamma )$), topologized by the standard $%
\epsilon $-neighborhoods $U_{\epsilon }\subset $ $\mathbb{R}%
_{+}^{|e(\Gamma )|}$. The automorphism group $Aut(\Gamma )$ acts naturally
on such a space via the homomorphism $Aut(\Gamma )\rightarrow \mathfrak{G}%
_{e(\Gamma )}$, where $\mathfrak{G}_{e(\Gamma )}$ denotes the symmetric group
over $|e(\Gamma )|$ elements, and the resulting quotient space $\mathbb{R}%
_{+}^{|e(\Gamma )|}/Aut(\Gamma )$ is a differentiable orbifold.

\subsection{The space of 1-skeletons of Regge polytopes.}
Let $Aut_{\partial }(P_{l})\subset Aut(P_{l})$, denote the subgroup of ribbon
graph automorphisms of the (trivalent) $1$-skeleton $\Gamma $ of $%
|P_{T_{l}}|\rightarrow {M}$ that preserve the (labeling of the) boundary
components of $\Gamma $. Then, the space $K_{1}RP_{g,N_{0}}^{met}$ of $1$%
-skeletons of conical Regge polytopes $|P_{T_{l}}|\rightarrow {M}$, with $%
N_{0}(T)$ labelled boundary components, on a surface $M$ of genus $g$ can be
defined by \cite{mulase}
\begin{equation}
K_{1}RP_{g,N_{0}}^{met}=\bigsqcup_{\Gamma \in RGB_{g,N_{0}}}\frac{\mathbb{R}%
_{+}^{|e(\Gamma )|}}{Aut_{\partial }(P_{l})},  \label{DTorb}
\end{equation}
where the disjoint union is over the subset of all trivalent ribbon graphs
(with labelled boundaries) satisfying the topological stability condition $%
2-2g-N_{0}(T)<0$, and which are dual to generalized triangulations. It
follows, (see \cite{mulase} theorems 3.3, 3.4, and 3.5), that the set $%
K_{1}RP_{g,N_{0}}^{met}$ is locally modelled on a stratified space
constructed from the components (rational orbicells) $\mathbb{R}%
_{+}^{|e(\Gamma )|}/Aut_{\partial }(P_{l})$ by means of a (Whitehead)
expansion and collapse procedure for ribbon graphs, which amounts to
collapsing edges and coalescing vertices, (the Whitehead move in $%
|P_{T_{l}}|\rightarrow {M}$ is the dual of the familiar flip move \cite{ambjorn} for
triangulations). Explicitly, if $l(t)=tl$ is the length of an edge $\rho
^{1}(j)$ of a ribbon graph $\Gamma _{l(t)}\in $ $K_{1}RP_{g,N_{0}}^{met}$,
then, as $t\rightarrow 0$, we get the metric ribbon graph $\widehat{\Gamma }$
which is obtained from $\Gamma _{l(t)}$ by collapsing the edge $\rho ^{1}(j)$%
. By exploiting such construction, we can extend the space $%
K_{1}RP_{g,N_{0}}^{met}$ to a suitable closure $\overline{K_{1}RP}%
_{g,N_{0}}^{met}$ \cite{looijenga}, (this natural topology on $K_{1}RP_{g,N_{0}}^{met}$
shows that, at least in two-dimensional quantum gravity, the set of Regge
triangulations with \emph{fixed connectivity} does not explore the full
configurational space of the theory). The open cells of 
$K_{1}RP_{g,N_{0}}^{met}$, being associated with trivalent graphs, have
dimension provided by the number $N_{1}(T)$ of edges of $|P_{T_{l}}|%
\rightarrow {M}$, \emph{i.e.} 
\begin{equation}
\dim \left[ K_{1}RP_{g,N_{0}}^{met}\right] =N_{1}(T)=3N_{0}(T)+6g-6.
\end{equation}
There is a natural projection 
\begin{gather}
p:K_{1}RP_{g,N_{0}}^{met}\longrightarrow \mathbb{R}_{+}^{N_{0}(T)} \\
\Gamma \longmapsto p(\Gamma )=(l_{1},...,l_{N_{0}(T)}),  \notag
\end{gather}
where $(l_{1},...,l_{N_{0}(T)})$ denote the perimeters of the polygonal
2-cells $\{\rho ^{2}(j)\}$ of $|P_{T_{l}}|\rightarrow {M}$. With respect to
the topology on the space of metric ribbon graphs, the orbifold 
$K_{1}RP_{g,N_{0}}^{met}$ endowed with such a projection acquires the
structure of a cellular bundle. For a given sequence $\{l(\partial (\rho
^{2}(k)))\}$, the fiber 
\begin{equation}
p^{-1}(\{l(\partial (\rho ^{2}(k)))\})=\left\{ |P_{T_{l}}|\rightarrow {M}\in
K_{1}RP_{g,N_{0}}^{met}:\{l_{k}\}=\{l(\partial (\rho ^{2}(k)))\}\right\}
\end{equation}
is the set of all generalized conical Regge polytopes with the given set of
perimeters. If we take into account the $N_{0}(T)$ constraints associated
with the perimeters assignments, it follows that the fibers $p^{-1}(\{l(\partial (\rho ^{2}(k)))\})$ have dimension provided by 
\begin{equation}
\dim \left[ p^{-1}(\{l(\partial (\rho ^{2}(k)))\}\right] =2N_{0}(T)+6g-6,
\end{equation}
which again corresponds to the real dimension of the moduli space $\mathfrak{M}_{g},_{N_{0}}$ of $N_{0}$-pointed Riemann surfaces of genus $g$.

\subsection{Orbifold labelling and dynamical triangulations.}
Let us denote by 
\begin{equation}
\Omega _{T_{a}}\doteq \frac{\mathbb{R}_{+}^{|e(\Gamma )|}}{Aut_{\partial
}(P_{T_{a}})}  \label{omega}
\end{equation}
the rational cell associated with the 1-skeleton of the conical polytope $%
|P_{T_{a}}|\rightarrow {M}$ dual to a dynamical triangulation $%
|T_{l=a}|\rightarrow M$. The orbicell (\ref{omega}) contains the ribbon
graph associated with $|P_{T_{a}}|\rightarrow {M}$ and all (trivalent)
metric ribbon graphs $|P_{T_{L}}|\rightarrow {M}$ with the same
combinatorial structure of $|P_{T_{a}}|\rightarrow {M}$ but with all
possible length assignments $\{l(\rho ^{1}(h))\}_{h=1}^{N_{1}(T)}$ associated
with the corresponding set of edges $\{\rho ^{1}(h)\}_{1}^{N_{1}(T)}$. The
orbicell $\Omega _{T_{a}}$ is naturally identified with the convex
polytope (of dimension $(2N_{0}(T)+6g-6)$) in $\mathbb{R}_{+}^{N_{1}(T)}$
defined by 
\begin{equation}
\left\{ \{l(\rho ^{1}(j))\}\in \mathbb{R}_{+}^{N_{1}(T)}:%
\sum_{j=1}^{q(k)}A_{(k)}^{j}(T_{a})l(\rho ^{1}(j))\,=\frac{\sqrt{3}}{3}%
aq(k),\;k=1,...,N_{0}\;\right\} ,  \label{strata}
\end{equation}
where $A_{(k)}^{j}(T_{a})$ is a $(0,1)$ indicator matrix, depending on the
given dynamical triangulation $|T_{l=a}|\rightarrow M$, with $%
A_{(k)}^{j}(T_{a})=1$ if the edge $\rho ^{1}(j)$ belongs to $\partial (\rho
^{2}(k))$, and $0$ otherwise, and $\frac{\sqrt{3}}{3}aq(k)$ is the perimeter
length $l(\partial (\rho ^{2}(k)))$ in terms of the corresponding curvature
assignment $q(k)$. Note that $|P_{T_{a}}|\rightarrow {M}$ appears as the
barycenter of such a polytope.

Since the cell decomposition (\ref{DTorb}) of the space of trivalent metric
ribbon graphs $K_{1}RP_{g,N_{0}}^{met}$ depends only on the combinatorial
type of the ribbon graph, we can use the equilateral polytopes $%
|P_{T_{a}}|\rightarrow {M}$, dual to dynamical triangulations, as the set
over which the disjoint union in (\ref{DTorb}) runs. Thus we can write 
\begin{equation}
K_{1}RP_{g,N_{0}}^{met}=\bigsqcup_{\mathcal{DT}(N_{0})}\Omega _{T_{a}},
\end{equation}
where 
\begin{equation}
\mathcal{DT}_{g}\left( N_{0}\right) \doteq \left\{ |T_{l=a}|\rightarrow
M\;:(\sigma ^{0}(k))\;k=1,...,N_{0}(T)\right\}
\end{equation}
denote the set of distinct generalized dynamically triangulated surfaces
of genus $g$, with a given set of $N_{0}(T)$ ordered labelled vertices.  

Note that, even if the set $\mathcal{DT}_{g}\left( N_{0}\right) $ can be
considered (through barycentrical dualization) a well-defined subset of $%
K_{1}RP_{g,N_{0}}^{met}$, it is not an orbifold over $\mathbb{N}$ \cite{mulase2}. For this
latter reason, the analysis of the metric stuctures over (generalized)
polytopes requires the use of the full orbicells $\Omega _{T_{a}}$ and we
cannot limit our discussion to equilateral polytopes.

\subsection{The ribbon graph parametrization of the
moduli space} 
We start by recalling that the moduli space $\mathfrak{M}%
_{g},_{N_{0}}$ of genus $g$ Riemann surfaces with $N_{0}$ punctures is a
dense open subset of a natural compactification (Knudsen-Deligne-Mumford )
in a connected, compact complex orbifold denoted by $\overline{\mathfrak{M}}%
_{g},_{N_{0}}$. This latter is, by definition, the moduli space of stable $%
N_{0}$-pointed curves of genus $g$, where a stable curve is a compact
Riemann surface with at most ordinary double points such that all of its parts are
hyperbolic. The closure $\partial \mathfrak{M}_{g},_{N_{0}}$ of $\mathfrak{M}%
_{g},_{N_{0}}$ in $\overline{\mathfrak{M}}_{g},_{N_{0}}$ consists of stable
curves with double points, and gives rise to a stratification decomposing $%
\overline{\mathfrak{M}}_{g},_{N_{0}}$ into subvarieties. By definition, a
stratum of codimension $k$ is the component of $\overline{\mathfrak{M}}%
_{g},_{N_{0}}$ parametrizing stable curves (of fixed topological type)
with $k$ double points.

\begin{figure}[ht]
\begin{center}
\includegraphics[bb= 0 0 450 820,scale=.3]{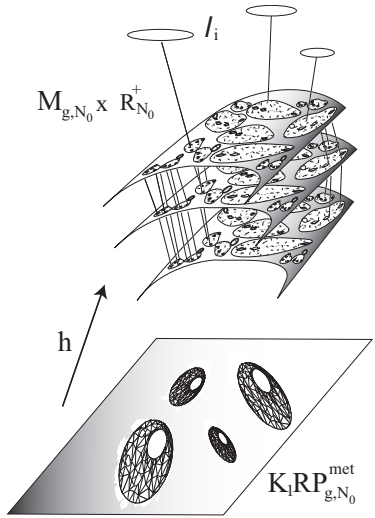}
\caption{The map $h$ associates to each ribbon graph an element of the decorated moduli space $\mathfrak{M}_{g,N_{0}}\times\Re^{+}_{N_{0}}$.}\label{fig8}
\end{center}
\end{figure}

The complex analytic geometry of the space of conical Regge polytopes which
we will discuss in the next section generalizes the well-known bijection (a
homeomorphism of orbifolds) between the space of metric ribbon graphs $%
K_{1}RP_{g,N_{0}}^{met}$ (which forgets the conical geometry) and the moduli
space $\mathfrak{M}_{g},_{N_{0}}$ of genus $g$ Riemann surfaces $((M;N_{0}),%
\mathcal{C})$ with $N_{0}(T)$ punctures \cite{mulase}, \cite{looijenga}. This bijection
results in a local parametrization of $\mathfrak{M}_{g},_{N_{0}}$ defined by 
\begin{gather}
h:K_{1}RP_{g,N_{0}}^{met}\rightarrow \mathfrak{M}_{g},_{N_{0}}\times {R}_{+}^{N}
\label{bijec} \\
\Gamma \longmapsto \lbrack ((M;N_{0}),\mathcal{C}),l_{i}]  \notag
\end{gather}
where $(l_{1},...,l_{N_{0}})$ is an ordered n-tuple of positive real numbers
and $\Gamma $ is a metric ribbon graphs with $N_{0}(T)$ labelled boundary
lengths $\{l_{i}\}$ (see figure \ref{fig8}). 

If $\overline{K_{1}RP}_{g,N_{0}}^{met}$ is the closure
of $K_{1}RP_{g,N_{0}}^{met}$, then the bijection $h$ extends to $\overline{%
K_{1}RP}_{g,N_{0}}^{met}\rightarrow \overline{\mathfrak{M}}_{g},_{N_{0}}\times {R%
}_{+}^{N_{0}}$ in such a way that a ribbon graph $\Gamma \in \overline{RGP}%
_{g,N_{0}}^{met}$ is mapped in two (stable) surfaces $M_{1}$ and $M_{2}$
with $N_{0}(T)$ punctures if and only if there exists an homeomorphism
between $M_{1}$ and $M_{2}$ preserving the (labelling of the) punctures, and
is holomorphic on each irreducible component containing one of the punctures.

According to Kontsevich \cite{kontsevich}, corresponding to each marked polygonal 2-cells $\{\rho ^{2}(k)\}$ of $|P_{T_{l}}|\rightarrow {M}$ 
there is a further (combinatorial) bundle map 
\begin{equation}
\mathcal{CL}_{k}\rightarrow K_{1}RP_{g,N_{0}}^{met}  \label{combundle}
\end{equation}
whose fiber over $(\Gamma ,\rho ^{2}(1),...,\rho ^{2}(N_{0}))$ is provided
by the boundary cycle $\partial \rho ^{2}(k)$, (recall that each boundary $\partial \rho ^{2}(k)$ 
comes with a positive orientation). 

To any such cycle
one associates \cite{looijenga}, \cite{kontsevich} the corresponding perimeter map $l(\partial (\rho^{2}(k)))=\sum $\ $l(\rho ^{1}(h_{\alpha }))$ 
which then appears as defining a natural connection on $\mathcal{CL}_{k}$. The piecewise smooth 2-form
defining the curvature of such a connection, 
\begin{equation}
\omega _{k}(\Gamma )=\sum_{1\leq h_{\alpha }<h_{\beta }\leq q(k)-1}d\left( 
\frac{l(\rho ^{1}(h_{\alpha }))}{l(\partial \rho ^{2}(k))}\right) \wedge
d\left( \frac{l(\rho ^{1}(h_{\beta }))}{l(\partial \rho ^{2}(k))}\right) ,
\label{chern}
\end{equation}
is invariant under rescaling and cyclic permutations of the $l(\rho
^{1}(h_{\mu }))$, and is a combinatorial representative of the Chern class
of the line bundle $\mathcal{CL}_{k}$.

\bigskip

It is important to stress that even if ribbon graphs can be thought of as
arising from Regge polytopes (with variable connectivity), the morphism (\ref
{bijec}) only involves the ribbon graph structure and the theory can be (and
actually is) developed with no reference at all to a particular underlying
triangulation. In such a connection, the role of dynamical triangulations
has been slightly overemphasized, they simply provide a convenient way of
labelling the different combinatorial strata of the mapping (\ref{bijec}),
but, by themselves they do not define a combinatorial parametrization of $%
\overline{\mathfrak{M}}_{g},_{N_{0}}$ for any finite $N_{0}$. However, it is
very useful, at least for the purposes of quantum gravity, to remember the
possible genesis of a ribbon graph from an underlying triangulation and be
able to exploit the further information coming from the associated conical
geometry. Such an information cannot be recovered from the ribbon graph
itself (with the notable exception of equilateral ribbon graphs, which can
be associated with dynamical triangulations), and must be suitably codified
by adding to the boundary lengths $\{l_{i}\}$ of the graph a further
decoration. This can be easily done by explicitly connecting Regge polytopes
to punctured Riemann surfaces.

\section{Punctured Riemann surfaces and Regge\hyphenation{po-ly-topes}
polytopes.}

\begin{figure}[ht]
\begin{center}
\includegraphics[bb= 0 0 600 640,scale=.3]{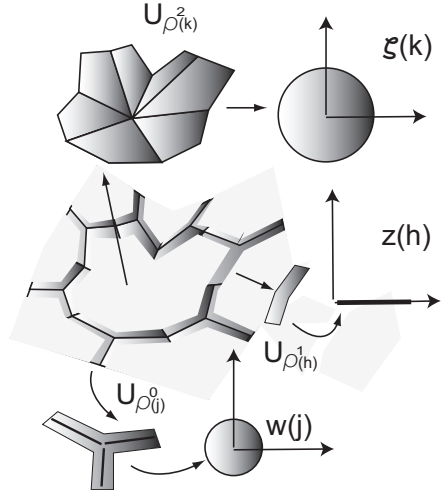}
\caption{The local presentation of uniformizing coordinates.}\label{fig9}
\end{center}
\end{figure}

As suggested by (\ref{cmetr}), the polyhedral metric associated with the
vertices $\{\sigma ^{0}(i)\}$ of a (generalized) Regge triangulation $%
|T_{l}|\rightarrow M$, can be conveniently described in terms of complex
function theory. We can extend the ribbon graph
uniformization of \cite{mulase} and associate with the polytope $|P_{T_{l}}|\rightarrow {M}$ a
complex structure $((M;N_{0}),\mathcal{C})$ (a punctured Riemann surface)
which is, in a well-defined sense, dual to the structure (\ref{cmetr})
generated by $|T_{l}|\rightarrow M$. Let $\rho ^{2}(k)$ be the generic
two-cell $\in |P_{T_{l}}|\rightarrow {M}$ barycentrically dual to the vertex 
$\sigma ^{0}(k)\in |T_{l}|\rightarrow M$. To the generic edge $\rho
^{1}(h) $ of $\rho ^{2}(k)$ we associate a complex uniformizing
coordinate $z(h)$ defined in the strip 
\begin{equation}
U_{\rho ^{1}(h)}\doteq \{z(h)\in \mathbb{C}\;|\;0<\func{Re}z(h)<l(\rho
^{1}(h))\},
\end{equation}
$l(\rho ^{1}(h))$ being the length of the edge considered. The uniformizing
coordinate $w(j)$, corresponding to the generic $3$-valent vertex $\rho
^{0}(j)\in \rho ^{2}(k)$, is defined in the open set 
\begin{equation}
U_{\rho ^{0}(j)}\doteq \{w(j)\in \mathbb{C}\;|\;|w(j)|<\delta ,\;w(j)[\rho
^{0}(j)]=0\},
\end{equation}
where $\delta >0$ is a suitably small constant. Finally, the two-cell $\rho
^{2}(k)$ is uniformized in the unit disk 
\begin{equation}
U_{\rho ^{2}(k)}\doteq \{\zeta (k)\in \mathbb{C}\;|\;|\zeta (k)|<1,\;\zeta
(k)[\sigma ^{0}(k)]=0\},
\end{equation}
where $\sigma ^{0}(k)$ is the vertex $\in |T_{l}|\rightarrow M$ 
corresponding to the given two-cell (see figure \ref{fig9}).

The various uniformizations $\{w(j),U_{\rho ^{0}(j)}\}_{j=1}^{N_{2}(T)}$, $%
\{z(h),U_{\rho ^{1}(h)}\}_{h=1}^{N_{1}(T)}$, and $\{\zeta (k),U_{\rho
^{2}(k)}\}_{k=1}^{N_{0}(T)}$ can be coherently glued together by noting that
to each edge $\rho ^{1}(h)\in $ $\rho ^{2}(k)$ we can associate the
standard quadratic differential on $U_{\rho ^{1}(h)}$ given by 
\begin{equation}\label{foliat}
\phi (h)|_{\rho ^{1}(h)}=dz(h)\otimes dz(h).  
\end{equation}
Such $\phi (h)|_{\rho ^{1}(h)}$ can be extended to the remaining local
uniformizations $U_{\rho ^{0}(j)}$, and $U_{\rho ^{2}(k)}$, by exploiting
a classic result in Riemann surface theory according to which a quadratic
differential $\phi $ has a finite number of zeros $n_{zeros}(\phi )$ with
orders $k_{i}$ and a finite number of poles $n_{poles}(\phi )$ of order $%
s_{i}$ such that 
\begin{equation}
\sum_{i=1}^{n_{zero}(\phi )}k_{i}-\sum_{i=1}^{n_{pole}(\phi )}s_{i}=4g-4.
\label{quadrel}
\end{equation}
In our case we must have $n_{zeros}(\phi )=N_{2}(T)$ with $k_{i}=1$,
(corresponding to the fact that the $1$-skeleton of $|P_{l}|\rightarrow M$
is a trivalent graph), and $n_{poles}(\phi )=$\ $N_{0}(T)$ with $%
s_{i}=s\;\forall i$, for a suitable positive integer $s$. According to such
remarks (\ref{quadrel}) reduces to 
\begin{equation}\label{poles}
N_{2}(T)-sN_{0}(T)=4g-4. 
\end{equation}
From the Euler relation $N_{0}(T)-N_{1}(T)+N_{2}(T)=2-2g$, and $%
2N_{1}(T)=3N_{2}(T)$ we get $N_{2}(T)-2N_{0}(T)=4g-4$. This is consistent
with (\ref{poles}) if and only if $s=2$. Thus the extension $\phi $ of $%
\phi (h)|_{\rho ^{1}(h)}$ along the $1$-skeleton of $|P_{l}|\rightarrow M$
must have $N_{2}(T)$ zeros of order $1$ corresponding to the trivalent
vertices $\{\rho ^{0}(j)\}$\ of $|P_{l}|\rightarrow M$ and $N_{0}(T)$
quadratic poles corresponding to the polygonal cells $\{\rho ^{2}(k)\}$ of
perimeter lengths $\{l(\partial (\rho ^{2}(k)))\}$. Around a zero of order
one and a pole of order two, every (Jenkins-Strebel \cite{strebel}) quadratic differential 
$\phi$ has a canonical local structure which (along with (\ref{foliat})) is
given by \cite{mulase}\cite{strebel} 
\begin{equation}
(|P_{T_{l}}|\rightarrow {M)\rightarrow }\phi \doteq \left\{ 
\begin{tabular}{l}
$\phi (h)|_{\rho ^{1}(h)}=dz(h)\otimes dz(h),$ \\ 
$\phi (j)|_{\rho ^{0}(j)}=\frac{9}{4}w(j)dw(j)\otimes dw(j),$ \\ 
$\phi (k)|_{\rho ^{2}(k)}=-\frac{\left[ l(\partial (\rho ^{2}(k)))\right]
^{2}}{4\pi ^{2}\zeta ^{2}(k)}d\zeta (k)\otimes d\zeta (k),$%
\end{tabular}
\right.  \label{differ}
\end{equation}
where $\{\rho ^{0}(j),\rho ^{1}(h),\rho ^{2}(k)\}$ runs over the set of
vertices, edges, and $2$-cells of $|P_{T_{l}}|\rightarrow M$. Since $\phi
(h)|_{\rho ^{1}(h)}$, $\phi (j)|_{\rho ^{0}(j)}$, and $\phi (k)|_{\rho
^{2}(k)}$ must be identified on the non-empty pairwise intersections $%
U_{\rho ^{0}(j)}\cap U_{\rho ^{1}(h)}$, $U_{\rho ^{1}(h)}\cap U_{\rho
^{2}(k)}$ we can associate to the polytope $|P_{T_{l}}|\rightarrow {M}$ a
complex structure $((M;N_{0}),\mathcal{C})$ by coherently gluing, along the
pattern associated with the ribbon graph $\Gamma $, the local
uniformizations $\{U_{\rho ^{0}(j)}\}_{j=1}^{N_{2}(T)}$, $\{U_{\rho
^{1}(h)}\}_{h=1}^{N_{1}(T)}$, and \ $\{U_{\rho ^{2}(k)}\}_{k=1}^{N_{0}(T)}$.
Explicitly, let $\{U_{\rho ^{1}(j_{\alpha })}\}$, $\alpha =1,2,3$ be the
three generic open strips associated with the three cyclically oriented
edges $\{\rho ^{1}(j_{\alpha })\}$ incident on the generic vertex $\rho
^{0}(j)$. Then the uniformizing coordinates $\{z(j_{\alpha })\}$ are related
to $w(j)$ by the transition functions 
\begin{equation}
w(j)=e^{2\pi i\frac{\alpha -1}{3}}z(j_{\alpha })^{\frac{2}{3}},\hspace{0.2in}\hspace{0.1in}\alpha =1,2,3.  \label{glue1}
\end{equation}
Note that in such uniformization the vertices $\{\rho ^{0}(j)\}$ do not
support conical singularities since each strip $U_{\rho ^{1}(j_{\alpha })}$
is mapped by (\ref{glue1}) into a wedge of angular opening $\frac{2\pi }{3}$. 
This is consistent with the definition of $|P_{T_{l}}|\rightarrow {M}$
according to which the vertices $\{\rho ^{0}(j)\}\in |P_{T_{l}}|\rightarrow {M}$ are the barycenters of the flat $\{\sigma ^{2}(j)\}\in
|T_{l}|\rightarrow M$. Similarly, if $\{U_{\rho ^{1}(k_{\beta })}\}$, $\beta =1,2,...,q(k)$ 
are the open strips associated with the $q(k)$
(oriented) edges $\{\rho ^{1}(k_{\beta })\}$ boundary of the generic
polygonal cell $\rho ^{2}(k)$, then the transition functions between the
corresponding uniformizing coordinate $\zeta (k)$ and the $\{z(k_{\beta })\}$
are given by \cite{mulase}
\begin{equation}
\zeta (k)=\exp \left( \frac{2\pi i}{l(\partial (\rho ^{2}(k)))}\left(
\sum_{\beta =1}^{\nu -1}l(\rho ^{1}(k_{\beta }))+z(k_{\nu })\right) \right) ,\hspace{0.2in}\nu =1,...,q(k),  \label{glue2}
\end{equation}
with $\sum_{\beta =1}^{\nu -1}\cdot \doteq 0$, for $\nu =1$.
\subsection{A Parametrization of the conical geometry.} 
Note that for any closed curve $c:\mathbb{S}^{1}\rightarrow U_{\rho ^{2}(k)}$,
homotopic to the boundary of $\overline{U}_{\rho ^{2}(k)}$, we get 
\begin{equation}
\oint_{c}\sqrt{\phi (k)_{\rho ^{2}(k)}}=l(\partial (\rho ^{2}(k))).
\label{length}
\end{equation}
which shows that the geometry associated with $\phi (k)_{\rho ^{2}(k)}$ is
described by the cylindrical metric canonically associated with a quadratic
differential with a second order pole,\emph{i.e.} 
\begin{equation}
|\phi (k)_{\rho ^{2}(k)}|=\frac{\left[ l(\partial (\rho ^{2}(k)))\right] ^{2}%
}{4\pi ^{2}|\zeta (k)|^{2}}|d\zeta (k)|^{2}.  \label{flmetr}
\end{equation}
If we denote by 
\begin{equation}
\Delta _{k}^{\ast }\doteq \{\zeta (k)\in \mathbb{C}|\;0<|\zeta (k)|<1\},
\label{puncdisk}
\end{equation}
the punctured disk $\Delta _{k}^{\ast}\subset U_{\rho ^{2}(k)}$, then for 
each given deficit angle $\varepsilon (k)=2\pi -\theta (k)$ we can introduce on each $\Delta _{k}^{\ast}$ the conical metric 
\begin{eqnarray}
ds_{(k)}^{2} &\doteq &\frac{\left[ L(k)\right] ^{2}}{
4\pi ^{2}}\left| \zeta (k)\right| ^{-2\left( \frac{\varepsilon (k)}{2\pi }
\right) }\left| d\zeta (k)\right| ^{2}=  \label{metrica} \\
&=& \left|
\zeta (k)\right| ^{2\left( \frac{\theta (k)}{2\pi }\right) }
|\phi (k)_{\rho ^{2}(k)}|.  \nonumber
\end{eqnarray}
It follows that we can apply the explicit construction \cite{mulase} of the mapping
(\ref{bijec}) for defining the decorated Riemann surface corresponding
to a conical Regge polytope. Then, an obvious adaptation of theorem 4.2 of
\cite{mulase} provides

\begin{proposition}\label{gluing}
Let $\{p_{k}\}_{k=1}^{N_{0}}\in M$ denote the set of punctures corresponding
to the decorated vertices $\{\sigma ^{0}(k),\frac{\varepsilon (k)}{2\pi }%
\}_{k=1}^{N_{0}}$ of the triangulation $|T_{l}|\rightarrow M$ and let $%
\Gamma $ be the ribbon graph associated with the corresponding dual conical
polytope $(|P_{T_{l}}|\rightarrow {M)}$, then the map 
\begin{gather}
\Upsilon :(|P_{T_{l}}|\rightarrow {M)\longrightarrow }((M;N_{0}),\mathcal{C}%
);\{ds_{(k)}^{2}\})  \label{riemsurf} \\
\Gamma \longmapsto \bigcup_{\{\rho ^{0}(j)\}}^{N_{2}(T)}U_{\rho
^{0}(j)}\bigcup_{\{\rho ^{1}(h)\}}^{N_{1}(T)}U_{\rho ^{1}(h)}\bigcup_{\{\rho
^{2}(k)\}}^{N_{0}(T)}(U_{\rho ^{2}(k)},ds_{(k)}^{2}),  \notag
\end{gather}
defines the decorated, $N_{0}$-pointed, Riemann surface $((M;N_{0}),%
\mathcal{C})$ canonically associated with the conical Regge polytope $%
|P_{T_{l}}|\rightarrow {M}$.
\end{proposition}

In order to describe the geometry of the uniformization of 
$((M;N_{0}),\mathcal{C}))$ defined by $\{ds_{(k)}^{2}\}$,  let us consider the image in $((M;N_{0}),\mathcal{C}))$ of the generic triangle $\sigma ^{2}(h,j,k)\in |T_{l}|\rightarrow M$ of sides $\sigma ^{1}(h,j)$, $\sigma ^{1}(j,k)$, and $\sigma ^{1}(k,h)$. Similarly, let $W(h,j)$, $W(j,k)$, and $W(k,h)$ be the images of the respective barycenters, (see (\ref{Gamma})). Denote by 
$\widehat{L}(k)=|W(h,j)\rho^{0}(h,j,k)|$,  $\widehat{L}(h)=|W(j,k)\rho^{0}(h,j,k)|$, and
$\widehat{L}(j)=|W(k,h)\rho^{0}(h,j,k)|$, the lengths, in the metric $\{ds_{(k)}^{2}\}$, of the half-edges connecting the (image of the) vertex $\rho^{0}(h,j,k)$ of the ribbon graph $\Gamma$ with $W(h,j)$, $W(j,k)$, and $W(k,h)$. Likewise, let us denote by $l(\bullet,\bullet)$ the length of the corresponding side 
$\sigma ^{1}(\bullet,\bullet)$ of the triangle.   A direct computation involving the geometry of the medians of    
$\sigma ^{2}(h,j,k)$ provides 
\begin{equation}
\begin{tabular}{ccc}
$\widehat{L}^{2}(j)$ & $=$ & $\frac{1}{18}l^{2}(j,k)+\frac{1}{18}
l^{2}(h,j)-\frac{1}{36}l^{2}(k,h )$ \\ 
$\widehat{L}^{2}(k)$ & $=$ & $\frac{1}{18}l^{2}(k,h)+\frac{1}{18}
l^{2}(j,k)-\frac{1}{36}l^{2}(h,j)$ \\ 
$\widehat{L}^{2}(h)$ & $=$ & $\frac{1}{18}l^{2}(h,j)+\frac{1}{18}
l^{2}(k,h)-\frac{1}{36}l^{2}(j,k)$ \\ 
&  &  \\ 
$l^{2}(k,h)$ & $=$ & $8\widehat{L}^{2}(h)+8\widehat{L}
^{2}(k)-4\widehat{L}^{2}(j)$ \\ 
$l^{2}(h,j)$ & $=$ & $8\widehat{L}^{2}(j)+8\widehat{L}
^{2}(h)-4\widehat{L}^{2}(k)$ \\ 
$l^{2}(j,k )$ & $=$ & $8\widehat{L}^{2}(k)+8\widehat{L}
^{2}(j)-4\widehat{L}^{2}(h)$
\end{tabular}
,
\end{equation}

\begin{figure}[ht]
\begin{center}
\includegraphics[scale=.5]{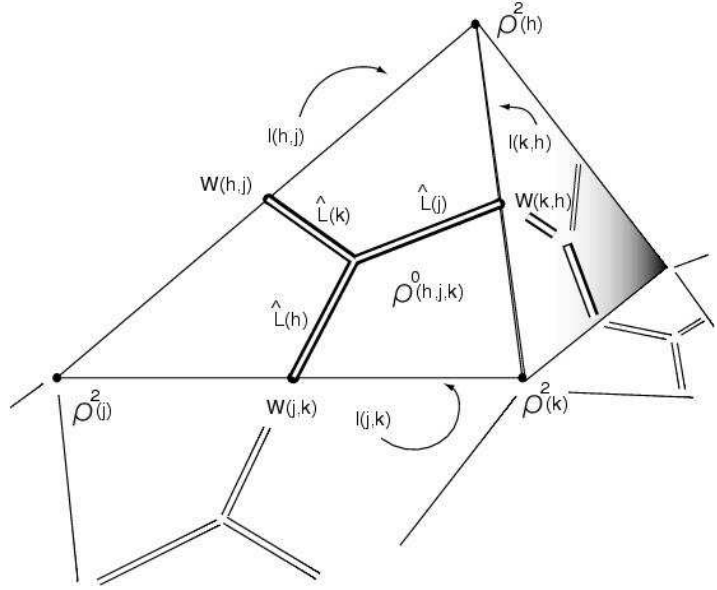}
\caption{The relation between the edge-lengths of the conical polytope and the edge-lenghts of the triangulation.}\label{fig3b}
\end{center}
\end{figure}

which allows to recover, as the indices $(h,j,k)$ vary, the metric geometry of  $|P_{T_{l}}|\rightarrow {M}$ and of its dual triangulation $|T_{l}|\rightarrow M$, from $((M;N_{0}),\mathcal{C});\{ds_{(k)}^{2}\})$ (see figure \ref{fig3b}).
 In this sense, the stiffening \cite{Thurston} of $((M;N_{0}),\mathcal{C})$ defined by the punctured Riemann surface 
\begin{gather}
((M;N_{0}),\mathcal{C});\{ds_{(k)}^{2}\})= \\
=\bigcup_{\{\rho ^{0}(h,j,k)\}}^{N_{2}(T)}U_{\rho ^{0}(h,j,k)}\bigcup_{\{\rho
^{1}(h,j)\}}^{N_{1}(T)}U_{\rho ^{1}(h,j)}\bigcup_{\{\rho
^{2}(k)\}}^{N_{0}(T)}(\Delta _{k}^{\ast },ds_{(k)}^{2}),  \nonumber
\end{gather}
is the uniformization of $((M;N_{0}),\mathcal{C})$ associated with the conical Regge polytope $|P_{l}|\rightarrow M$.

\begin{figure}[ht]
\begin{center}
\includegraphics[bb=40 110 550 670, scale=.5]{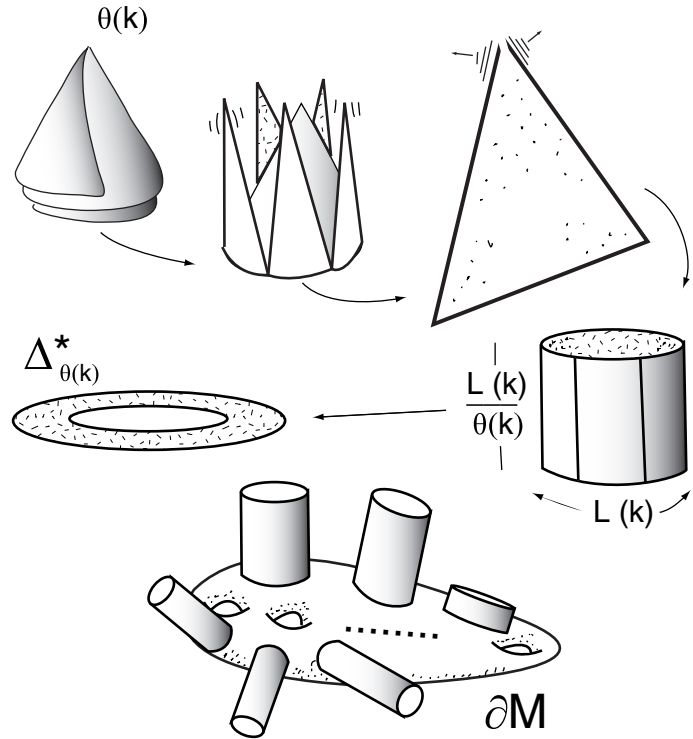}
\caption{Blowing up the conical geometry of the polytope into finite cylindrical ends generates a uniformized Riemann surface with cylindrical boundaries.}\label{fig5b}
\end{center}
\end{figure}

Although the correspondence between conical Regge polytopes and the above punctured Riemann surface is rather natural there is yet another uniformization representation of $|P_{l}|\rightarrow M$ which is of relevance in discussing conformal field theory on a given $|P_{l}|\rightarrow M$. The point is that the analysis of a CFT on a singular surface such as $|P_{l}|\rightarrow M$ calls for the imposition of suitable boundary conditions in order to take into account the conical singularities of the underlying Riemann surface $((M;N_{0}),\mathcal{C}, ds^2_{(k)})$. This is a rather delicate issue since conical metrics give rise to difficult technical problems in discussing the glueing properties of
the resulting conformal fields. In boundary conformal field theory, problems of this sort are taken care of (see e.g.[\cite{gawedzki}])  by (tacitly) assuming that a neighborhood of the possible boundaries is endowed with a cylindrical metric. In our setting such a prescription naturally calls into play the metric associated with the quadratic
differential $\phi $, and requires that we regularize into finite cylindrical ends the cones 
$(\Delta _{k}^{\ast },ds_{(k)}^{2})$.\ \  Such a  regularization is realized by noticing that if we introduce the annulus
\begin{equation}
\Delta _{\theta (k)}^{\ast }\doteq \left\{ \zeta (k)\in \mathbb{C}|e^{-\frac{
2\pi }{\theta (k)}}\leq |\zeta (k)|\leq 1\right\}\subset \overline{U_{\rho ^{2}(k)}},
\end{equation}
then the surface with boundary 
\begin{equation}
M_{\partial }\doteq ((M_{\partial };N_{0}),\mathcal{C})=\bigcup U_{\rho
^{0}(j)}\bigcup U_{\rho ^{1}(h)}\bigcup (\Delta _{\theta (k)}^{\ast },\phi
(k))
\end{equation}
defines the blowing up of the conical geometry of \ $((M;N_{0}),\mathcal{C}
,ds_{(k)}^{2})$ along the ribbon graph $\Gamma $ (see figure \ref{fig5b}).

The metrical geometry of \ $(\Delta _{\theta (k)}^{\ast },\phi (k))
$ is that of a flat cylinder with a circumference of length given by $L(k)$ and heigth given by $L(k)/\theta (k)$, (this latter being the slant radius of the
generalized Euclidean cone $(\Delta _{k}^{\ast },ds_{(k)}^{2})$ of base circumference $L(k)$ and vertex
conical angle $\theta (k)$).We also have \ \ 
\begin{eqnarray}
\partial M_{\partial } &=&\bigsqcup_{k=1}^{N_{0}}S_{\theta (k)}^{(+)}, \\
\partial \Gamma  &=&\bigsqcup_{k=1}^{N_{0}}S_{\theta (k)}^{(-)}  \notag
\end{eqnarray}
where the circles 
\begin{eqnarray}
S_{\theta (k)}^{(+)} &\doteq &\left\{ \zeta (k)\in \mathbb{C}||\zeta
(k)|=e^{-\frac{2\pi }{\theta (k)}}\right\} , \\
S_{\theta (k)}^{(-)} &\doteq &\left\{ \zeta (k)\in \mathbb{C}||\zeta
(k)|=1\right\}   \notag
\end{eqnarray}
respectively denote the inner and the outer boundary of the annulus $\Delta
_{\theta (k)}^{\ast }$. 
Note that by collapsing $S_{\theta (k)}^{(+)}$ to a point we get back the original cones $(\Delta _{k}^{\ast },ds_{(k)}^{2})$.
 Thus, the surface with boundary $
M_{\partial }$ naturally corresponds to the ribbon graph $\Gamma $
associated with the 1-skeleton $K_{1}(|P_{T_{l}}|\rightarrow {M})$ of the
polytope $|P_{T_{l}}|\rightarrow {M}$, decorated with the finite
cylinders $\{\Delta _{\theta (k)}^{\ast },|\phi (k)|\}$. In such a
framework the conical angles $\{\theta (k)=2\pi -\varepsilon (k)\}$ appears
as (reciprocal of the) moduli $m_{k}$ of the annuli $\{\Delta _{\theta
(k)}^{\ast }\}$, 
\begin{equation}
m(k)=\frac{1}{2\pi }\ln \frac{1}{e^{-\frac{2\pi }{\theta (k)}}}=\frac{1}{\theta (k)}
\end{equation}
(recall that the modulus of an annulus $r_{0}<|\zeta |<r_{1}$ is defined by $\frac{1}{2\pi }\ln \frac{r_{1}}{r_{0}}$). 
According to these remarks we can equivalently represent the conical Regge polytope $|P_{T_{l}}|\rightarrow {M}$ with the uniformization  
$((M;N_{0}),\mathcal{C});\{ds_{(k)}^{2}\})$ or with its blowed up version $M_{\partial }$. 

\chapter{The WZW model on Random Regge Triangulations}
The formulation of a conformal field theory and in particular of a WZW model over a triangulated Riemann surface 
has been a subject of great interests in the past years. As we have outilined in the
introduction, despite several attempts 
starting directly from a Chern-Simons theory (see as an example \cite{kawamoto}), 
many difficulties have arosen ranging from the dynamics of G-valued field to the non 
trivial dependance of the model from the underlying topology. In this chapter
we will follow a complete different approach \cite{carfora4} defining the WZW model 
directly from the triangulated Riemann surface; in order to accomplish this task we
will use the techniques introduced in the previous chapter which are more analytic in
spirit and for this reason they allow us to avoid the natural difficulties
raising from combinatorial calculus. Bearing in mind that our ultimate goal is to give
an holographic description of the CS/WZW relation and to intepret the functional
(\ref{holostate}), we will specialize our analysis to the SU(2) model;
we will also explicitly write the partition function for the theory at level $k=1$ 
which directly involves the $6j$ symbols of the quantum group at $q=\exp(i\frac{\pi}{3})$. 

\section{The WZW model on a Regge polytope}

Let $G$ be a connected and simply connected Lie group. In order to make
things simpler we shall limit our discussion to the case $G=SU(2)$, this
being the case of more direct interest to us. Recall \cite{gawedzki} that
the complete action of the Wess-Zumino-Witten model on a closed Riemann
surface $M$ of genus $g$ is provided by 
\begin{equation}
S^{WZW}(h)=\frac{\kappa }{4\pi \sqrt{-1}}\int_{M}tr\left( h^{-1}\partial
h\right) \left( h^{-1}\overline{\partial }h\right) +S^{WZ}(h),
\end{equation}
where $h:M\rightarrow SU(2)$ denotes a $SU(2)$-valued field on $M$, $\kappa $
is a positive constant (the level of the model),  $tr(\cdot )$ is the
Killing form on the Lie algebra (normalized so that the root has length $
\sqrt{2}$) and $S^{WZ}(h)$ is the topological Wess-Zumino term needed \cite
{witten2} in order to restore conformal invariance of the theory at the quantum level.
Explicitly, $S^{WZ}(h)$ can be characterized by extending the field $
h:M\rightarrow SU(2)$ to maps $\widetilde{h}:V_{M}\rightarrow SU(2)$ where $
V_{M}$ is a three-manifold with boundary such that $\partial V_{M}=M$, and
set 
\begin{equation}
S^{WZ}(h)=\frac{\kappa }{4\pi \sqrt{-1}}\int_{V_{M}}\widetilde{h}^{\ast
}\chi _{SU(2)},
\end{equation}
where $\widetilde{h}^{\ast }\chi _{SU(2)}$ denotes the pull-back to $V_{M}$
of the canonical 3-form on $SU(2)$

\begin{equation}
\chi _{SU(2)}\doteq \frac{1}{3}tr\left( h^{-1}dh\right) \wedge \left(
h^{-1}dh\right) \wedge \left( h^{-1}dh\right),  \label{treforma}
\end{equation}
(recall that for $SU(2)$, $\chi _{SU(2)}$ reduces to $4\mu _{S^{3}}$, where 
$\mu _{S^{3}}$ is the volume form on the unit 3-sphere $S^{3}$). As is well
known, $S^{WZ}(h)$ so defined depends on the extension $\widetilde{h}$ , the
ambiguity being parametrized by the period of the form $\chi _{SU(2)}$ over
the integer homology $H_{3}(SU(2))$. Demanding that the Feynman amplitude $
e^{-S^{WZW}(h)}$ is well defined requires that the level $\kappa$ is an
integer.

\begin{figure}[ht]
\begin{center}
\includegraphics[bb= 70 140 530 620, scale=.5]{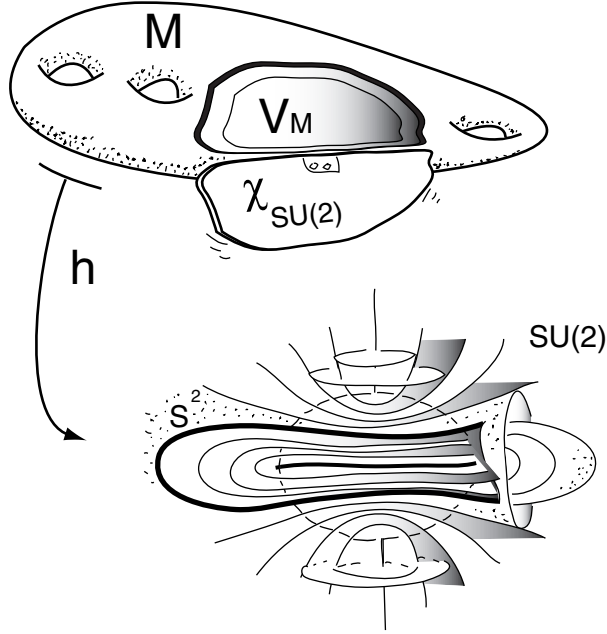}
\caption{The geometrical set up for the WZW model. The surface M opens up to show the associated handlebody. The group SU(2) is here shown as the 3-sphere foliated into (squashed) 2-spheres.}
\end{center}
\end{figure}

\subsection{Polytopes and the WZW model with boundaries}

From the results discussed in the previous chapter, it follows that a natural strategy for introducing
the WZW model on the Regge polytope $|P_{T_{l}}|\rightarrow {\ M}$ is to
consider maps  $h:M_{\partial }\rightarrow SU(2)$ on the associated surface
with cylindrical boundaries $M_{\partial }\doteq ((M_{\partial };N_{0}),
\mathcal{C})$. Such maps $h$ should satisfy suitable boundary conditions 
on the (inner and outer) boundaries $\{S_{\theta (k)}^{(\pm )}\}$  of the annuli $\{\Delta
_{\theta (k)}^{\ast }\}$, corresponding to the (given) values of the $SU(2)$ field on the boundaries of the cells of $|P_{T_{l}}|\rightarrow {\ M}$ and on their barycenters, (the field being free to fluctuate in the cells). Among all
possible boundary conditions, there is a choice
which is particularly simple and which allows us to reduce the study of WZW
model on each given Regge polytopes to the (quantum) dynamics of WZW fields on the
finite cylinders (annuli) $\{\Delta _{\theta (k)}^{\ast }\}$ decorating the
ribbon graph $\Gamma $ and representing the conical cells of $|P_{T_{l}}|\rightarrow {\ M}$. 
Such an approach corresponds to first study the WZW model on $|P_{T_{l}}|\rightarrow {\ M}$ as a CFT. Its (quantum) states will then depend on the boundary conditions on the $SU(2)$ field $h$ on $\{S_{\theta (k)}^{(\pm )}\}$; roughly speaking such a procedure turns out to be equivalent to a prescription assigning an irreducible representation of $SU(2)$ to each barycenter of the given polytope  $|P_{T_{l}}|\rightarrow {\ M}$. Such representations are parametrized by the boundary conditions which, by consistency, turn out to be necessarily quantized. They are also parametrized by elements of the geometry of $|P_{T_{l}}|\rightarrow {\ M}$, in particular by the deficit angles. 
\bigskip

\begin{figure}[ht]
\begin{center}
\includegraphics[bb= 60 180 500 700, scale=.5]{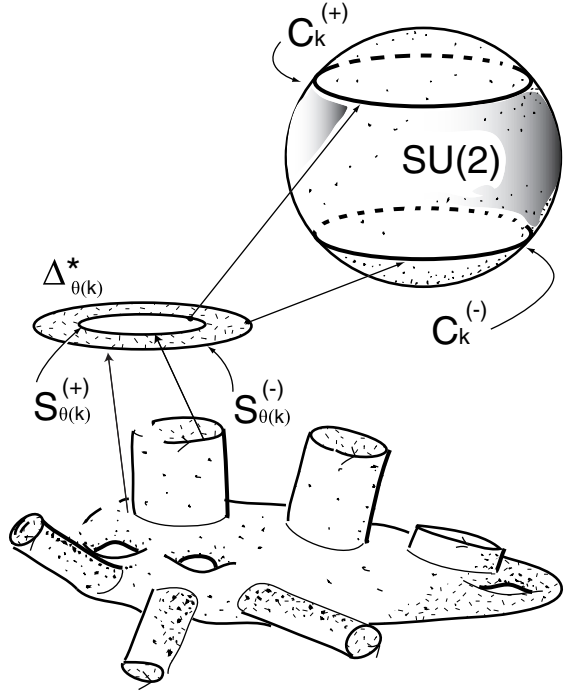}
\caption{The geometrical setup for SU(2) boundary conditions on each $(\Delta _{\theta (k)}^{\ast },\phi (k))$ decorating the 1-skeleton of $
|P_{T_{l}}|\rightarrow {M}$. For simplicity, the group SU(2) is incorrectly rendered; note that each circumference $C_k^{\pm}$ is actually a two-sphere, (or degenerates to a point).}\label{fig7b}
\end{center}
\end{figure}

In order to carry over such a program, let us associate with each  inner boundary $S_{\theta
(i)}^{(+)}$ the $SU(2)$ Cartan generator 
\begin{equation}
\Lambda _{i}\doteq \frac{\lambda (i)}{\kappa }\mathbf{\sigma }_{3}\text{, with\ }\mathbf{
\sigma }_{3}=\left( 
\begin{array}{cc}
1 & 0 \\ 
0 & -1
\end{array}
\right) 
\end{equation}
where, for later convenience, $\lambda (i)\in \mathbb{R}$ has been normalized to the level $\kappa $, \
and let 
\begin{equation}
C_{i}^{(+)}\doteq \left\{ \gamma e^{2\pi \sqrt{-1}\Lambda _{i}}\gamma ^{-1}\;|\;\gamma
\in SU(2)\right\} .
\end{equation}
denote the (positively oriented) two-sphere $S^{2}_{\theta (i)}$ in $SU(2)$ representing the associated conjugacy class, (note that $C_{i}^{(+)}$ degenerates to a single point for the center of $SU(2)$). Such a prescription basically prevent out-flow of momentum across the boundary and has been suggested, in the framework of D-branes theory in \cite{alekseev}, (see also \cite{gawedzki}). Similarly, to the outer boundary $S_{\theta
(i)}^{(-)}$ we associate the conjugacy class $C_{i}^{(-)}=\overline{C_{i}^{(+)}}$ describing the conjugate two-sphere $\overline{S^{2}_{\theta (i)}}$  (with opposite orientation) in $SU(2)$ associated with  $S^{2}_{\theta (i)}$.   \ Given such data, 
we consider maps $h:M_{\partial
}\rightarrow SU(2)$ that satisfy the fully symmetric boundary conditions (see figure \ref{fig7b}) 
\cite{gawedzki2},
\begin{equation}
h(S_{\theta (i)}^{(\pm)})\subset C_{i}^{(\pm)}.
\label{achoice}
\end{equation}

Note that since $C_{i}^{(+)}$ and $C_{i}^{(-)}$ carry opposite orientations, the functions $h(S_{\theta (i)}^{(\pm)})$ are normalized to 
$h(S_{\theta (i)}^{(-)})h(S_{\theta (i)}^{(+)})={\mathbf e}$, (the identity $\in{SU(2)}$). The advantage of considering this subset of maps $h:M_{\partial
}\rightarrow SU(2)$ is that when
restricted to the boundary $\partial M_{\partial}$, ({\em i.e.}, to the inner conjugacy classes $C_{i}^{(+)}$), the 3-form $\chi _{SU(2)}$
(\ref{treforma}) becomes exact, and one can write 
\begin{equation}
\left. \chi _{SU(2)}\right| _{C_{i}}=d\omega _{i},
\end{equation}
where the 2-form $\omega _{i}$ is provided by 
\begin{equation}
\omega _{i}=tr(\gamma ^{-1}d\gamma )e^{2\pi \sqrt{-1}\Lambda _{i}}(\gamma
^{-1}d\gamma )e^{-2\pi \sqrt{-1}\Lambda _{i}}.
\end{equation}
In such a case, we can extend \cite{gawedzki} the map $h:M_{\partial
}\rightarrow SU(2)$ to a map ${\widehat{h}}:((M;N_{0}),\mathcal{C})\rightarrow SU(2)$ from the closed surface $((M;N_{0}),\mathcal{C})$ to $SU(2)$ in such a way that 
$\widehat{h}(\delta _{\theta (i)})\subset C_{i}^{(+)}$, where
\begin{equation}
\delta _{\theta (i)}\doteq \left\{ \zeta (i)\in \mathbb{C}|\;|\zeta (i)|\leq
e^{-\frac{2\pi }{\theta (i)}}\right\} 
\end{equation}
is the disk capping the cylindrical end $\{\Delta _{\theta (i)}^{\ast
},|\phi (i)|\}$, (thus $\partial \delta _{\theta (i)}=S_{\theta (k)}^{(+)}$ and 
$\Delta _{\theta
(i)}^{\ast }\cup \delta _{\theta (i)}\ \ \simeq \overline{U}_{\rho
^{2}(i)}$). In this connection note
that the boundary conditions $h(S_{\theta (i)}^{(+)})\subset C_{i}^{(+)}$
define elements of the loop group
\begin{equation}
\mathcal{L}_{(i)}SU(2)\doteq Map(S_{\theta (i)}^{(+)},SU(2))\simeq
Map(S^{1},SU(2)).
\end{equation}


\noindent Similarly, any other extension \ $h^{\prime }_{i}=\widehat{h}_{i}g$, ($g\in SU(2)$),
of $h$ over the capping disks $\delta _{\theta (i)}$, can be considered as an element of the group $Map(\delta
_{\theta (i)},SU(2))$. In the same vein, we can interpret $\widetilde{h}
_{i}=(\widehat{h}_{i},h_{i}^{\prime })$ as a map from the spherical double (see below) $
S_{i}^{2}$ of $\delta _{\theta (i)}$ into $SU(2)$, \emph{i.e.}, as an
element of the group $Map(S_{i}^{2},SU(2))$. It follows that each possible
extension of \ the boundary condition $h(S_{\theta (i)}^{(+)})$ fits into
the exact sequence of groups 
\begin{equation}
1\rightarrow Map(S_{i}^{2},SU(2))\rightarrow Map(\delta _{\theta
(i)},SU(2))\rightarrow Map(S_{\theta (i)}^{(+)},SU(2))\rightarrow 1.
\label{exseq}
\end{equation}
In order to discuss the properties of such  extensions we can proceed as follows, (see
\cite{gawedzki} for the analysis of these and
related issues in the general setting of boundary CFT).


Let us denote by $
V_{M}$,\ \ with $\partial V_{M}=((M,N_{0});\mathcal{C})$, the 3-dimensional
handlebody associated with the surface $((M,N_{0});\mathcal{C})$, and
corresponding to the mapping $\widehat{h}:((M,N_{0});\mathcal{C})\rightarrow
SU(2)\simeq S^{3}$ thought of as an \emph{immersion} in the 3-sphere. Since
the conjugacy classes $C_{i}^{(+)}$ are 2-spheres and the homotopy group $
\pi _{2}(SU(2))$ is trivial, we can further extend the maps $\widehat{h}$ to
a smooth function $\ \widehat{H}:V_{M}\rightarrow SU(2)$, (thus, by
construction \ $\widehat{H}(\delta _{\theta (i)})\subset $\ $C_{i}^{(+)}$).
Any such an extension can be used to pull-back to the handlebody $V_{M}$ the
Maurer-Cartan 3-form $\chi _{SU(2)}$ and it is natural to define the
Wess-Zumino term associated with $((M,N_{0});\mathcal{C})$ according to 

\begin{equation}
S_{|P_{T_{l}}|}^{WZ}(\widehat{h},\widehat{H})\doteq \frac{\kappa }{4\pi \sqrt{-1}}\int_{V_{M}}
\widehat{H}^{\ast }\chi _{SU(2)}-\frac{\kappa }{4\pi \sqrt{-1}}
\sum_{j=1}^{N_{0}}\int_{\delta _{\theta (j)}}\left. \widehat{h}\right|
_{\delta _{\theta (j)}}^{\ast }\omega _{j}.
\label{WZterm}
\end{equation}
In general, \ such a definition of \ $S_{|P_{T_{l}}|}^{WZ}(\widehat{h},\widehat{H})$\ depends
on the particular extensions $(\widehat{h},\widehat{H})$ we are considering,
and if we denote by $(h^{\prime }=\widehat{h}g,H^{\prime })$, $g\in SU(2)$,\
a different extension, then, by reversing the orientation of the handlebody $
V_{M}$ and of the capping disks $\delta _{\theta (j)}$ over which $
S_{|P_{T_{l}}|}^{WZ}(h^{\prime },H^{\prime })$ is
evaluated, the difference between the resulting WZ terms can be written as 
\begin{gather}
 S_{|P_{T_{l}}|}^{WZ}(\widehat{h},\widehat{H}
)-S_{|P_{T_{l}}|}^{WZ}(h^{\prime },H^{\prime })=
\label{diff} \\
=\frac{\kappa }{4\pi \sqrt{-1}}\left( \int_{V_{M}}\widehat{H}^{\ast }\chi
_{SU(2)}+\int_{V_{M}^{(-)}}H^{\prime \ast }\chi _{SU(2)}\right) -  \notag \\
-\frac{\kappa }{4\pi \sqrt{-1}}\sum_{j=1}^{N_{0}}\left( \int_{\delta
_{\theta (j)}}\left. \widehat{h}\right| _{\delta _{\theta (j)}}^{\ast
}\omega _{j}+\int_{\delta _{\theta (j)}^{(-)}}\left. h^{\prime }\right|
_{\delta _{\theta (j)}}^{\ast }\omega _{j}\right) .  \notag
\end{gather}
Note that 
\begin{equation}
(V_{M},\widehat{H})\cup (V_{M}^{(-)},H^{\prime })=(\widetilde{V}_{M},
\widetilde{H})
\end{equation}
is the 3-manifold (ribbon graph) double of $V_{M}$ \ endowed with the
extension $\widetilde{H}\doteq (\widehat{H},H^{\prime })$ \ \ and 
\begin{equation}
(\delta _{\theta (j)},\widehat{h_{j}})\cup (\delta _{\theta
(j)}^{(-)},h_{j}^{\prime })=(S_{j}^{2},\widetilde{h_{j}}),
\end{equation}
are the 2-spheres defined by doubling the capping disks $\delta _{\theta (j)}
$, decorated with the extension $\widetilde{h}_{j}\doteq (\widehat{h}
_{j},h_{j}^{\prime })\in C_{j}^{(+)}$. By construction $(\widetilde{V}_{M},
\widetilde{H})$ is such that $\partial (\widetilde{V}_{M},\widetilde{H}
)=\cup _{j=1}^{N_{0}}(S_{j}^{2},\widetilde{h_{j}})$ so that we can
equivalently write (\ref{diff}) as
\begin{gather}
 S_{|P_{T_{l}}|}^{WZ}(\widehat{h},\widehat{H}
)-S_{|P_{T_{l}}|}^{WZ}(h^{\prime },H^{\prime })=
\frac{\kappa }{4\pi \sqrt{-1}}\int_{\widetilde{V}_{M}}\widetilde{H}^{\ast
}\chi _{SU(2)}-  \label{diff2} \\
-\frac{\kappa }{4\pi \sqrt{-1}}\sum_{j=1}^{N_{0}}\int_{S_{j}^{2}}\widetilde{h
}^{\ast }\omega _{j}.  \notag
\end{gather}
To such an expression we add and subtract 
\begin{equation}
\frac{\kappa }{4\pi \sqrt{-1}}\sum_{j=1}^{N_{0}}\int_{B_{j}^{3}}\widetilde{
H_{j}}^{\ast }\chi _{SU(2)}
\end{equation}
where $B_{j}^{3}$ are 3-balls such that $\partial B_{j}^{3}=S_{j}^{2(-)}$,
(the boundary orientation is inverted so that we can glue such $B_{j}^{3}$
to the corresponding boundary components of $\widetilde{V}_{M}$), and $
\widetilde{H_{j}}$ are corresponding extensions of $\widetilde{H}$ with $
\widetilde{H_{j}}|_{S_{j}^{2}}=\widetilde{h}_{j}$. Since $\widetilde{V}
_{M}\cup $\ $B_{j}^{3}$ results in a closed 3-manifold $W^{3}$, we
eventually get 
\begin{gather}
S_{|P_{T_{l}}|}^{WZ}(\widehat{h},\widehat{H}
)-S_{|P_{T_{l}}|}^{WZ}(\eta )(h^{\prime },H^{\prime })=
\frac{\kappa }{4\pi \sqrt{-1}}\int_{W^{3}}\widetilde{H}^{\ast }\chi _{SU(2)}-
\\
-\frac{\kappa }{4\pi \sqrt{-1}}\sum_{j=1}^{N_{0}}\left( \int_{B_{j}^{3}}
\widetilde{H_{j}}^{\ast }\chi _{SU(2)}-\int_{\partial B_{j}^{3}}\widetilde{h}
^{\ast }\omega _{j}\right) ,  \notag
\end{gather}
where we have rewritten the integrals over $S_{j}^{2}$\ appearing in (\ref
{diff2}) as integrals over $\partial B_{j}^{3}=S_{j}^{2(-)}$, (hence the
sign-change). \ This latter expression shows that inequivalent extensions
are parametrized by the periods of $(\chi _{SU(2)},\omega _{j})$ over the
relative integer homology groups $H_{3}(SU(2),\cup _{j=1}^{N_{0}}C_{j})$.
Explicitly, the first term provides 
\begin{equation}
\frac{\kappa }{4\pi \sqrt{-1}}\int_{W^{3}}\widetilde{H}^{\ast }\chi _{SU(2)}=
\frac{\kappa }{4\pi \sqrt{-1}}\int_{\widetilde{H}(W^{3})}\chi _{SU(2)}=\frac{
\kappa }{4\pi \sqrt{-1}}\int_{S^{3}}\chi _{SU(2)}.
\end{equation}
Since $\int_{S^{3}}\chi _{SU(2)}=8\pi ^{2}$, we get $\frac{\kappa }{4\pi 
\sqrt{-1}}\int_{W^{3}}\widetilde{H}^{\ast }\chi _{SU(2)}=-2\pi \kappa \sqrt{-1}$
. Each addend in the second group of terms yields 
\begin{gather}
\frac{\kappa }{4\pi \sqrt{-1}}\left( \int_{B_{j}^{3}}\widetilde{H_{j}}^{\ast
}\chi _{SU(2)}-\int_{\partial B_{j}^{3}}\widetilde{h}^{\ast }\omega
_{j}\right) =  \label{remterm} \\
=\frac{\kappa }{4\pi \sqrt{-1}}\left( \int_{\widetilde{H}_{j}(B_{j}^{3})}
\chi _{SU(2)}-\int_{\widetilde{h}(\partial B_{j}^{3})}\omega _{j}\right) . 
\notag
\end{gather}


The domain of integration $\widetilde{h}(\partial B_{j}^{3})$ is the
2-sphere \ $C_{j}\subset SU(2)$ associated with the given conjugacy class,
whereas $\widetilde{H}_{j}(B_{j}^{3})$ is one of the two 3-dimensional balls
in $SU(2)$ with boundary $C_{j}$. In the defining representation of $
SU(2)\doteq \{x_{0}\mathbf{I}+\sqrt{-1}\sum x_{k}\mathbf{\sigma }_{k}|\;x_{0}^{2}+\sum
x_{k}^{2}=1\}$, the conjugacy classes $C_{j}$ are defined by $x_{0}=\cos 
\frac{2\pi \lambda (j)}{\kappa }$ with $0\leq \frac{2\pi \lambda (j)}{\kappa 
}\leq \pi $, whereas the two 3-balls $\widetilde{H}_{j}(B_{j}^{3})$ bounded
by $C_{j}$ are defined by $x_{0}\geq \cos \frac{2\pi \lambda (j)}{\kappa }$
and $x_{0}\leq \cos \frac{2\pi \lambda (j)}{\kappa }$. An explicit
computation \cite{gawedzki} over the ball $x_{0}\geq \cos \frac{2\pi \lambda (j)}{\kappa }$
shows that (\ref{remterm}) is provided by $-4\pi \lambda (j)\sqrt{-1}$, and
by $4\pi \sqrt{-1}(\frac{\kappa }{2}-\lambda (j))$ for $x_{0}\leq \cos \frac{
2\pi \lambda (j)}{\kappa }$, \ respectively. From these remarks it follows
that 
\begin{equation}
S_{|P_{T_{l}}|}^{WZ}(\widehat{h},\widehat{H}
)-S_{|P_{T_{l}}|}^{WZ}(h^{\prime },H^{\prime })\in
2\pi \sqrt{-1}\mathbb{Z}
\end{equation}
as long as $\kappa $ is an integer, and $0\leq \lambda (j)\leq \frac{\kappa 
}{2}$ with $\lambda (j)$ integer or half-integer; in such a case the
exponential of \ the WZ term $S_{|P_{T_{l}}|}^{WZ}(\widehat{h},
\widehat{H})$ is independent from the chosen extensions $(\widehat{h},
\widehat{H})$, and we can unambiguosly write $S_{|P_{T_{l}}|}^{WZ}(\widehat{h})$.
\bigskip
It follows from such remarks that we can define the $SU(2)$ WZW action on 
$|P_{T_{l}}|\rightarrow {M}$ according to
\begin{equation}
S_{|P_{T_{l}}|}^{WZW}(\widehat{h})\doteq \frac{\kappa }{4\pi \sqrt{-1}}
\int_{((M;N_{0}),\mathcal{C})}tr\left( \widehat{h}^{-1}\partial \widehat{h}
\right) \left( \widehat{h}^{-1}\overline{\partial }\widehat{h}\right)
+S_{|P_{T_{l}}|}^{WZ}(\widehat{h}).  \label{WZWact}
\end{equation}
where the WZ term $S_{|P_{T_{l}}|}^{WZ}(\eta )$ is provided by (\ref{WZterm}). It is worthwhile stressing that the condition $0\leq \lambda (j)\leq \frac{\kappa}{2}$ plays here the role of a quantization condition on the possible set of boundary conditions allowable for the WZW model on $|P_{T_{l}}|\rightarrow {M}$. Qualitatively, the situation is quite similar to the dynamics of branes on group manifolds, where in order to have stable, non point-like branes, we need a non vanishing $B$-field generating a NSNS 3-form $H$, (see e.g. \cite{schomerus}), here provided by $\omega_{j}$ and $\chi_{SU(2)}$, respectively. In such a setting, stable branes on $SU(2)$ are either point-like (corresponding to elements in the center $\pm e$ of $SU(2)$), or 2-spheres associated with a discrete set of radii. In our approach, such branes appear as the geometrical loci describing boundary conditions for WZW fields evolving on singular Euclidean surfaces. It is easy to understand the connection between the two formalism: in our description of the $\kappa$-level $SU(2)$ WZW model on $|P_{T_{l}}|\rightarrow {M}$ we can interpret the $SU(2)$ field as parametrizing an immersion of $|P_{T_{l}}|\rightarrow {M}$ in $S^{3}$ (of radius $\simeq \sqrt{\kappa}$). In particular, the annuli $\Delta^{*}_{\theta(i)}$ associated with the  ribbon graph boundaries $\{\partial\Gamma_i\}$ can be thought of as sweeping out in $S^{3}$ closed strings which couples with the branes defined by  $SU(2)$ conjugacy classes.

\section{The Quantum Amplitudes at k=1}

We are now ready to discuss the quantum properties of the fields $\widehat{h}$ involved in the above characterization of the $SU(2)$ WZW action on $|P_{T_{l}}|\rightarrow {M}$. Such properties follow by exploiting the action of the (central extension of the) loop group $Map(S_{\theta (i)}^{(+)},SU(2))$ generated,
on the infinitesimal level, by the conserved currents 
\begin{gather}
J(\zeta (i))\doteq -\kappa \partial _{(i)}\widehat{h}_{i}\widehat{h}_{i}^{-1}
\\
\overline{J}(\overline{\zeta }(i))\doteq \kappa \widehat{h}_{i}^{-1}
\overline{\partial }_{(i)}\widehat{h}_{i},  \notag
\end{gather}
where $\partial _{(i)}\doteq \partial _{\zeta (i)}$. By writing $J(\zeta
(i))=J^{a}(\zeta (i))\mathbf{\sigma }_{a}$, we can introduce the
corresponding modes $J_{n}^{a}(i)$, from the Laurent expansion in each disk $
\delta _{\theta (i)}$, 
\begin{equation}
J^{a}(\zeta (i))=\sum_{n\in \mathbb{Z}}\zeta (i)^{-n-1}J_{n}^{a}(i),
\end{equation}
(and similarly for the modes $\overline{J}_{n}^{a}(i)$). The operator
product expansion of the currents $J^{a}(\zeta (i))J^{a}(\zeta ^{\prime }(i))
$, (with $\zeta (i)$ and $\zeta ^{\prime }(i)$ both in $\delta _{\theta (i)}$
) yields \cite{gawedzki} the commutation relations of an affine $\widehat{\mathfrak{su}}(2)$
algebra at the level $\kappa $, \emph{i.e.} 
\begin{equation}
\left[ J_{n}^{a}(i),J_{m}^{b}(i)\right] =\sqrt{-1}\varepsilon
_{abc}J_{n+m}^{c}(i)+\kappa n\delta _{ab}\delta _{n+m,0}.
\end{equation}
According to a standard procedure, we can then construct the Hilbert space $
\mathcal{H}_{(i)}$ associated with the  WZW fields $\widehat{h}_{i}$ by
considering unitary irreducible highest weight representations of the two
commuting copies of the current algebra $\widehat{\mathfrak{su}}(2)$ generated
by $J^{a}(\zeta (i))|_{S_{\theta (i)}^{(+)}}$ and $\overline{J^{a}}(
\overline{\zeta }(i))|_{S_{\theta (i)}^{(+)}}$. Such representations are
labelled by the level $\kappa $ and by the irreducible representations  of $
SU(2)$ with spin $0\leq \lambda (i)\leq \frac{\kappa }{2}$.
Note in particular that for $\kappa =1$ every highest weight representation of \ $\widehat{\mathfrak{su}}
(2)_{\kappa =1}$ also provides a representation of Virasoro algebra $Vir$
with central charge $c=1$. In such a case the representations of $\widehat{
\mathfrak{su}}(2)_{\kappa =1}$ can be decomposed into $\mathfrak{su}(2)\oplus Vir$,
and, up to Hilbert space completion, we can write 
\begin{equation}
\mathcal{H}_{(i)}=\bigoplus_{0\leq \lambda (i)\leq \frac{1}{2},0\leq n\leq
\infty }\left( V_{\mathfrak{su}(2)}^{(n+\lambda (i))}\otimes \overline{V}_{\mathfrak{
su}(2)}^{(n+\lambda (i))}\right) \otimes \left( \mathcal{H}_{(n+\lambda
(i))^{2}}^{Vir}\otimes \overline{\mathcal{H}}_{(n+\lambda
(i))^{2}}^{Vir}\right)   \label{Hilb1}
\end{equation}
where $V_{\mathfrak{su}(2)}^{(n+\lambda (i))}$ denotes the $(2\lambda (i)+1)$
-dimensional spin $\lambda (i)$ representation of $\mathfrak{su}(2)$, and $
\mathcal{H}_{(n+\lambda (i))^{2}}^{Vir}$ is the (irreducible highest weight)
representation of the Virasoro algebra of weight $(n+\lambda (i))^{2}$.
Since $0\leq \lambda (i)\leq \frac{1}{2}$, it is convenient to set 
\begin{equation}
j_{i}\doteq n+\lambda (i)\in \frac{1}{2}\mathbb{Z}^{+}
\end{equation}
(with $0\in \mathbb{Z}^{+}$), and rewrite (\ref{Hilb1}) as 
\begin{equation}
\mathcal{H}_{(i)}=\bigoplus_{j_{i},\widehat{j}_{i}\in \frac{1}{2}\mathbb{Z}
^{+}}\left( V_{\mathfrak{su}(2)}^{j_{i}}\otimes \overline{V}_{\mathfrak{su}(2)}^{
\widehat{j}_{i}}\right) \otimes \left( \mathcal{H}_{j_{i}^{2}}^{Vir}\otimes 
\overline{\mathcal{H}}_{\widehat{j}_{i}^{2}}^{Vir}\right) ,
\end{equation}
with $j_{i}+\widehat{j}_{i}$\ $\in \mathbb{Z}^{+}$, \cite{gaberdiel2}.\ \ Owing to
this particularly simple structure of the representation spaces $\mathcal{H}_{(i)}$, 
we shall limit our analysis to the case $\kappa =1$.

 Since
the boundary of $\partial M$ of the surface $M$ is defined by the disjoint
union $\bigsqcup S_{\theta (i)}^{(+)}$ and the boundary $\partial \Gamma $ \
of the ribbon graph $\Gamma $ \ is provided by $\bigsqcup S_{\theta
(i)}^{(-)}$, it follows that\ we can associate to both $\partial M$ and $
\partial \Gamma $ the Hilbert space 
\begin{equation}
\mathcal{H}(\partial M)\simeq \mathcal{H}(\partial \Gamma
)=\bigotimes_{i=1}^{N_{0}}\mathcal{H}_{(i)}.
\end{equation}
Let us denote by $\left| \widehat{h}(S_{\theta (i)}^{(+)})\right\rangle \in 
\mathcal{H}_{(i)}$ the Hilbert space state vector associated with the
boundary condition $\widehat{h}(S_{\theta (i)}^{(+)})$ on the $i$-th\
boundary component $S_{\theta (i)}^{(+)}$ of $M_{\partial }$. According to
the analysis of the previous section, the ribbon graph double $\widetilde{V}
_{M}$ generates a Schottky \ $M^{D}$\ double of the surface with cylindrical
boundaries $M_{\partial }$, ($M^{D}$ is the closed surface obtained by
identifying $M_{\partial }$ with another copy $M_{\partial }^{\prime }$ of $
M_{\partial }$ with opposite orientation along their common boundary $
\bigsqcup S_{\theta (i)}^{(+)}$). Such $M^{D}$\ carries an orientation
reversing involution 
\begin{equation}
\Upsilon :M^{D}\rightarrow M^{D},\;\Upsilon ^{2}=id
\end{equation}
that interchanges $M_{\partial }$ and $M_{\partial }^{\prime }$ and which
has the boundary $\bigsqcup S_{\theta (i)}^{(+)}$ as its fixed point set.
The request of preservation of conformal symmetry along $\bigsqcup S_{\theta
(i)}^{(+)}$ under the anticonformal involution $\Upsilon $ requires that the
state $\left| \widehat{h}(S_{\theta (i)}^{(+)})\right\rangle $ must satisfy
the glueing condition $(\mathbb{L}_{n}-\overline{\mathbb{L}}_{-n})\left| 
\widehat{h}(S_{\theta (i)}^{(+)})\right\rangle =0$, where, for $n\neq 0$, 
\begin{equation}
\mathbb{L}_{n}=\frac{1}{2+\kappa }\sum_{m=-\infty }^{\infty
}J_{n-m}^{a}J_{m}^{a},
\end{equation}
and similarly for $\overline{\mathbb{L}}_{-n}$. \ The glueing conditions
above can be solved mode by mode, and to each irreducible representation of
the Virasoro algebra $\mathcal{H}_{j_{i}^{2}}^{Vir}$ and its conjugate $
\overline{\mathcal{H}}_{\widehat{j}_{i}^{2}=j_{i}^{2}}^{Vir}$, labelled by
the given $j_{i}\doteq n+\lambda (i)\in \frac{1}{2}\mathbb{Z}^{+}$,\ we can
associate a set of conformal Ishibashi states parametrized by the $\mathfrak{su}
(2)$ representations $V_{\mathfrak{su}(2)}^{j_{i}}\otimes V_{\mathfrak{su}
(2)}^{j_{i}}$. Such states are usually denoted by 
\begin{equation}
\left. \left| j_{i};m,n\right\rangle \right\rangle ,\;\;m,n\in
(-j_{i},-j_{i}+1,...,j_{i}-1,j_{i}),
\end{equation}
and one can write \cite{gaberdiel}    
\begin{equation}
\left| \widehat{h}(S_{\theta (i)}^{(+)})\right\rangle =\frac{1}{2^{\frac{1}{4
}}}\sum_{j_{i};m,n}D_{m,n}^{j_{i}}(\widehat{h}(S_{\theta (i)}^{(+)}))\left.
\left| j_{i};m,n\right\rangle \right\rangle ,
\end{equation}
where 
\begin{align}
D_{m,n}^{j_{i}}(\widehat{h}(S_{\theta (i)}^{(+)}))& =\sum_{l=\max
(0,n-m)}^{\min (j_{i}-m,j_{i}+n)}\frac{\left[
(j_{i}+m)!(j_{i}-m)!(j_{i}+n)!(j_{i}-n)!\right] ^{\frac{1}{2}}}{
(j_{i}-m-l)!(j_{i}+n-l)!l!(m-n+l)!}\times  \\
& \times a^{j_{i}+n-l}\;d^{j_{i}-m-l}\;b^{l}\;c^{m-n+l},  \notag
\end{align}
is the $V_{\mathfrak{su}(2)}^{j_{i}}$-representation matrix associated with the $SU(2)$ element
\begin{equation}
\widehat{h}(S_{\theta (i)}^{(+)})=\left( 
\begin{array}{cc}
a & b \\ 
c & d
\end{array}
\right) \in C_{i}^{(+)},
\end{equation}
in the $C_{i}^{(+)}$ conjugacy class.

\subsection{The Quantum Amplitudes for the cylindrical ends}

With the above preliminary remarks along the way, let us consider explicitly the structure of the quantum amplitude associated with the  WZW model defined by the
action $S_{|P_{T_{l}}|}^{WZW}(\widehat{h})$. Formally, such an amplitude is provided by the
functional integral 
\begin{equation}
\left| \partial{M} ,\otimes_{i}\widehat{h}(S_{\theta (i)}^{(+)})\right\rangle =\int_{\{\widehat{
h}|_{S_{\theta (i)}^{(\pm )}}\in C_{i}^{(\pm
)}\}}e^{-S_{|P_{T_{l}}|}^{WZW}(\widehat{h})}D\widehat{h},
\label{amplit}
\end{equation}
where the integration is over all the maps $\widehat{h}$ satisfying the boundary conditions $\{\widehat{
h}|_{S_{\theta (i)}^{(\pm )}}\in C_{i}^{(\pm )}\}$, and where $D\widehat{
h}$ is the local product $\prod_{\zeta \in ((M;N_{0}),\mathcal{C})}d
\widehat{h}(\zeta )$ over $((M;N_{0}),\mathcal{C})$ of the $SU(2)$ Haar
measure. As the notation suggests,  the formal expression (\ref{amplit})
takes value in the Hilbert space $\mathcal{\ H}$. Let us recall that the fields $\widehat{
h}$ are constrained over the disjoint boundary components of $\partial
\Gamma $ \ to belong to the conjugacy classes $\{\widehat{h}|_{S_{\theta
(i)}^{(-)}}\in C_{i}^{(-)}\}$. This latter remark implies that the maps $\widehat{h}$ \ \ fluctuate
on the $N_{0}$ finite cylinders $\{\Delta
_{\theta (i)}^{\ast }\}$ wheras on the ribbon graph $\Gamma$ they are represented by boundary operators which mediate the changes in the boundary conditions on adjacent boundary components $\{\partial{\Gamma}_{i}\}$ of $\Gamma$. In order to exploit such a  factorization property of (\ref{amplit})
the first step is the computation of the amplitude, (for each given index $
i$),  for the cylinder $\Delta _{\theta (i)}^{\ast }$ with in and
out boundary conditions $\widehat{h}|_{S_{\theta (i)}^{(\pm )}}\in
C_{i}^{(\pm )}$, 
\begin{equation}
Z_{\Delta _{\theta (i)}^{\ast }}\doteq \int_{\widehat{h}|_{S_{i}^{(\pm )}}\in
C_{i}^{(\pm )}}e^{-S^{WZW}(\widehat{h};\Delta _{\theta (i)}^{\ast })}D
\widehat{h}
\end{equation}
where $S^{WZW}(\widehat{h};\Delta _{\theta (i)}^{\ast })$
is the restriction to $\Delta _{\theta (i)}^{\ast }$ of $
S_{|P_{T_{l}}|}^{WZW}(\widehat{h})$. If we introduce the Virasoro
operator $\mathbb{L}_{0}(i)$ defined by 
\begin{equation}
\mathbb{L}_{0}(i)=\frac{2}{2+\kappa }\sum_{m=0}^{\infty
}J_{-m}^{a}(i)J_{m}^{a}(i).
\end{equation}
and notice that $\mathbb{L}_{0}(i)+\overline{\mathbb{L}}_{0}(i)-\frac{c}{12}$
,  defines the Hamiltonian of the WZW theory on the cylinder $\Delta
_{\theta (i)}^{\ast }$, ($c=\frac{3\kappa }{2+\kappa }$ being the central
charge of the SU(2) WZW\ theory), then we can explicitly write
\begin{equation}
Z_{\Delta _{\theta (i)}^{\ast }}(\{C_{i}^{(\pm )}\})=\langle \widehat{h}(S_{\theta
(i)}^{(-)})|e^{-\frac{2\pi }{\theta (i)}(\mathbb{L}_{0}(i)+\overline{\mathbb{
L}}_{0}(i)-\frac{c}{12})}|\widehat{h}(S_{\theta (i)}^{(+)})\rangle ,
\label{annz}
\end{equation}
where 
$\langle \widehat{h}(S_{\theta (i)}^{(-)})|$ and $|\widehat{h}(S_{\theta
(i)}^{(+)})\rangle $ respectively denote the Hilbert space vectors associated with the boundary
conditions $h(S_{\theta (i)}^{(-)})$ and $h(S_{\theta (i)}^{(+)})$ and normalized to $\langle \widehat{h}(S_{\theta
(i)}^{(-)})||\widehat{h}(S_{\theta (i)}^{(+)})\rangle =1$
, (a normalization that follows from the fact that $\widehat{h}(S_{\theta (i)}^{(-)})$ and $
\widehat{h}(S_{\theta (i)}^{(+)})$ belong, by hypotheses, to the
conjugated 2-spheres $C_{i}^{(-)}$ and  $C_{i}^{(+)}$ in $SU(2)$).

\begin{figure}[ht]
\begin{center}
\includegraphics[scale=.4]{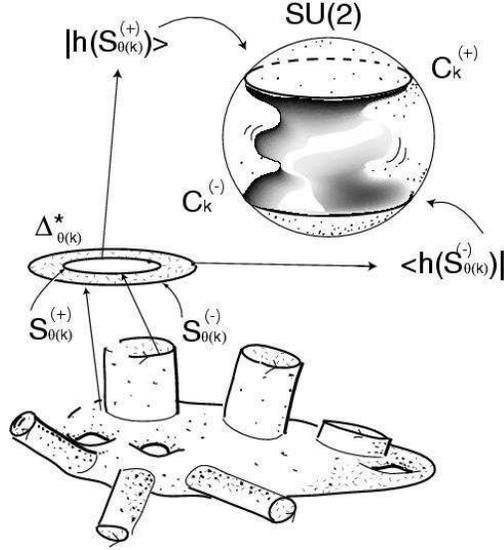}
\caption{A pictorial rendering of the set up for computing the quantum amplitudes for the cylindrical ends associated with the surface $\partial M$.}
\end{center}
\end{figure}

The computation of the annulus partition function (\ref{annz}) has been explicitly carried out \cite{gaberdiel} for the boundary $SU(2)$ CFT at level $\kappa=1$. We restrict our analysis to this particular case and  
if we apply the results of \cite{gaberdiel}, (see in particular eqn. (4.1) and the accompanying analysis) we get
\begin{multline}
Z_{\Delta _{\theta (i)}^{\ast }}(\{C_{i}^{\pm }\})= \\
=\frac{1}{\sqrt{2}}\sum_{j_i\in \frac{1}{2}\mathbb{Z}_{+}}\sum_{m,n}
(-1)^{m-n}D_{-m,-n}^{j_i}(\widehat{h}^{-1}(S_{\theta (i)}^{(-)}))D_{m,n}^{j_i}(\widehat{h}(S_{\theta
(i)}^{(+)}))\chi _{j_i^{2}}(e^{-\frac{4\pi }{\theta (i)}}),  
\end{multline}
where 
\begin{equation}
\chi _{j_i^{2}}(e^{-\frac{4\pi }{\theta (i)}})=\frac{e^{-\frac{4\pi }{\theta
(i)}j_i^{2}}-e^{-\frac{4\pi }{\theta (i)}(j_i+1)^{2}}}{\eta (e^{-\frac{4\pi }{
\theta (i)}})}.
\end{equation}
is the character of
the Virasoro highest weight representation, and 
\begin{equation}
\eta (q)\doteq q^{\frac{1}{24}}\prod_{n=1}^{\infty }(1-q^{n}),
\end{equation}
is the
Dedekind $\eta $-function. 

By diagonalizing we can consider $
h^{-1}(S_{\theta (i)}^{(-)})h(S_{\theta (i)}^{(+)})$ as an element of the maximal torus in $SU(2)$, i.e., we can write   
\begin{equation}
h^{-1}(S_{\theta (i)}^{(-)})h(S_{\theta (i)}^{(+)})=\left( 
\begin{array}{cc}
e^{4\pi \sqrt{-1}\lambda(i)} & 0 \\ 
0 & e^{-4\pi \sqrt{-1} \lambda(i)}
\end{array}
\right) ,
\end{equation}
and a representation-theoretic computation \cite{gaberdiel} eventually provides
\begin{equation}
Z_{\Delta _{\theta (i)}^{\ast }}(\{C_{i}^{\pm }\})=\frac{1}{\sqrt{2}}
\sum_{j\in \frac{1}{2}\mathbb{Z}_{+}}\cos (8\pi j_i\lambda(i))\frac{e^{-\frac{
4\pi }{\theta (i)}j_i^{2}}}{\eta (e^{-\frac{4\pi }{\theta (i)}})}.
\label{stringy}
\end{equation}
(Note that ${\alpha}$ in \cite{gaberdiel} corresponds to our $4\pi \sqrt{-1}\lambda(i)$, hence the presence of $\cos(8\pi j_i\lambda(i))$ in place of their $\cosh(2j_i\alpha(i))$). 

\bigskip

An important point to stress is that, according to the above analysis, the
partition function $Z_{\Delta _{\theta (i)}^{\ast }}(\{C_{i}^{\pm }\})$ can
be interpreted as the superposition over all possible $j_{i}$ channel
amplitudes 
\begin{equation}
\partial \Gamma _{i}\longmapsto A(j_{i})\doteq \frac{1}{\sqrt{2}}\cos
(8\pi j_{i}\lambda (i))\frac{e^{-\frac{4\pi }{\theta (i)}j_{i}^{2}}}{\eta (e^{-
\frac{4\pi }{\theta (i)}})}  \label{cellampl}
\end{equation}
that can be associated to the boundary component $\partial \Gamma _{i}$ of
the ribbon graph $\Gamma $. Such amplitudes can be interpreted as the
various $j_{i}=(n+\lambda (i))$, ($0\leq \lambda (i)\leq \frac{1}{2}$),
Virasoro (closed string) modes propagating along the cylinder $\Delta
_{\theta (i)}^{\ast }$.

\subsection{The Ribbon graph insertion operators}

In order to complete the picture, we need to discuss how the $N_{0}$
amplitudes $\{A(j_{i})\}$ defined by (\ref{cellampl}) interact along $\Gamma 
$. Such an interaction is described by boundary operators which mediate the
change in the boundary conditions $|\widehat{h}(S_{\theta (p)}^{(+)})\rangle
_{\partial \Gamma _{p}}$ and $|\widehat{h}(S_{\theta (q)}^{(+)})\rangle
_{\partial \Gamma _{q}}$ between any two adjacent boundary components $
\partial \Gamma _{p}$ and $\partial \Gamma _{q}$, (note that the adjacent
boundaries of the ribbon graph are associated with adjacent cells $\rho
^{2}(p)$, $\rho ^{2}(q)$ of \ $|P_{T_{l}}|\rightarrow M$, and thus to the
edges $\sigma ^{1}(p,q)$ of the triangulation $|T_{l}|\rightarrow M$). In
particular, the coefficients of the operator product expansion (OPE),
describing the short-distance behavior of the boundary operators on adjacent 
$\partial \Gamma _{p}$ and $\partial \Gamma _{q}$, will keep tract of the
combinatorics associated with $|P_{T_{l}}|\rightarrow {M}$.

To this end, let us consider generic pairwise adjacent 2-cells $\rho ^{2}(p)$
, $\rho ^{2}(q)$ and $\rho ^{2}(r)$ in $|P_{T_{l}}|\rightarrow M$, and the
associated cyclically ordered 3-valent vertex $\rho ^{0}(p,q,r)\in
|P_{T_{l}}|\rightarrow M$. Let $\{U_{\rho ^{0}(p,q,r)},w\}$ the coordinate
neighborhood of such a vertex, and $\{U_{\rho ^{1}(p,q)},z\}$, $\{U_{\rho
^{1}(q,r)},z\}$, and $\{U_{\rho ^{1}(r,p)},z\}$ the neighborhoods of the
corresponding oriented edges, (the $z$'s appearing in distinct $\{U_{\rho
^{1}(\circ ,\bullet )},z\}$ are distinct). Consider the edge $\rho ^{1}(p,q)$
and two (infinitesimally neighboring) points $z_{1}=x_{1}+\sqrt{-1}y_{1}$
and $z_{2}=x_{2}+\sqrt{-1}y_{2}$, $\func{Re}z_{1}=\func{Re}z_{2}$, in the
corresponding $U_{\rho ^{1}(p,q)}$, with $x_{1}=x_{2}$. Thus, for $
y_{1}\rightarrow 0^{+}$ we approach $\partial \Gamma _{p}\cap \rho ^{1}(p,q)$
, whereas for $y_{2}\rightarrow 0^{-}$ we approach a point $\ \in \partial
\Gamma _{q}\cap \rho ^{1}(q,p)$.

\begin{figure}[ht]
\begin{center}
\includegraphics[bb= 50 160 520 550, scale=.6]{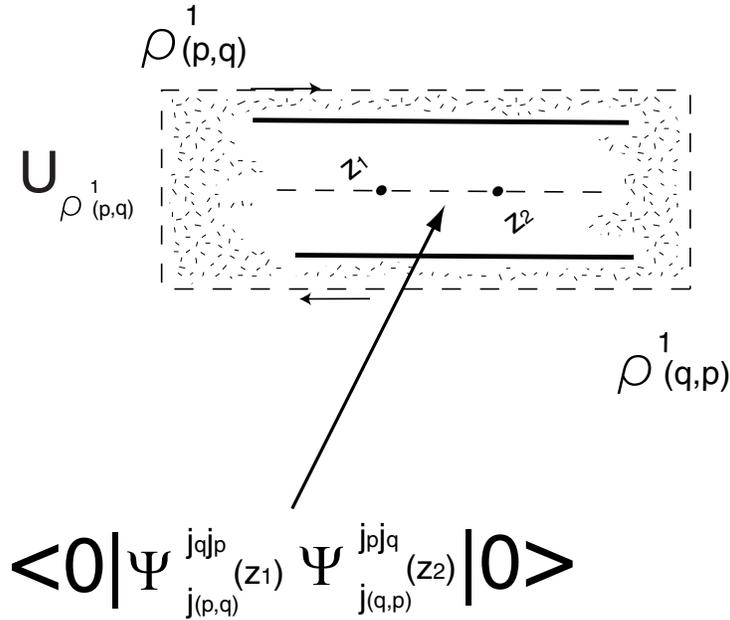}
\caption{The insertion of boundary operators $\psi _{j_{(p,q)}}^{j_{q}j_{p}}$
 in the complex coordinate neighborhood $U_{\rho ^{1}(p,q)}$, giving rise to the two-point function in the corresponding oriented edge $\rho^{1}(p,q)$.}\label{fig13}
\end{center}
\end{figure}

Associated with the edge $\rho ^{1}(p,q)\ $
we have the two adjacent boundary conditions $|\widehat{h}(S_{\theta
(p)}^{(+)})\rangle _{\partial \Gamma _{p}}$, and $|\widehat{h}(S_{\theta
(q)}^{(+)})\rangle _{\partial \Gamma _{q}}$, respectively describing the
given values of the field $\widehat{h}$ on the two boundary components $
\partial \Gamma _{p}\cap \rho ^{1}(p,q)$ and $\partial \Gamma _{q}\cap \rho
^{1}(q,p)$ of $\rho ^{1}(p,q)$. At the points $z_{1},z_{2}\in $ $U_{\rho
^{1}(p,q)}$ we can consider the insertion of boundary operators $\psi
_{j_{(p,q)}}^{j_{q}j_{p}}(z_{1})$ and $\psi _{j_{(q,p)}}^{j_{p}j_{q}}(z_{2})$
mediating between the corresponding boundary conditions, \emph{i.e.} 
\begin{eqnarray}
&&\psi _{j_{(p,q)}}^{j_{q}j_{p}}(z_{1})|\widehat{h}(S_{\theta
(p)}^{(+)})\rangle _{\partial \Gamma _{p}}\underset{y_{1}\rightarrow 0^{+}}{=
}|\widehat{h}(S_{\theta (q)}^{(+)})\rangle _{\partial \Gamma _{q}},  \notag
\\
&& \\
&&\psi _{j_{(q,p)}}^{j_{p}j_{q}}(z_{2})|\widehat{h}(S_{\theta
(q)}^{(+)})\rangle _{\partial \Gamma _{q}}\underset{y_{2}\rightarrow 0^{-}}{=
}|\widehat{h}(S_{\theta (p)}^{(+)})\rangle _{\partial \Gamma _{p}}.  \notag
\end{eqnarray}
Note that $\psi _{j_{(p,q)}}^{j_{q}j_{p}}$ carries the single primary isospin label $
j_{(p,q)}$\ (also indicating the oriented edge $\rho ^{1}(p,q)$ where we are
inserting the operator), and the two additional isospin labels $j_{p}$ and $j_{q}$
indicating the two boundary conditions at the two portions of $\partial
\Gamma _{p}$ and $\partial \Gamma _{q}$\ adjacent to the insertion edge $
\rho ^{1}(p,q)$. \ Likewise, by considering the oriented edges $\rho
^{1}(q,r)$ and $\rho ^{1}(r,p)$, \ we can introduce the operators $\psi
_{j_{(r,q)}}^{j_{q}j_{r}}$, $\psi _{j_{(q,r)}}^{j_{r}j_{q}}$, $\psi
_{j_{(p,r)}}^{j_{r}j_{p}}$, and $\psi _{j_{(r,p)}}^{j_{p}j_{r}}$. In full
generality, we can rewrite the above definition explicitly in terms of the
adjacency matrix $B(\Gamma )$ of the ribbon graph $\Gamma $, 
\begin{equation}
B_{st}(\Gamma )=\left\{ 
\begin{tabular}{lll}
$1$ & if & $\rho ^{1}(s,t)$ is an edge of $\Gamma $ \\ 
&  &  \\ 
$0$ &  & otherwise
\end{tabular}
\right. ,
\end{equation}
according to 
\begin{equation}
\psi _{j_{(p,q)}}^{j_{q}j_{p}}(z_{1})|\widehat{h}(S_{\theta
(p)}^{(+)})\rangle _{\partial \Gamma _{p}}\underset{y_{1}\rightarrow 0^{+}}{=
}B_{pq}(\Gamma )|\widehat{h}(S_{\theta (q)}^{(+)})\rangle _{\partial \Gamma
_{q}}.
\end{equation}
Any\ such boundary operator, say $\psi _{j_{(p,q)}}^{j_{q}j_{p}}$, is a
primary field (under the action of Virasoro algebra) of conformal dimension $
H_{j_{(p,q)}}$, and they are all characterized \cite{sagnotti}, \cite{lewellen}, \cite{felder} by the following properties
dictated by conformal invariance (in the corresponding coordinate
neighborhood $U_{\rho ^{1}(p,q)}$, see figure \ref{fig13}) 
\begin{gather}
\langle 0|\psi _{j_{(p,q)}}^{j_{q}j_{p}}(z_{1})|0\rangle =0,\;\;\langle 
\widehat{h}(S_{\theta (p)}^{(-)})|\mathbb{I}^{j_{p}j_{p}}|\widehat{h}
(S_{\theta (p)}^{(+)})\rangle \;\!=a^{j_{p}j_{p}},  \notag \\
\label{twopoints} \\
\langle 0|\psi _{j_{(p,q)}}^{j_{q}j_{p}}(z_{1})\psi
_{j_{(q,p)}}^{j_{p}j_{q}}(z_{2})|0\rangle
=b_{j_{(p,q)}}^{j_{q}j_{p}}|z_{1}-z_{2}|^{-2H_{j_{(p,q)}}}\delta
_{j_{(p,q)}j_{(q,p)}},  \notag
\end{gather}
where $\mathbb{I}^{j_{p}j_{p}}$ is the identity operator, and where $
a^{j_{p}j_{p}}$ and $b_{j_{(p,q)}}^{j_{q}j_{p}}$ are normalization factors.
In particular, the parameters $b_{j_{(p,q)}}^{j_{q}j_{p}}$ define the
normalization of the two-points function. Note that \cite{sagnotti} for $SU(2)$ the $
b_{j_{(p,q)}}^{j_{q}j_{p}}$ are such that $
b_{j_{(p,q)}}^{j_{q}j_{p}}=b_{j_{(q,p)}}^{j_{p}j_{q}}(-1)^{2j_{(p,q)}}$, and
are (partially) constrained by the OPE of the $\psi _{j_{(p,q)}}^{j_{q}j_{p}}
$. As customary in boundary CFT, we leave such a normalization factors
dependence explicit in what follows.

\begin{figure}[ht]
\begin{center}
\includegraphics[bb= 50 130 550 670, scale=.5]{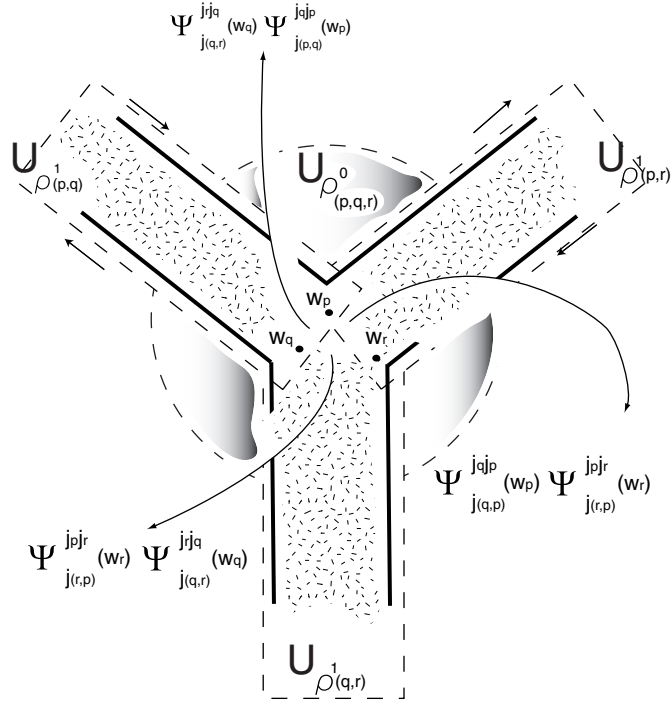}
\caption{The OPEs between the boundary operators around a given vertex $\rho^{0}(p,q,r)$ in the corresponding complex coordinates neighborhoods $U_{\rho ^{0}(p,q,r)}$, $U_{\rho ^{1}(p,q)}$, etc..}\label{fig14}
\end{center}
\end{figure}

In order to discuss the properties of \ the $\psi _{j_{(p,q)}}^{j_{q}j_{p}}$
, let us extend the (edges) coordinates $z$\ \ to the unit disk \ $U_{\rho
^{0}(p,q,r)}$ associated to the generic vertex $\rho ^{0}(p,q,r)$, and
denote by 
\begin{eqnarray}
w_{p} &=&\frac{\varepsilon }{3}e^{\frac{1}{2}\pi \sqrt{-1}}\in U_{\rho
^{0}(p,q,r)}\cap U_{\rho ^{1}(p,q)}  \notag \\
w_{q} &=&\frac{\varepsilon }{2}e^{\frac{7}{6}\pi \sqrt{-1}}\in U_{\rho
^{0}(p,q,r)}\cap U_{\rho ^{1}(q,r)}  \label{wcoord} \\
w_{r} &=&\varepsilon e^{\frac{11}{6}\pi \sqrt{-1}}\in U_{\rho
^{0}(p,q,r)}\cap U_{\rho ^{1}(r,p)}  \notag
\end{eqnarray}
the coordinates of three points in an $\varepsilon $- neighborhood ($
0<\varepsilon <1$) of the vertex $w=0$, (fractions of $\varepsilon $ are
introduced for defining a radial ordering; note also that by exploting the
coordinate changes (\ref{glue1}), one can easily map such points in the upper half
planes associated with the edge complex variables $z$ corresponding to $
U_{\rho ^{1}(p,q)}$, $U_{\rho ^{1}(q,r)}$, and $U_{\rho ^{1}(r,p)}$, and
formulate the theory in a more conventional fashion). To
these points we associate the insertion of boundary operators $\psi
_{j_{(r,p)}}^{j_{p}j_{r}}(w_{r})$, $\psi _{j_{(q,r)}}^{j_{r}j_{q}}(w_{q})$, $
\psi _{j_{(p,q)}}^{j_{q}j_{p}}(w_{p})$ which pairwise mediate among the
boundary conditions $|\widehat{h}(S_{\theta (p)}^{(+)})\rangle $, $|\widehat{
h}(S_{\theta (q)}^{(+)})\rangle $, and $|\widehat{h}(S_{\theta
(r)}^{(+)})\rangle $. The behavior of such insertions at the vertex $\rho
^{0}(p,q,r)$, (\emph{i.e.}, as $\varepsilon \rightarrow 0$), is described by
the following OPEs (see \cite{lewellen}, \cite{sagnotti})
\begin{gather}
\psi _{j_{(r,p)}}^{j_{p}j_{r}}(w_{r})\psi _{j_{(q,r)}}^{j_{r}j_{q}}(w_{q})= 
\notag \\
\label{OPE} \\
=\sum_{j}C_{j_{(r,p)}j_{(q,r)}j}^{j_{p}j_{r}j_{q}}|w_{r}-w_{q}|^{H_{j}-H_{j_{(r,p)}}-H_{j_{(q,r)}}}(\psi _{j}^{j_{p}j_{q}}(w_{q})+...),
\notag
\end{gather}
\begin{gather}
\psi _{j_{(q,r)}}^{j_{r}j_{q}}(w_{q})\psi _{j_{(p,q)}}^{j_{q}j_{p}}(w_{p})= 
\notag \\
\\
=\sum_{j}C_{j_{(q,r)}j_{(p,q)}j}^{j_{r}j_{q}j_{p}}|w_{q}-w_{p}|^{H_{j}-H_{j_{(q,r)}}-H_{j_{(p,q)}}}(\psi _{j}^{j_{r}j_{p}}(w_{p})+...),
\notag
\end{gather}
\begin{gather}
\psi _{j_{(p,q)}}^{j_{q}j_{p}}(w_{p})\psi _{j_{(r,p)}}^{j_{p}j_{r}}(w_{r})= 
\notag \\
\\
=\sum_{j}C_{j_{(p,q)}j_{(r,p)}j}^{j_{q}j_{p}j_{r}}|w_{p}-w_{r}|^{H_{j}-H_{j_{(p,q)}}-H_{j_{(r,p)}}}(\psi _{j}^{j_{q}j_{p}}(w_{r})+...),
\notag
\end{gather}
where the dots stand for higher order corrections in $|w_{\circ }-w_{\bullet
}|$, the $H_{J_{...}}$ are the conformal weights of the corresponding
boundary operators, and the $C_{j_{(r,p)}j_{(q,r)}j}^{j_{p}j_{r}j_{q}}$\ are
the OPE structure constants (see figure \ref{fig14}).

\begin{figure}[ht]
\begin{center}
\includegraphics[bb= 50 270 520 530, scale=.6]{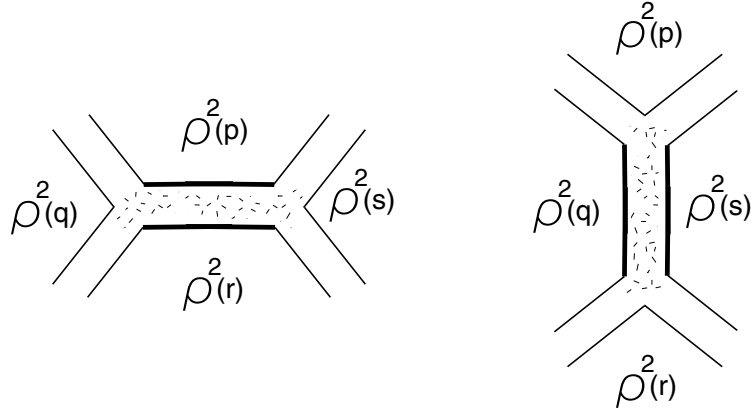}
\caption{The dual channels in evaluating the correlation function of the four boundary operators corresponding to the four boundary components involved.}\label{fig15}
\end{center}
\end{figure}

As is well known \cite{sagnotti}, the parameters $
b_{j_{(p,q)}}^{j_{q}j_{p}}$ and the constants $
C_{j_{(r,p)}j_{(q,r)}j}^{j_{p}j_{r}j_{q}}$ are not independent. In our
setting this is a consequence of the fact that to the oriented vertex $\rho
^{0}(p,q,r)$ we can associate a three-point function which must be invariant
under cyclic permutations, \emph{i.e.} 
\begin{gather}
\langle \psi _{j_{(r,p)}}^{j_{p}j_{r}}(w_{r})\psi
_{j_{(q,r)}}^{j_{r}j_{q}}(w_{q})\psi _{j_{(p,q)}}^{j_{q}j_{p}}(w_{p})\rangle
=\langle \psi _{j_{(p,q)}}^{j_{q}j_{p}}(w_{r})\psi
_{j_{(r,p)}}^{j_{p}j_{r}}(w_{q})\psi _{j_{(q,r)}}^{j_{r}j_{q}}(w_{p})\rangle
=  \notag \\
\\
=\langle \psi _{j_{(q,r)}}^{j_{r}j_{q}}(w_{r})\psi
_{j_{(p,q)}}^{j_{q}j_{p}}(w_{q})\psi _{j_{(r,p)}}^{j_{p}j_{r}}(w_{p})\rangle
.  \notag
\end{gather}
By using the boundary OPE (\ref{OPE}), each term can be computed in two
distinct ways, \emph{e.g.}, by denoting with $\underbrace{}$\ an OPE
pairing, we must have 
\begin{equation}
\langle \underbrace{\psi _{j_{(r,p)}}^{j_{p}j_{r}}(w_{r})\psi
_{j_{(q,r)}}^{j_{r}j_{q}}(w_{q})}\psi
_{j_{(p,q)}}^{j_{q}j_{p}}(w_{p})\rangle =\langle \psi
_{j_{(r,p)}}^{j_{p}j_{r}}(w_{r})\underbrace{\psi
_{j_{(q,r)}}^{j_{r}j_{q}}(w_{q})\psi _{j_{(p,q)}}^{j_{q}j_{p}}(w_{p})}
\rangle 
\end{equation}
which (by exploiting (\ref{twopoints})) in the limit $w\rightarrow 0$\ \
provides 
\begin{equation}
C_{j_{(r,p)}j_{(q,r)}j_{(p,q)}}^{j_{p}j_{r}j_{q}}b_{j_{(q,p)}}^{j_{p}j_{q}}=C_{j_{(q,r)}j_{(p,q)}j_{(r,p)}}^{j_{r}j_{q}j_{p}}b_{j_{(r,p)}}^{j_{p}j_{r}},
\label{Cval}
\end{equation}
(note that the Kronecker $\delta $ in (\ref{twopoints}) implies that 
$j_{(q,p)}=j_{(p,q)}$, etc. ). From the OPE evaluation of the remaining
two three-points function one similarly obtains 
\begin{eqnarray}
C_{j_{(p,q)}j_{(r,p)}j_{(q,r)}}^{j_{q}j_{p}j_{r}}b_{j_{(r,q)}}^{j_{q}j_{r}}
&=&C_{j_{(r,p)}j_{(q,r)}j_{(p,q)}}^{j_{p}j_{r}j_{q}}b_{j_{(p,q)}}^{j_{q}j_{p}},
\notag \\
&& \\
C_{j_{(q,r)}j_{(p,q)}j_{(r,p)}}^{j_{r}j_{q}j_{p}}b_{j_{(p,r)}}^{j_{r}j_{p}}
&=&C_{j_{(p,q)}j_{(r,p)}j_{(q,r)}}^{j_{q}j_{p}j_{r}}b_{j_{(q,r)}}^{j_{r}j_{q}}.
\notag
\end{eqnarray}
Since 
\begin{eqnarray}
b_{j_{(q,p)}}^{j_{p}j_{q}} &=&b_{j_{(p,q)}}^{j_{q}j_{p}}(-1)^{2j_{(p,q)}}, 
\notag \\
b_{j_{(r,p)}}^{j_{p}j_{r}} &=&b_{j_{(p,r)}}^{j_{r}j_{p}}(-1)^{2j_{(p,r)}}, \\
b_{j_{(q,r)}}^{j_{r}j_{q}} &=&b_{j_{(r,q)}}^{j_{q}j_{r}}(-1)^{2j_{(r,q)}}, 
\notag
\end{eqnarray}
one eventually gets 
\begin{eqnarray}
C_{j_{(r,p)}j_{(q,r)}j_{(p,q)}}^{j_{p}j_{r}j_{q}}b_{j_{(q,p)}}^{j_{p}j_{q}}
&=&(-1)^{2j_{(q,p)}}C_{j_{(p,q)}j_{(r,p)}j_{(q,r)}}^{j_{q}j_{p}j_{r}}b_{j_{(q,r)}}^{j_{r}j_{q}},
\notag \\
&&  \notag \\
C_{j_{(p,q)}j_{(r,p)}j_{(q,r)}}^{j_{q}j_{p}j_{r}}b_{j_{(r,q)}}^{j_{q}j_{r}}
&=&(-1)^{2j_{(r,q)}}C_{j_{(q,r)}j_{(p,q)}j_{(r,p)}}^{j_{r}j_{q}j_{p}}b_{j_{(r,p)}}^{j_{p}j_{r}},
\\
&&  \notag \\
C_{j_{(q,r)}j_{(p,q)}j_{(r,p)}}^{j_{r}j_{q}j_{p}}b_{j_{(p,r)}}^{j_{r}j_{p}}
&=&(-1)^{2j_{(p,r)}}C_{j_{(r,p)}j_{(q,r)}j_{(p,q)}}^{j_{p}j_{r}j_{q}}b_{j_{(p,q)}}^{j_{q}j_{p}},
\notag
\end{eqnarray}
which are the standard relation between the OPE parameters and the normalization of the 2-points function for boundary $SU(2)$ insertion operators, \cite{sagnotti}. Such a lengthy (and slightly pedantic) analysis is necessary to show that our association of boundary insertion operators $\psi
_{j_{(r,p)}}^{j_{p}j_{r}}$,  to the edges of the ribbon graph $\Gamma $ is actually consistent with $SU(2)$ boundary CFT, in particular that geometrically the correlator $\langle \psi
_{j_{(r,p)}}^{j_{p}j_{r}}(w_{r})\psi _{j_{(q,r)}}^{j_{r}j_{q}}(w_{q})\psi
_{j_{(p,q)}}^{j_{q}j_{p}}(w_{p})\rangle $ \ \ is associated with the three
mutually adjacent boundary components $\partial \Gamma _{p}$, $\partial
\Gamma _{q}$, and $\partial \Gamma _{r}$ of the ribbon graph $\Gamma $. More
generally, let us consider four mutually adjacent boundary components $
\partial \Gamma _{p}$, $\partial \Gamma _{q}$, $\partial \Gamma _{r}$, and $
\partial \Gamma _{s}$. Their adjacency relations can be organized in two
distinct ways labelled by the distinct two vertices they generate: if $
\partial \Gamma _{p}$ is adjacent to $\partial \Gamma _{r}$ then \ we have
the two vertices $\rho ^{0}(p,q,r)$ and $\rho ^{0}(p,r,s)$ connected by the
edge $\rho ^{1}(p,r)$; conversely, if $\partial \Gamma _{q}$ is adjacent to $
\partial \Gamma _{s}$ then \ we have the two vertices $\rho ^{0}(p,q,s)$ and 
$\rho ^{0}(q,r,s)$ connected by the edge $\rho ^{1}(q,s)$. \ It follows that
the correlation function of the corresponding four boundary operators, $
\langle \psi _{j_{(s,p)}}^{j_{p}j_{s}}\psi _{j_{(r,s)}}^{j_{s}j_{r}}\psi
_{j_{(q,r)}}^{j_{r}j_{q}}\psi _{j_{(p,q)}}^{j_{q}j_{p}}\rangle $, can be
evaluated by exploiting the ($(S)$-channel) factorization associated with
the coordinate neighborhood $\{U_{\rho ^{1}(r,p)},z^{(S)}\}$, or,
alternatively, by exploiting the ($(T)$-channel) factorization associated
with $\{U_{\rho ^{1}(q,s)},z^{(T)}\}$ (see figure \ref{fig15}). 

From the observation that both such
expansions must yield the same result, it is possible \cite{felder} to directly relate the
OPE coefficients $C_{j_{(r,p)}j_{(q,r)}j_{(p,q)}}^{j_{p}j_{r}j_{q}}$ with
the fusion matrices $F_{j_{r}j_{(p,q)}}\left[\begin{array}{cc}
j_p & j_q\\
j_{(r,p)} & j_{(q,r)}\end{array}\right]$ which express the crossing duality between
four-points conformal blocks. Recall that for WZW models the fusion ring can
be identified with the character ring of the quantum deformation $\mathcal{U}
_{Q}(\mathbf{g})$ of \ the enveloping algebra of $\mathbf{g}$\ evaluated at
the root of unity given by $Q=e^{\pi \sqrt{-1}/(\kappa +h^{\vee })}$\ (where 
$h^{\vee }$ is the dual Coxeter number and $\kappa $ is the level of the
theory). In other words, for WZW models, the fusion\ \ matrices are the $6j$
-symbols of the corresponding (quantum) group. From such remarks, it follows
that in our case (\emph{i.e.}, for $\kappa =1$, $h^{\vee }=2$) the structure
constants $C_{j_{(r,p)}j_{(q,r)}j_{(p,q)}}^{j_{p}j_{r}j_{q}}$ are suitable
entries \cite{gaume} of the $6j$-symbols of the quantum group $SU(2)_{Q=e^{\frac{\pi }{3}
\sqrt{-1}}}$, $\emph{i.e.}$
\begin{equation}
C_{j_{(r,p)}j_{(q,r)}j_{(p,q)}}^{j_{p}j_{r}j_{q}}=\left\{ 
\begin{array}{ccc}
j_{(r,p)} & j_{p} & j_{r} \\ 
j_{q} & j_{(q,r)} & j_{(p,q)}
\end{array}
\right\} _{Q=e^{\frac{\pi }{3}\sqrt{-1}}}  \label{seigei}
\end{equation}

\subsection{The partition function.}

The final step in our construction is to uniformize the local coordinate
representation of the ribbon graph $\Gamma $ with the cylindrical metric $
\{|\phi (i)|\}$, defined by the quadratic differential $\{\phi (i)\}$. In
such a framework, there is a natural prescription for associating to the
resulting metric ribbon graph $(\Gamma ,\{|\phi (i)|\})$ a factorization of
the correlation functions of $\ $the $N_{1}$ insertion operators $\{\psi
_{j_{(p,q)}}^{j_{q}j_{p}}\}$, (recall that $N_{1}$ is the number of edges of 
$\Gamma $). Explicitly, for the generic vertex \ $\rho ^{0}(p,q,r)$, let \ $
z_{p}^{(0)}\in U_{\rho ^{1}(p,q)}\cap U_{\rho ^{0}(p,q,r)}$, \ $
z_{q}^{(0)}\in U_{\rho ^{1}(q,r)}\cap U_{\rho ^{0}(p,q,r)}$, and \ $
z_{r}^{(0)}\in U_{\rho ^{1}(r,p)}\cap U_{\rho ^{0}(p,q,r)}$ respectively
denote the coordinates of the points $w_{p}$, $w_{q}$, and $w_{r}$ (see (\ref
{wcoord})) in the respective edge uniformizations, and for notational
purposes, let us set, (in an $\varepsilon $-neighborhood of $z_{\rho
^{0}(p,q,r)}=0\in U_{\rho ^{0}(p,q,r)}$), 
\begin{eqnarray}
\psi _{j_{(r,p)}}^{j_{p}j_{r}}(\rho ^{0}(p,q,r)) &\doteq &\psi
_{j_{(r,p)}}^{j_{p}j_{r}}(z_{r}^{(0)}),  \notag \\
\psi _{j_{(q,r)}}^{j_{r}j_{q}}(\rho ^{0}(p,q,r)) &\doteq &\psi
_{j_{(q,r)}}^{j_{r}j_{q}}(z_{q}^{(0)}), \\
\psi _{j_{(p,q)}}^{j_{q}j_{p}}(\rho ^{0}(p,q,r)) &\doteq &\psi
_{j_{(p,q)}}^{j_{q}j_{p}}(z_{p}^{(0)}).  \notag
\end{eqnarray}
Let us consider, (in the limit $\varepsilon \rightarrow 0$ ),\ the
correlation function 
\begin{gather}
\left\langle \bigotimes_{i=1}^{N_{0}(T)}\partial \Gamma _{i};\otimes
j_{i}\right\rangle \doteq   \notag \\
\\
\doteq \left\langle \prod_{\{\rho ^{0}(p,q,r)\}}^{N_{2}(T)}\psi
_{j_{(r,p)}}^{j_{p}j_{r}}(\rho ^{0}(p,q,r))\psi
_{j_{(q,r)}}^{j_{r}j_{q}}(\rho ^{0}(p,q,r))\psi
_{j_{(p,q)}}^{j_{q}j_{p}}(\rho ^{0}(p,q,r))\right\rangle ,  \notag
\end{gather}
where \ the product runs over the $N_{2}(T)$ vertices $\{\rho ^{0}(p,q,r)\}$
\ of $\Gamma $.\ We can factorize it along the $N_{1}(T)$ channels generated
by the edge cordinate neighborhoods $\{U_{\rho ^{1}(p,q)}\}$ according to 
\begin{gather}
\left\langle \bigotimes_{i=1}^{N_{0}(T)}\partial \Gamma _{i};\otimes
j_{i}\right\rangle =  \notag \\
\\
=\sum_{\{j_{(r,p)}\}}\prod_{\{\rho ^{0}(p,q,r)\}}^{N_{2}(T)}\underset{\rho
^{0}(p,q,r)}{\left\langle \psi _{j_{(r,p)}}^{j_{p}j_{r}}\psi
_{j_{(q,r)}}^{j_{r}j_{q}}\psi _{j_{(p,q)}}^{j_{q}j_{p}}\right\rangle }
\prod_{\{\rho ^{1}(p,r)\}}^{N_{1}(T)}\underset{\rho ^{1}(p,r)}{\left\langle
\psi _{j_{(r,p)}}^{j_{p}j_{r}}\psi _{j_{(p,r)}}^{j_{r}j_{p}}\right\rangle } 
\notag
\end{gather}
where we have set 
\begin{gather}
\underset{\rho ^{0}(p,q,r)}{\left\langle \psi _{j_{(r,p)}}^{j_{p}j_{r}}\psi
_{j_{(q,r)}}^{j_{r}j_{q}}\psi _{j_{(p,q)}}^{j_{q}j_{p}}\right\rangle }\doteq 
\notag \\
\\
\doteq \left\langle \psi _{j_{(r,p)}}^{j_{p}j_{r}}(\rho ^{0}(p,q,r))\psi
_{j_{(q,r)}}^{j_{r}j_{q}}(\rho ^{0}(p,q,r))\psi
_{j_{(p,q)}}^{j_{q}j_{p}}(\rho ^{0}(p,q,r))\right\rangle ,  \notag
\end{gather}
\begin{equation}
\underset{\rho ^{1}(p,r)}{\left\langle \psi _{j_{(r,p)}}^{j_{p}j_{r}}\psi
_{j_{(p,r)}}^{j_{r}j_{p}}\right\rangle }\doteq \left\langle \psi
_{j_{(r,p)}}^{j_{p}j_{r}}(\rho ^{0}(p,q,r))\psi
_{j_{(p,r)}}^{j_{r}j_{p}}(\rho ^{0}(p,r,s))\right\rangle ,
\end{equation}
and where the summation runs over all $N_{1}(T)$ primary highest weight
representation $\widehat{\mathfrak{su}}(2)_{\kappa =1}$,\ labelling the
intermediate edge channels $\{j_{(r,p)}\}$. Note that according to (\ref
{twopoints}) we can write 
\begin{equation}
\underset{\rho ^{1}(p,r)}{\left\langle \psi _{j_{(r,p)}}^{j_{p}j_{r}}\psi
_{j_{(p,r)}}^{j_{r}j_{p}}\right\rangle }
=b_{j_{(r,p)}}^{j_{p}j_{r}}L(p,r)^{-2H_{j_{(r,p)}}},
\label{critedge}
\end{equation}
(recall that $j_{(r,p)}=j_{(p,r)}$), where $L(p,r)$ denotes the length of
the edge $\rho ^{1}(p,r)$ in the uniformization $(U_{\rho ^{1}(p,r)},\{|\phi
(i)|\})$. Moreover, \ since (see (\ref{Cval})) 
\begin{equation}
\underset{\rho ^{0}(p,q,r)}{\left\langle \psi _{j_{(r,p)}}^{j_{p}j_{r}}\psi
_{j_{(q,r)}}^{j_{r}j_{q}}\psi _{j_{(p,q)}}^{j_{q}j_{p}}\right\rangle }
=C_{j_{(r,p)}j_{(q,r)}j_{(p,q)}}^{j_{p}j_{r}j_{q}}b_{j_{(q,p)}}^{j_{p}j_{q}},
\end{equation}
we get for the boundary operator correlator associated with the 
ribbon graph $\Gamma $ the expression 
\begin{gather}
\left\langle \bigotimes_{i=1}^{N_{0}(T)}\partial \Gamma _{i};\otimes
j_{i}\right\rangle =  \notag \\
\\
=\sum_{\{j_{(r,p)}\}}\prod_{\{\rho
^{0}(p,q,r)
\}}^{N_{2}(T)}C_{j_{(r,p)}j_{(q,r)}j_{(p,q)}}^{j_{p}j_{r}j_{q}}b_{j_{(q,p)}}^{j_{p}j_{q}}\prod_{\{\rho ^{1}(p,r)\}}^{N_{1}(T)}b_{j_{(r,p)}}^{j_{p}j_{r}}L(p,r)^{-2H_{j_{(r,p)}}}.
\notag
\end{gather}


By identifying each $C_{j_{(r,p)}j_{(q,r)}j_{(p,q)}}^{j_{p}j_{r}j_{q}}$ with
the corresponding $6j$-symbol, and observing that each normalization factor $
b_{j_{(q,p)}}^{j_{p}j_{q}}$ occurs exactly twice, we eventually obtain 
\begin{gather}
\left\langle \bigotimes_{i=1}^{N_{0}(T)}\partial \Gamma _{i};\otimes
j_{i}\right\rangle =\sum_{\{j_{(r,p)}\}}\prod_{\{\rho ^{0}(p,q,r)\}}^{N_{2}(T)}
\left\{ 
\begin{array}{ccc}
j_{(r,p)} & j_{p} & j_{r} \\ 
j_{q} & j_{(q,r)} & j_{(p,q)}
\end{array}
\right\}_{Q=e^{\frac{\pi }{3}\sqrt{-1}}}\cdot  \notag 
\label{graphA} \\
\cdot\prod_{\{\rho
^{1}(p,r)\}}^{N_{1}(T)}\left( b_{j_{(r,p)}}^{j_{p}j_{r}}\right)
^{2}L(p,r)^{-2H_{j_{(r,p)}}}.  \notag
\end{gather}
As the notation suggests, such a correlator has a residual
dependence on the representation labels $\{j_{i}\}$. In other words, it can
be considered as an element of the tensor product $\mathcal{H}(\partial
\Gamma )=$ $\otimes _{i=1}^{N_{0}(T)}\mathcal{H}_{(i)}$. It is then natural
to interpret its evaluation over the amplitudes $\{A(j_{i})\}$ defined by (
\ref{cellampl}) as the partition function $Z^{WZW}(|P_{T_{l}}|,\{\widehat{h}
(S_{\theta (i)}^{(+)})\})$ associated with the quantum amplitude (\ref{amplit}),
and describing the $SU(2)$ WZW model (at level $\kappa =1$) on a random
Regge polytope $|P_{T_{l}}|\rightarrow M$. By inserting the $N_{0}(T)$
amplitudes $\{A(j_{i})\}$ into (\ref{graphA}), and summing over all possible
representation indices $\{j_{p}\}$ we immediately get 
\begin{gather}
Z^{WZW}(|P_{T_{l}}|,\{\widehat{h}(S_{\theta (i)}^{(+)})\})=  \notag \\
\notag \\
=\left( \frac{1}{\sqrt{2}}\right) ^{N_{0}(T)}\sum_{\{j_{p}\in \frac{1}{2}
\mathbb{Z}_{+}\}}\sum_{\{j_{(r,p)}\}}\prod_{\{\rho
^{0}(p,q,r)\}}^{N_{2}(T)}\left\{ 
\begin{array}{ccc}
j_{(r,p)} & j_{p} & j_{r} \\ 
j_{q} & j_{(q,r)} & j_{(p,q)}
\end{array}
\right\} _{Q=e^{\frac{\pi }{3}\sqrt{-1}}}\cdot   \label{FinPart} \\
\notag \\
\cdot \prod_{\{\rho ^{1}(p,r)\}}^{N_{1}(T)}\left(
b_{j_{(r,p)}}^{j_{p}j_{r}}\right) ^{2}L(p,r)^{-2H_{j_{(r,p)}}}\cos
(8\pi j_{p}\lambda (i))\frac{e^{-\frac{4\pi }{\theta (i)}j_{p}^{2}}}{\eta (e^{-
\frac{4\pi }{\theta (i)}})},  \notag
\end{gather}
where the summation $\sum_{\{j_{p}\in \frac{1}{2}\mathbb{Z}_{+}\}}$ is over
all possible $N_{0}(T)$ channels $j_{p}$ describing the Virasoro (closed
string) modes propagating along the cylinders $\{\Delta _{\theta (p)}^{\ast
}\}_{p=1}^{N_{0}(T)}$.
This is the partition function of our WZW model on a random Regge triangulation. The WZW fields are still present through their boundary labels $\lambda(i)$, (which can take the values 
$0,1/2$), wheras the metric geometry of the polytope enters
explicitly both with the edge-length terms $
L(p,r)^{-2H_{j_{(r,p)}}}$ and with the conical angle factors $\frac{e^{-
\frac{4\pi }{\theta (i)}j_{p}^{2}}}{\eta (e^{-\frac{4\pi }{\theta (i)}})}$. The expression of 
$Z^{WZW}(|P_{T_{l}}|,\{\widehat{h}
(S_{\theta (i)}^{(+)})\})$, also shows the mechanism through which the $SU(2)$ fields couple with
simplicial curvature: the coupling amplitudes $\{A(j_{i})\}$ can be
interpreted as describing a closed string emitted by $\partial \Gamma
_{i}\simeq S_{\theta (i)}^{(-)}$, or rather by the $\overline{S_{\theta
(i)}^{2}}$\ \ brane image of this boundary component in $SU(2)$, and
absorbed by the brane $S_{\theta (i)}^{2}$ image of the outer boundary $
S_{\theta (i)}^{(+)}$, (the curvature carrying vertex). This exchange of
closed strings between $2$-branes in \ $SU(2)\simeq S^{3}$ describes the
interaction of the quantum $SU(2)$ field with the classical gravitational
background associated with the edge-length assignments $\{L(p,r)\}$, and
with the deficit angles $\{\varepsilon (i)\doteq 2\pi -\theta (i)\}$. 

\chapter{Holography and asymptotically flat space-times}
One of the key success of the holographic principle is the chance to reconstruct spacetime starting from datas living on the boundary. The classical
example is the AdS/CFT correspondence \cite{ADS/CFT} where an equivalence between partition functions of a theory living in the bulk and in the boundary is imposed. 
More precisely the latter, in asymptotically AdS space-times, is strictly related to a gauge theory carrying indices of the asymptotic symmetry group (the conformal group).

This kind of relation has been mainly studied in the AdS case (and with less success also in dS spacetimes \cite{DS/CFT}) but no real attempt until now
has been done in aymptotically flat space-times. Thus, the purpose of this chapter will be the study of the asymptotic symmetry group of asymptotically flat
space-times (the Bondi-Metzner-Sachs group) as a candidate for the gauge group of the holographic boundary theory. Furthermore we will show the differences
araising from the AdS case emphasizing in particular the failure of a unique geometric reconstruction of an asymptotically flat space-time.
 
\section{Emergence of the BMS group in asymptotically flat spacetimes}
In this section we briefly revisit the derivation of the BMS group and some of its 
properties. As said, the BMS group represents the ASG of asymptotically 
flat spacetimes. More properly it is a transformation pseudo-group of 
asymptotic isometries of the region close to infinity of the asymptotically 
flat (lorentzian) spacetimes \footnote{For the euclidean case see section \ref{wave}.}. There is however a derivation proposed by Penrose as a group of transformations 
living intrinsically {\it on} $\Im$. In this case the BMS group {\it is} 
the transformation group on the boundary $\Im$.\footnote{Recall that $\Im$ is the 
disjoint union of $\Im^+$ (future null infinity) with $\Im^-$ (past null
infinity). In the rest of the paper we will refer to $\Im^+$ but because of the
symmetry 
the same conclusions will hold on $\Im^-$  too in all cases 
unless differently specified.}

From the point of view of the holographic principle we are interested in the 
theory living on the boundary and its symmetry group. We therefore 
prefer to consider this ``boundary description'' of the BMS group as more relevant 
and fundamental for our purposes. Moreover, it keeps into 
account the degenerate nature of $\Im$, which one has to face up when 
choosing a null screen. For completeness, however, we review various derivations.
\subsection{BMS as asymptotic symmetry group}
The BMS group was originally discovered \cite{Bondi}, \cite{sachsa} by studying gravitational radiation 
emitted by bounded systems in asymptotically flat spacetimes; it is the 
group leaving invariant the asymptotic form of the metric describing 
these processes. Quite generally, one can choose $(u,r,\theta,\phi)$ coordinates 
close to null infinity and check that the components of the metric tensor behave 
like those of the Minkowski metric in null polar coordinates in the limit 
$r \rightarrow \infty$.

A BMS transformation $(\alpha,\Lambda)$ is then
\begin{gather} 
\bar{u} =[K_\Lambda (x)]^{-1} (u+ \alpha(x)) + O(1/r) \\
\bar{r} = K_\Lambda(x) r +J(x,u) +O(1/r)\\ 
\bar{\theta}= (\Lambda x)_\theta + H_\theta (x,u) r^{-1} + O(1/r)\\
\bar{\phi}=(\Lambda x)_\phi + H_\phi (x,u) r^{-1} +O(1/r)
\end{gather}

\noindent where $x$ is a point on the two sphere $S^2$ with coordinates $(\theta,\phi)$, 
$\Lambda$ represents a Lorentz transformation acting on $S^2$ as a conformal 
transformation and $K_\Lambda(x)$ is the corresponding conformal factor.
Furthermore $\alpha$ 
is a scalar function on $S^2$ associated with the so called supertranslation 
subgroup. It represents the ``size'' of the group as we will see below. 

The other functions are uniquely determined by $(\alpha,\Lambda)$ imposing
\begin{gather}
(\alpha_1,\Lambda_1)(\alpha_2,\Lambda_2)=(\alpha_1+\Lambda_1\alpha_2, \Lambda_1
\Lambda_2)\label{semid}\\
(\Lambda_1\alpha_2)(x)=[K_\Lambda (x)]^{-1} \alpha_2(\Lambda_1^{-1}x)\label{groupstr}
\end{gather}

\noindent One immediately notices from (\ref{semid}) the structure of semidirect product. Therefore the 
BMS group $B= N \ltimes L $ is the semidirect product of the infinite 
dimensional supertranslation group $N$ with (the connected component of the
homogeneous) Lorentz transformations group $L$.\footnote{We will consider
$SL(2,\mathbb{C})$, the universal covering group of $L$.}

An important point to keep in mind is that the ASG thus defined is {\it universal} 
since one gets the same group for all asymptotically flat spacetimes. This is 
quite surprising. In addition, the group is infinite dimensional due to the presence 
of extra symmetries which reflect the presence of gravity in the bulk.

It is also possible \cite{tamburino} to derive the BMS group B working in the 
unphysical spacetime and imposing differential and topological requirements 
on $\Im$, avoiding then asymptotic series expansions. In 
a sense, this is a finite version of the original BMS derivation, since 
one constructs a so called conformal Bondi frame in some finite
neighborhood of $\Im$ and this finite region corresponds to an infinite region 
of the original physical spacetime.

Even if one is not working with an asymptotic expansion, we prefer to 
consider this derivation from a slightly different perspective with respect to 
the one of Penrose, who considers the emergence of the BMS working 
intrinsically on $\Im$. Actually, 
we are still working asymptotically, 
 even if, in the unphysical space, 
 the ``infinite is brought to finite''. Recall, however, that had we chosen another 
conformal frame we would have obtained an isomorphic group. In other words, covariance 
is preserved by differentiable transformations acting on $\Im$. This indeed 
motivated Penrose to work directly on the geometrical properties of
$\Im$ as we will see below and it seems more convenient to investigate the 
holographic principle in this context.

Choosing $x^0=u$, $x^A=(\theta,\phi)$ as coordinates on $\Im^+$ and $x^1=r$ 
defining the inverse luminosity distance, the (unphysical) metric $g^{\mu \nu}$ in the conformal Bondi frame
is thus
$$g^{\mu \nu}=\left[\begin{array}{ccc}
0 & g^{01} & 0\\
g^{01} & g^{11} & g^{1A}\\
0 & g^{1A} & g^{AB}
\end{array}\right]$$ 
\noindent Using the freedom of gauge choice of the conformal factor 
 and imposing global and asymptotic requirements on $\Im$ one can write
the metric (in a neighbourhood of $\Im$) as 
$$g^{\mu \nu}=\left[\begin{array}{ccc}
0 & 1 & 0\\
1 & 0 & 0\\
0 & 0 & q^{AB}
\end{array}\right]$$ 
\noindent where $q^{AB}$ is the metric on the $S^2$ therefore time independent.
One can eventually compute the generators $\xi^\mu$ of asymptotic infinitesimal transformations 
$x^\mu \rightarrow x^\mu +\xi^\mu$
by solving 
\begin{equation}
\xi^{(\mu ; \nu)}- (\Omega,_\rho \xi^\rho / \Omega) g^{\mu \nu} =0.
\end{equation} 
One finds 
\begin{gather}
\xi^A=f^A(x^B)\\
\xi^0 = \frac{1}{2}u f_{;A}^A+ \alpha (x^B),\; \xi^1 =0.
\end{gather}
\noindent Setting the supertranslations $\alpha(x^A)$ to zero we get the
(orthochronous) 
Lorentz group while setting the $f^A$ to zero we get the group of supetranslations as 
expected.

These are exactly the Killing vectors found by Sachs \cite{sachsa} in studying radiation at null
infinity. However, one interprets
 the notion of asymptotic symmetry as follows: one declares an
infinitesimal asymptotic symmetry to be described by a vector field $\xi^a$ 
(more precisely an equivalence of vector fields in the physical spacetime) 
such that the Killing equation $L_\xi g_{\mu \nu} =0$ is satisfied to as 
good an approximation as possible as one moves towards $\Im$.

One can also consider another derivation of the BMS group proposed by 
Geroch \cite{geroch}. In a nutshell, this procedure considers the ASG as the 
group of consometries of $\Im$, i.e. conformally invariant 
structures associated with $\Im$. More properly, these
structures live on the so called 
``asymptotic geometry'', a 3 dimensional manifold diffeomorphic to $\Im$ endowed with a 
tensor structure. We prefer to consider as truly intrinsic the derivation 
of Penrose which we discuss below. Note however that Geroch approach has received 
a lot of attention and has been adopted in particular by Ashtekar and (many) others to endow 
$\Im$ with a symplectic structure to study then fluxes and angular momenta of 
radiation at null infinity.
\subsection{Penrose derivation of the BMS group}
The derivation of Penrose is based on conformal techniques and the null nature 
of $\Im$ is explicitly taken into account. The underlying idea is the 
following: consider a motion in the physical spacetime; this will naturally 
generate a motion in the unphysical spacetime and in turn a conformal motion 
on the boundary. The latter can persist, even if the starting physical spacetime 
has no symmetry at all providing thus a definition of ASG. However the degenerate 
metric on $\Im$ does not itself endow sufficient structure to define
 the BMS group.\footnote{More precisely, one considers a future/past 
3 asymptotically simple spacetime, with null $\Im^+$,$\Im^-$ and strong
asymptotic Einstein condition holding on it \cite{nonlinear}.} 

The natural structure living on $\Im$ is that of a inner (degenerate) conformal 
metric, the topology being $\mathbb{R} \times S^2$; the ``$\mathbb{R}$'' represent the 
null geodesic $\Im$ generators with ``cuts'' given by two dimensional spacelike
hypersurfaces each with $S^2$ topology. Choosing a Bondi coordinate
 system \cite{nonlinear}, one can indeed write the degenerate 
 metric on $\Im^+$ as
\begin{equation} \label{degmetric}
ds^2=0.du^2+d\theta^2+\sin^2\theta d\phi^2.
\end{equation}
\noindent Using stereographic coordinates for the two sphere
 $(\zeta=e^{i\phi}cot(\theta/2))$  one has
\begin{equation}
ds^2=0.du^2+4d\zeta d \bar{\zeta}(1+\zeta \bar{\zeta})^{-2}.
\end{equation} 
\noindent and recalling that all holomorphic
 bijections of the Riemann sphere are of the 
form
\begin{equation}
f(\zeta)=\frac{a\zeta +b}{c \zeta +d}
\end{equation}
\noindent with $ad-bc=1$, one immediately concludes that the metric (\ref{degmetric}) is preserved
under the transformations
\begin{gather} \label{nugroup}
u \rightarrow F(u,\zeta , \bar{\zeta}) \\
\zeta \rightarrow \frac{a\zeta +b}{c \zeta +d}
\end{gather}
\noindent These {\it coordinate} transformations define the so called
Newman-Unti (NU)
group \cite{nugroup}, namely the group of non reflective motions of $\Im^+$ preserving its 
intrinsic degenerate conformal metric. 

The NU is still a very large group. One can restrict it by requiring to preserve 
additional structure on  $\Im$. One actually enlarges the notion
of angles and endows $\Im$ with ``strong conformal geometry''. In addition to ordinary 
angles one considers null angles: finite angles are formed by two 
different directions in $\Im$ at a point in $\Im$ which are not coplanar with the 
null direction in $\Im$, while null angles are formed by two directions at a 
point in $\Im$ which are coplanar with the null direction.

One can show that the set of strong conformal geometry preserving transformations 
restricts the $u$ transformations of the NU group to the following form
\begin{equation}
u \rightarrow  K(u+ \alpha(\zeta,\bar{\zeta}))
\end{equation}
\noindent with
\begin{equation}
K=\frac{1+\zeta \bar{\zeta}}{(a\zeta+b)(a^* \zeta^*+b^*)+(c 
\zeta +d)(c^* \zeta^* +d^*)}
\end{equation}
\noindent the same appearing in (\ref{groupstr}) and $\alpha(\zeta,\bar{\zeta})$ an
arbitrary function defined on the two
sphere. But then this set of transformations together with the conformal transformations 
on the two sphere define precisely the BMS group as shown before. 

Note that in terms of this intrinsic description, these transformations have to be interpreted 
not as coordinate transformations but as {\it point} transformations mapping 
$\Im$ into itself. In other words, a conformal transformation of $\Im$ induces 
conformal transformations between members of families of asymptotic
2-spheres when moving 
along the affine parameter $u$. 

This construction further motivates the mapping between $\Im$ and the so called 
cone space\cite{bramson}, which we are going to discuss in the following as a possible abstract 
space where the holographic data might live.

Finally one has to remember that the global structure of
the BMS group in {\it four} dimensions cannot be generalized to a generic dimension as
$BMS_d=N_{d-2}\ltimes SL(2,\mathbb{C})$ where $N_{d-2}$ is the abelian group of scalar
functions from $S^{d-2}$ to the real axis; an example is the three dimensional
case
where \cite{Ashtekar3} it has been shown that $BMS_3=N_1\ltimes Diff(S^1)$. In what 
follows we are always going to work in {\it four} dimensions.

\subsection{BMS subgroups and angular momentum}

We review a bit more in detail the BMS subgroups. One 
has the subgroup N given by
\begin{gather} \label{supertr}
u \rightarrow u+ \alpha(\zeta,\bar{\zeta}) \\
\zeta \rightarrow \zeta
\end{gather}
\noindent known as supertranslations. It is an infinite
dimensional abelian subgroup; note that 
\begin{equation}
\frac{ BMS}{N}\simeq  SO(3,1)\simeq
PSL(2,\mathbb{C}),
\end{equation}
which follows from the fact that the BMS group is the semidirect product of N 
with SL(2,C). Choosing for the conformal 
factor $K$ on the sphere
\begin{equation}
K=\frac{A+B\zeta +B^* \bar{\zeta} +C \zeta \bar{\zeta}}{1+\zeta \bar{\zeta}}
\end{equation}
one has the subgroup $T^4$ of translations which one can prove to be 
the unique 4-parameter normal subgroup of N.

On the other hand, the property of a supertranslation to be translation 
free is {\it not} Lorentz invariant. Therefore there are several Poincar\'e groups at 
$\Im$, one for each supertranslation which is not a translation, and none of
them is preferred. This causes the well known difficulties in asymptotically 
flat spacetimes in defining the angular momentum, the origin basically being 
``free'' (because of the presence of gravity). We recall for completeness the reason: in Minkowski space-time, the angular momentum is described
 by a skew symmetric tensor
which is well-defined up to a choice of an origin. Whereas this last condition
is equivalent to fix 4 parameters, in the case of $\Im$ this condition requires 
an infinite number of parameters to fix the ``orbital'' part of the 
angular momentum  since the translation group $T^4$ is
substituted by the supertranslations N.

 Although many ways to circumvent this
problem have been proposed, no really satisfactory solution has emerged until
now. In \cite{Ashtekar2}, \cite{penrose} one ends up with a reasonable 
definition of
angular momentum in asymptotically stationary flat space-times, where
the space of good cross sections (i.e. sections with null asymptotic shear) is not empty; one can 
then select a Poincar\'e subgroup from the BMS group and define accordingly the angular momentum. We are going to examine 
in more detail the notion of good/bad sections in the following so as to discuss bulk 
entropy production from the point of view of boundary symmetries.

\section{Bulk entropy and boundary symmetries}

\subsection{Bulk entropy and BMS boundary symmetries}
As previously recalled the BMS group is defined as those mappings acting on 
$\Im$ which preserve both the degenerate metric and the null angles.\\
In the case of null infinity, one can associate
 a {\it complex} function 
$\sigma(r,u,\zeta,\bar{\zeta})$, which in physical terms is a measure of the 
shear of the null cones which intersect $\Im^+$ at constant $u$. To define the 
shear one chooses a spinor field $O^A$ whose 
flagpole directions point along the the null geodesics of the congruence. The 
complex shear $\sigma$ is then defined as follows
\begin{equation}
O^A O^B \nabla_{AA'} O_B = \sigma \bar{O_{A'}}
\end{equation}
The argument of the shear $\sigma$ defines the plane of maximum shear and its modulus 
the magnitude of the shear. Now in the case of mild divergence
 of null geodesics (as with the Bondi-type hypersurfaces we are considering) one has
\begin{equation}
\sigma = \frac{\sigma^0}{r^2} + O(\frac{1}{r^3})
\end{equation}
The quantity $\sigma^0$ is $r$ independent and it is a measure of
 the {\it asymptotic} 
shear of the congruence of null geodesics intersecting $\Im^+$ at constant $u$. The 
$r$ independence is in agreement with the peeling-off \cite{rindler} properties of the 
radiation.

One can also read the shear from the asymptotic expansion of the metric. Consider 
for example the metric originally proposed by Sachs \cite{sachsa} 
\begin{equation}
ds^2= e^{ 2 \beta} V r^{-1} du^2 -2 e^{2 \beta} dudr+r^2 h_{ab}(dx^a-U^a du)(
dx^b-U^bdu)
\end{equation}
with $a,b$ indices running over 
angular coordinates and $V,\beta,U^a, h_{ab}$ are functions of 
the coordinates $(u,r,\theta,\phi)$ to be expanded in $1/r$ powers.
 The shear appears in the expansion of the function $\beta$
\begin{equation}
\beta = -\sigma(u,\zeta,\bar{\zeta}) \sigma^* (2r)^{-2}+O(r^{-4})
\end{equation}

Now when $\sigma^0 =0$, i.e. the asymptotic shear of the congruence of null 
geodesics vanish at infinity, one has ``good'' cross sections. On the other
hand, when non vanishing, one has ``bad'' sections. The latter corresponds to 
null geodesics ending up with complicated 
crossover 
regions in the bulk. Good cross sections do not exist in general spacetimes. However, a very 
special situation occurs in stationary spacetimes: in this case one can find
asymptotic shear free sections and the space of such cuts is isomorphic to Minkowski space
time, where a good section corresponds trivially to the lightcone originating 
from a point in the bulk. Of course  in the case of stationary spacetimes points 
of the isomorphic Minkowski space are not in one to one correspondence with 
points of the physical curved spacetime; the behaviour in the bulk of 
null geodesics will be however quite mild (compared to 
bad sections) to end up in an almost clean 
vertex. 

The intersection of the congruence of null geodesics originating from the bulk
 with $\Im$  is a connected two dimensional spacelike surface so we
  can apply the covariant entropy bound and  
  deduce that bad cross sections will in general correspond to more entropic 
configurations from bulk point of view. Indeed  in the case of 
bad cross sections lightrays ``percolate'' more than 
in the case of good sections, producing therefore more entropy.

One can say more and relate the notion of good/bad cuts to BMS boundary symmetries, having 
in mind a tentative holographic description. Indeed,
 under BMS supertranslations the transformation rule among asymptotic 
shears is
\begin{equation}
\sigma^{o'}(u',\zeta,\bar{\zeta})=\sigma^o (u'-\alpha,\zeta,{\bar{\zeta}}) +(\edth)^2 
\alpha(\theta, \phi)
\end{equation}
where the operator $\edth$ on the r.h.s is the so called ``edth'' operator (for a definition see
\cite{rindler}). One is then interested in finding transformations which produce new good
cuts. For Minkowski spacetime and (remarkably!) again stationary spacetimes one can map good cuts
into good cuts by means of translations and in these cases the BMS group can
be reduced to the Poincar\'e group by asking for the subgroup of the BMS transformations
which map good cuts into good cuts. In the general case, however, there are 
no good, i.e. asymptotic shear free sections, and from
BMS point of view this corresponds to not Lorentz 
free supertranslations. 

Applying therefore the covariant entropy bound one finds that bad sections correspond to more entropic 
configurations in the bulk and (not Lorentz free) supetranslations on the 
boundary. Time dependence produces more irregularities in the bulk 
giving therefore more entropy according to previous considerations; this is 
interestingly reflected in complicated supertranslations acting on the null
 boundary. 
 
If holographic data are stored then in the
  $S^2$ spheres on $\Im$ some of them will contain more/less information corresponding 
 to more/less entropy in the bulk. We return to this point in Section 7.

As said, asymptotic vanishing shear allows to reduce the BMS goup to Poincar\'e. One might 
think to start from the stationary case then for simplicity. There is still 
however a remnant of supertranslations. Suppose indeed to consider a system which 
emits a burst of radiation and it is stationary before and after the burst. The 
corresponding Poincar\'e subgroups will be different and they will have in common 
{\it only} their translation group. They will be related by means of a non trivial 
supertranslation in general. This is quite different from what 
happens in the AdS case as we are going to see in the next Section. 

\subsection{Difficulties in reconstructing spacetime and comparison with 
the AdS case} 
We now continue the previous analysis and make some comparisons with the 
AdS case.

Let us consider again the asymptotic shear. The previous picture tells 
us that the ASG can be reduced to Poincar\'e in some specific points along 
the boundary where the asymptotic shear does indeed vanish. This means that small 
shapes are preserved asymptotically as we follow the lines generating the null 
congruence using to construct the lightsheets. However, lightsheets acquire
in general shear in the asymptotic region. This is due tidal forces which
are responsible for the bending of light rays. But these are 
in turn described by the (asymptotic)
Weyl tensor and  this quantity (more properly the 
rescaled one in the unphysical metric) enters into the 
definitions of the so called Bondi news functions \cite{sachsa} which measure the amount 
of gravitational radiation at infinity. There is however another tensor, namely the Bach 
tensor (recall we are in four dimensions) $B_{\alpha \beta}$ which does not vanish in the presence of non zero Bondi news. In the case 
of asymptotically flat spacetimes it is {\it not} zero asymptotically. This is 
in {\it sharp} contrast with asymptotically AdS cases, where it vanishes
 asymptotically, the 
Bondi news being zero in that case. Actually the condition 
$B_{\alpha \beta} =0$ on 
$\Im$ {\it is} used in the definition \cite{magnon} of asymptotically AdS spacetimes.

This has however deeper consequences for holography. In AdS case this allows to 
reduce (enormously) the diffemorphism group on the boundary precisely to the conformal 
group. There is then (as already noticed in \cite{magnon}) a discontinuity in taking 
the limit $\Lambda\rightarrow0$ of the cosmological constant. In 
asymptotically flat spacetimes it means that 
one cannot propagate the boundary data to reconstruct the bulk in a {\it unique}
 way.
And this was the essence of the Fefferman-Graham theorems for the AdS case
\cite{fefferman}. 
We therefore see a remarkable difference. As a consequence, it
 also seems quite unlikely that 
GKPW (Gubser-Klebanov-Polyakov-Witten) prescription 
relating bulk-boundaries partition functions 
holds in this case. It seems also difficult to recover a S-matrix for 
asymptotically flat spacetimes starting from AdS/CFT and taking then the 
large radius limit of AdS.

As observed before, in general backgrounds the asymptotic shear does not vanish and therefore 
we remain with a big group on the boundary. Notice
 that using relativistic generalization 
of Navier-Stokes theory one can show that precisely the Bach current \cite{glass}
 can be used to describe 
entropy production in the bulk in the 
case of non stationary spacetimes. We see therefore that all the times (basically the majority) we cannot
 reduce BMS supertranslations to translations we have 
more entropy production in the bulk according to the covariant entropy bound and 
the production of this entropy can be measured in a quantitative way 
just by using the Bach current, a quantity which translates the effect of the 
bending of light before the system reaches equilibrium.

The fact that boundary symmetries cannot be reduced as in AdS case suggests not only that 
the propagation in the bulk is not unique (therefore we don't see the possibility of 
a naive holographic RG) as in AdS case but also that a degree of non 
locality will be present-because of the impossibility of reducing 
the big symmetry group in general- in the candidate 
boundary theory, where fields will carry in general 
representations of the BMS group.

This motivates the following analysis of 
wave equations for the BMS group.

\section{Representations of the BMS group}    

As said our target is to write (covariant) wave equations as commonly done 
in physics for other groups \cite{Group2}. We therefore first review  very 
briefly in this section the 
representation theory of the BMS group (See Appendix A.1 and A.2 
for details). We recall the situation for the Poincar\'e group to compare then similarities and 
differences with respect to the BMS case. We give the ``kets'' 
 to show explicitly the labelling of the corresponding states. Theory and 
 definitions 
 used in induced representations of semidirect product groups are reviewed 
 in Appendix A.

\begin{center}
{\large Poincar\'e group}
\end{center}

In this case we deal
with $P=T^4 \ltimes SL(2,C)$ whose little groups and orbits are well known and are summarized in the following table (see \cite{Simms} and
\cite{Group})\footnote{The Lie algebra of the 2-d Euclidean group
is:
$$[L_3,E_\alpha]=i\epsilon_{\alpha\beta}E_\beta,$$
$$[E_\alpha,E_\beta]=0.$$
The two Casimirs of the group are $E^2$ and $C_2=exp(2\pi iL_3)$ where $C_2=\pm
1$ (integer and half-integer values of $L_3$). The two $E(2)$ are actually the same group with different
representations depending if the value of the Casimir operator $E^2$ is
different (first case) or equal (second case) to zero}:

\vspace{0.5cm}

\begin{tabular}{|c|c|c|} 
\hline
Little group & orbit invariant & representation label \\
\hline
$SU(2)$ & $p^2=m^2$, sgn($p_0$) & discrete spin $j$ (dim=2j+1)\\
\hline
$SU(1,1)$ & $p^2=-m^2$, & discrete spin $j^\prime$\\
\hline
$E(2)$ & $p^2=0$, sgn($p_0$) & $\infty$-dimensional,\\
\hline
$E(2)$& $p^2=0$, sgn($p_0$) & 1-dimensional $\lambda$. \\ 
\hline
\end{tabular}

\vspace{0.5cm}

We first notice that the
generators of $T^4$ can be simultaneously diagonalized and for this reason the
orbit is the spectrum of energy. We have to impose some physical
restrictions, namely we call unphysical those representations related to negative
square mass and negative sign of $p_0$. Unfortunately, this is not enough since
we have to deal with a continuous
spin coming from $E(2)$. This case is excluded by hand and so we end up with
two spin quantum numbers, i.e. $j$ from $SU(2)$ and $\lambda$ from $E(2)$;
therefore the general ket for the Poincar\'e group is 
\begin{equation}\label{ketp}
\mid {\bf p}, j>,\;\;\;\mid {\bf p},\lambda>,
\end{equation}
respectively for massive and massless states.

\vspace{0.3cm}

\begin{center}
{\large BMS group}
\end{center}

In this case one has additional freedom since one is {\it free} to choose the topology 
for supertranslations. This is due to the fact that $\Im$ is not a Riemannian 
manifold: it is degenerate precisely in the directions along which supertranslations 
act. Having in mind Penrose description of the BMS as an {\it exact} symmetry 
acting of $\Im$,  arbitrary supertranslations 
functions describe indeed symmetry transformations along $\Im$. The standard
choice \cite{Mc1}, \cite{Mc4}
made in the literature is Hilbert or nuclear topology. The former should be associated with 
bounded systems (for which indeed the BMS group turns out to be the 
asymptotic symmetry group as originally discovered), while the latter
 with unbounded (See section 7.3 for the role 
of unbounded systems).

We first consider the Hilbert topology-i.e. we endow $N$ with the 
ordinary $L^2$
inner product on $S^2$. 
Following \cite{Mc1} , \cite{Mc2}, we remember that the supertranslations
 space can be decomposed in a translational and a supertranslational part 
$$N=A\oplus B,$$
where only $A$ is invariant under the action of $G=SL(2,\mathbb{C})$ and
$T^4=\frac{N}{B}$. Furthermore there is also this chain of isomorphisms:
$$N\sim\hat{N}\sim{N^\prime}\sim N,$$
where $\hat{N}$ is the character space and $N^\prime$ is the dual space of $N$. 
This means that given a supertranslation $\alpha$ we can
associate to it a character $\chi(\alpha)\doteq e^{if(\alpha)}$,
where the function $f(\alpha)=<\phi,\alpha>$ and where $\phi\in N$. 

The dual space can be decomposed as $N^\prime=B^0\oplus A^0$, where $B^0$ and
$A^0$ are respectively the space of all linear functionals  vanishing on $B$ and
$A$ and where only $A^0$ is G-invariant. Also the following relations are
G-invariant i.e.
\begin{equation}\label{iso} 
(N/A)^\prime \sim A^0\;\;\; N^\prime/A^0 \sim A^\prime.
\end{equation}

In view of the isomorphism between $N^\prime$ and $N$, we can expand the
supermomentum $\phi$ in spherical harmonics as
$$\phi=\sum_{l=0}^1\sum_{m=-l}^l p_{lm}Y_{lm}+\sum_{l>1}\sum_{m=-l}^l
p_{lm}Y_{lm},$$
where the first term lies in $B^0$ and the second in $A^0$.\footnote{One can interpret 
the piece belonging to $A^0$ as composed of spectrum generating 
operators, while those in $B^0$ act on the vacuum in the context of a 
holograhic description. We thank J. de Boer for the remark.}Relying on
(\ref{iso}), we can think of the coefficients $p_{lm}$ with
$l=0,1$ as the components of the Poincar\'e momentum. Thus, we call the dual
space of $N$ the supermomentum space and define a projection map:
$$\pi:N^\prime\to N^\prime/A^0,$$
assigning to each supertranslation $\phi$ a 4-vector $\pi (\phi) =(p_0,p_1,p_2,p_3)$.

At the end of the day one ends up with \cite{Mc2}

\vspace{0.3cm}

\begin{center}
\begin{tabular}{|c|c|c|} 
\hline
Little group & orbit invariant & representation label \\
\hline
$SU(2)$ & $p^2=m^2$, sgn($p_0$) & discrete spin $j$ (dim=2j+1)\\
\hline
$\Gamma$ & $p^2=m^2$, sgn($p_0$)& discrete spin $s$\\
\hline
$\Gamma$ & $p^2=0$, sgn($p_0$)& discrete spin $s$\\
\hline
$\Gamma$ & $p^2=-m^2$, & discrete spin $s$\\
\hline
$\Theta$ & $p^2=m^2$, sgn($p_0$) & discrete spin $s$\\
\hline
$\tilde{C}_n$ & $p^2=m^2$, sgn($p_0$) & finite dimensional $k$, \\ 
\hline
$\tilde{C}_n$ & $p^2=0$, sgn($p_0$) & finite dimensional $k$, \\ 
\hline
$\tilde{C}_n$ & $p^2=-m^2$, & finite dimensional $k$, \\ 
\hline
$\tilde{D}_n$ & $p^2>0$, sgn($p_0$) (for $p_0>0) $& finite dimensional $d_n$,\\ 
\hline
$\tilde{D}_n$ & $p^2<0$ & finite dimensional $d_n$,\\ 
\hline
$\tilde{T}$ & $p^2>0$, sgn($p_0$) & finite dimensional $t$. \\ 
\hline
$\tilde{O}$ & $p^2>0$, sgn($p_0$) & finite dimensional $o$. \\ 
\hline
$\tilde{I}$ & $p^2>0$, sgn($p_0$) & finite dimensional $i$. \\ 
\hline
\end{tabular}
\end{center}

\vspace{0.3cm}

\noindent Therefore the
general kets of the BMS group for massive and non massive
particles\footnote{In the BMS group the massive and the massless kets are both labelled
by discrete quantum numbers related to faithful representations of (almost the
same) compact groups whereas in the Poincar\'e case massless states are labelled by the
discrete number of the unfaithful representation of the non compact group $E(2).$} are:
\begin{equation}\label{ketBH}
\mid {\bf p}, j, s, k, d_n, t, o, i>,\;\;\;\mid {\bf p}, s, k>,
\end{equation}
where the new quantum numbers were originally 
interpreted as possible internal symmetries of 
bounded states \cite{Mc2},\cite{Mc5},\cite{komar} and the BMS group was indeed proposed to 
substitute the usual Poincar\'e group to label elementary particles 
due to the absence of non compact little groups and 
therefore of continuous spins. 

Choosing a different topology for supertranslations, however, one registers the 
appearence of non compact little groups in the BMS representations theory too 
\cite{Girardello}. We believe that precisely for this reason the hope to use 
BMS group to label elementary particles was abandoned. However another interpretation 
of these numbers has been suggested as we are going to see soon.

Consider then the ket for the nuclear (or finer) topology. First of all, recall that in this case it is
impossible to have an exaustive answer since not much is known about discrete
subgroups. Nevertheless we have \cite{Mc4} 

\vspace{0.5cm}

\begin{center}
\begin{tabular}{|c|c|c|} 
\hline
Little group & orbit invariant & representation label \\
\hline
$\Gamma$ & $p^2=m^2,0, -m^2$, sgn($p_0$)& discrete 1 dim. spin $s$\\
\hline
$SU(2)$ &$p^2=m^2$, sgn($p_0$) &discrete $2j+1$-dim. spin $j$ \\
\hline
$\Delta$ &$p^2=0$, sgn($p_0$) &finite dim. $\delta$ or $\infty$ dim.\\
\hline
$S_1$ &$p^2=0$, sgn($p_0$) &finite dim. $s_1$\\
\hline
$S_2$ &$p^2=0$, sgn($p_0$) &finite dim. $s_2$\\
\hline
$S_3$ &$p^2=0$, sgn($p_0$) &finite dim. $s_3$\\
\hline
$S_4$ &$p^2=0$, sgn($p_0$) &finite dim. $s_4$\\
\hline
$S_5$ &$p^2=0$, sgn($p_0$) &finite dim. $s_5$\\
\hline
\end{tabular}
\end{center}

\vspace{0.5cm}
We omit the study of little groups with  $m^2<0$ since in principle they have no physical relevance.  
Thus the general ket for the BMS group in the nuclear topology is:
\begin{equation}\label{Nket}
|{\bf p}, j,s,\left\{t_n\right\}>, \;\;\;
|{\bf p}, j, s,\delta, \left\{s_n\right\}> 
\end{equation}
where the first case refers to faithful representations with $m^2>0$ and the
index $\left\{t_n\right\}$ stands for all the representation numbers of finite
groups; the second ket instead refers
to the massless case and $\left\{s_n\right\}$ stands for all
 the representation
numbers of the non connected groups.

Note that it is because of the infinite dimensionality of
 supermomentum space that one has non-connected or even discrete little groups, since 
 one can have a lot of invariant vectors in this case. This is
 quite unfamiliar, since angular momenta are normally associated
  with connected groups 
 of rotations. From an experimental point of view this also renders problematic
the measurement of these ``Bondi spins'' as they are normally called. Indeed 
only in the case of the little group $SU(2)$ BMS representations contain a 
single Poincar\'e spin, otherwise they contain a mixture of Poincar\'e spins. Note 
also a curious fact: for $m^2>0 $
 all bosons with the same mass appear in the same multiplet while all fermions 
 with the same mass appear in the same multiplet though corresponding to different 
 representation.

\section{Wave equations}\label{wave}

In this section we derive the covariant wave equations for the BMS group. 
As remarked before and in the following, nuclear (or even finer) 
topology is expected to be associated with unbounded systems, perhaps with 
infinite energy too. This would require in general Einstein equations in 
distributional sense and a different notion of conformal infinity. We therefore 
restrict to the Hilbert topology describing bounded systems in the bulk.

Canonical wave equations have been suggested in \cite{Mc1}, though 
physicists normally use covariant wave equations. To derive them we use the 
theorem contained in \cite{Asorey} which shows how to get irreducible covariant 
wave functions starting from (irreducible) canonical ones. The framework- based 
on fiber bundle techniques, is quite general and 
elegant. For definitions and notations see 
Appendix B, which we suggest to read before this section. 

We are also going to use sort of diagrams in the discussion which, although not 
completely rigorous, may help to handle the formalism easier.

Consider then the following diagram 

$$\xymatrix{G=N\ltimes SL(2,\mathbb{C})
\ar[d]^\pi\ar[r]^{\;\;\;\;\;\;\;\;\;\;\Sigma} & GL(V)\ar[d]^{*} \\
N & F(N,V)\ar[l]^{\pi^\prime}}$$

The representations induced in the above way are usually referred as
\emph{covariant}. Since the bundle is topologically trivial i.e. $G=N\times
SL(2,\mathbb{C})$ we are free to choose a global section $s_0:N\to G$ and the
natural choice is $s_0(n)=(n,e)$; it is possible to see
from (\ref{gamma}) that $\gamma_0((n,k),n^\prime)=(0,k)$ which implies that
 the matrix $A(g,n^\prime)=\Sigma(\gamma(g,n^\prime))$ (see
Appendix B for definitions) does not depend on the choice of the supertranslation $n^\prime$ whereas
this fails for induced representations. The only problem for covariant representations
is that in general they are not irreducible even if $\Sigma$ is, but
as said a method \cite{Asorey} to compare covariant and induced
representations was formulated. Consider indeed the following ``diagram'':
$$\xymatrix{G\sim\hat{G}=\hat{N}\ltimes SL(2,\mathbb{C})\ar[d]^\pi\ar[r]^{\;\;\;\;\;\;\;\;\;\;\Sigma} & GL(V)\ar@{-->}[d]^{*} \\
\hat{N} & \hat{F}(\hat{N},V)\ar[l]^{\pi^\prime}},$$
where $\hat{N}$ is the character space of $N$ and $\hat{F}$ is the bundle
$\hat{N}\ltimes SL(2,\mathbb{C})\times_{SL(2,\mathbb{C})}V$. Thus we can introduce a $\hat{T}$
representation acting on the sections of $\Gamma(\hat{F})$:
$$\hat{T}(n,k)\hat{\psi}(k\chi)=(k\chi)(n)k\hat{\psi}(\chi),$$
which is a transposition of 
\begin{equation}\label{psisec}
\hat{T}(g)\hat{\psi}(g\chi)=g\hat{\psi}(\chi).
\end{equation}

Using the natural section (in the character space) $\hat{s}_0(\chi)=(\chi,e)$,
the action of the group $G$ on the function $\hat{f}:\hat{N}\to V$ is given from
(\ref{psisec}) by
\begin{equation}\label{psi4}
\left(\hat{T}_{s_0}(n,k)\hat{f}\right)(k\chi)=(k\chi)(n)\Sigma(k)\hat{f}(\chi),
\end{equation}

which we can refer to as the \emph{covariant wave equation}.
The relation between induced and covariant
representations can be made now choosing a fixed character on an orbit $\Omega$
(physically speaking going on shell)
and denoting with $\sigma$ the representation of $K\chi_0$ subdued by $\Sigma$.

The essence of the Theorem contained in \cite{Asorey} is that if $W$ is the canonical representation of $G$ in $\Gamma(F)$ induced from
$\chi_0\sigma$ than there exists an isomorphism of bundles
$\rho:F\to\hat{F}_\Omega$ such that the map
$R:\Gamma(F)\to\Gamma(\hat{F}_\Omega)$ defined by $R\psi(\chi)=\rho(\psi(\chi))$
satisfies
$$R\circ W=\hat{T}_\Omega\circ R,$$
which states the equivalence between $W$ and $\hat{T}_{\Omega}$. Notice
that with $\hat{T}_\Omega$ and with $\hat{F}_\Omega$ we simply refer to the
(on shell) restriction on the $SL(2,\mathbb{C})$-orbit. The above framework 
can be applied for a given group $G=N\ltimes H$ as follows:
\begin{enumerate}
\item identify all little groups $H_\chi\subset H$ and their orbits (labelled 
by Casimir invariants)
\item construct a representation induced from $H_\chi$ and choose a section for
the bundle $G\left((G/G_\chi),\pi, G_\chi\right)$,
\item construct the canonical wave equations for each little group.
\item construct the covariant wave equation starting from a representation of
$H$ upon the choice of a section of the fiber bundle $G(N,\pi,H)$. Write the
covariant wave equation in the dual space.
\item relate the induced and the covariant wave equations restricting the 
latter to the orbit of a little group and then defining a linear transformation $V$
acting on functions given by
$$ V=U_{\lambda}(s^{-1}(p)),$$
where $U$ is the representation of $H$ and $s$ is the section chosen at point
(2).
\item since the representation $U$ is in $H$ and it is reducible, it is
necessary to impose some constraints (i.e. projections) on the wave functions corresponding to the
reduction of the representation $U$ to one of the little groups. 
\end{enumerate}
\begin{center}
{\large Poincar\'e group}
\end{center}

We review  first how the above construction applies to the Poincar\'e group
$T^4\ltimes SL(2,\mathbb{C})$. For
further details we refer to the exsisting literature \cite{Group},
\cite{Group2} and \cite{Landsman}. As discussed before, little groups and 
orbits of the Poincar\'e group are well known and in particular for a
massive particle the orbit is the mass hyperboloid $SL(2,\mathbb{C})/SU(2)\sim\Re^3$ and a fixed
point is given by the 4-vector $(m,0,0,0)$. A representation for $SU(2)$ is the
usual $2j+1$-dimensional $D^j$ and a global section (see appendix of \cite{Mc2})
for $SU(2)$ can be chosen
remembering the (unique) polar decomposition for 
an element $g\in SL(2,\mathbb{C})$ given by
$g=\rho u$ where $\rho$ is a positive definite hermitian matrix and $u\in
SU(2)$. Thus a section $\eta$ can be written as:
\begin{equation}\label{sections}
\eta(\rho)=\eta(\rho[SU(2)])=(\rho,0).
\end{equation}
This concludes the first two points in our construction; a canonical wave
equation can be immediately written starting from (\ref{psi2}) as:
\begin{equation}\label{inducedP}
U^{m,+,j}\left(g\right)f(gp)=e^{ip\cdot
a}D^j[\rho^{-1}_{\Lambda_g\Lambda_p}\Lambda_g\rho_{\Lambda_p}]f(p),
\end{equation}  
where $g=(a,\Lambda)$, the exponential term is the character
 of $a\in T^4$, the $\cdot$ represents the Minkowskian internal product, 
$p$ is a point over the orbit, $U$ is a representation of
the little group over an Hilbert space $\mathcal{H}_j$ and $f(p)$ is a function in
$L^2(SL(2,\mathbb{C})\ltimes N/SU(2)\ltimes N)\otimes\mathcal{H}_j$. Notice also
that the representation is labelled by the $SU(2)$ quantum number $j$ and by the
indices $m,+$ that allow to select a unique point over the orbit through the
identification $m=K$ and $sgn(K)>0$.\\
The covariant wave equation can be written for functions $f(p)\in
L^2(T^4)\otimes\mathcal{H}_\lambda$ as:
\begin{equation}\label{covp}
\left(U^\lambda(g)f\right)(a)=U_\lambda(\Lambda)f(\Lambda^{-1}(a-a^\prime)),
\end{equation}
where $g=(a^\prime,\Lambda)$ and $U_\lambda$ is a representation of
$SL(2,\mathbb{C})$. In the dual space (\ref{covp}) becomes:
\begin{equation}\label{fcovp}
\left(U^\lambda(g)\hat{f}\right)(p)=e^{ip\cdot a}U_\lambda(\Lambda)\hat{f}(\Lambda^{-1}p).
\end{equation}
This concludes the fourth point of our construction; the reduction to the orbit
of the $SU(2)$ little group can be achieved requiring the mass condition for
each $\hat{f}(p)$:
$$\theta(p_0)(p^2-m^2)\hat{f}(p)=0,$$
which amounts to restrict \footnote{Here and in the following we denote with $\prime$ the 
measure restricted to the orbit.} the measure
$d^4(p)$ to $d\mu^\prime(p)=2\pi\delta(p^2-m^2)\theta(p^0)d^4 p$. This means that instead of
dealing with functions in $L^2(T^4,d^4 p)\otimes\mathcal{H}_\lambda$ we consider
elements of the Hilbert space $\mathcal{H}^{m,+,\lambda}=L^2(T^4,d\mu^\prime (p))\otimes\mathcal{H}_\lambda$.
In order to relate the canonical and the covariant wave equations, we 
introduce the operator
$$V=U_\lambda(\rho^{-1}_p),$$
where $\rho^{-1}_p=\eta(p)^{-1}$. This acts on functions as
$$(V\hat{f})(p)=U_\lambda\left(\rho^{-1}_p\right)\hat{f}(p).$$ 
Thus if we define the space $\mathcal{H}_\eta^{m,+,\lambda}$ to coincide as a
vector space with $\mathcal{H}^{m,+,\lambda}$ but equipped with the inner
product
$$(f,f^\prime)_\eta=\int
d\mu^\prime(p)\left(U_\lambda(\eta(p))^{-1}f(p),U_\lambda(\eta(p)^{-1})f^\prime(p)\right)_{\mathcal{H}_\lambda},$$
we see that the map $V$ is indeed unitary and substituting it in
(\ref{fcovp}) through $V^{-1}U^{m,+,\lambda}V=U_\eta^{m,+,\lambda}$ we
get:
\begin{equation}\label{covariantpoinc}
\left(U^{m,+,\lambda}(g)f\right)(p)=e^{i<\phi,a>}U_\lambda(\rho^{-1}_p\Lambda\rho_{\Lambda^{-1}p})f(\Lambda^{-1}p),
\end{equation}
which is coincident with the canonical wave equation even if 
$U_\lambda$ is still a representation of $SL(2,\mathbb{C})$. This means that in
general our wave function has more components that those needed and for this
reason, we have to introduce a suitable orthoprojection modding out the unwanted
components. This can be done introducing a matrix $\pi$ such that
$$\pi f(\bar{p})=f(\bar{p}),$$
where $\bar{p}$ is a fixed point over the orbit. Another way to express the projection operation is to consider the point
$f^\prime(\bar{p})=(U(\Lambda)f)(\bar{p})$ and the equation
(\ref{covariantpoinc}) in order to obtain the more familiar expression
$$\pi(p)f(p)=f(p),$$
where $p=L^{-1}_\Lambda\bar{p}$ and where $\pi(p)=D^{-1}(\Lambda)\pi D(\Lambda)$. Using the decomposition
$\Lambda=\rho_p\gamma$ that we introduced in order to construct the section for
the $SU(2)$-orbit and using the fact that $\pi$ commutes with $D(\gamma)$ with
$\gamma\in SU(2)$, then $\pi(p)=D^{-1}(\rho_p)\pi D(\rho_p)$; so we have 
that $\pi(p)$ transforms as a covariant matrix operator. 

Let us consider
explicitly the case of a massive particle with spin $\frac{1}{2}$; to 
preserve parity we consider the representation $D^{(\frac{1}{2},0)}\oplus
D^{(0,\frac{1}{2})}$, the matrix $\pi$ projecting away the unwanted components
is: 
$$\pi= diag[1,0,0,0]=\frac{1}{2}(\gamma_0+I),$$
where $\gamma_0$ is the usual Dirac's $\gamma$ matrix. Using the covariant
transformation of this matrix, we find that
$$(D^{(\frac{1}{2},0)}\oplus
D^{(0,\frac{1}{2})})^{-1}(\Lambda_p)\pi \left(D^{(\frac{1}{2},0)}\oplus
D^{(0,\frac{1}{2})}(\Lambda_p)\right)=\frac{1}{2m}(\gamma_\mu p^\mu+m),$$
where $p=L_\Lambda \bar{p}$ and where $\bar{p}=(m,0,0,0)$ is a fixed point in
the $SU(2)$-orbit. Thus the equation 
$\pi(p)f(p)=f(p)$ for $f(p)\in
L^2(\Re^3,d\mu(p))\otimes\mathcal{H_j}$ becomes:
$$(\gamma_\mu p^\mu-m)f(p)=0,$$
which is the well known Dirac equation. In a similar way we can find the 
well known equations for all $SU(2)$ spins. 

Let us briefly remark that
for massless particles the situation is more complicated since in
this case we have to deal with the non compact $E(2)$ group. The
representation of $SL(2,\mathbb{C})$ cannot be fully decomposed into
representations of $E(2)$ and thus the orthoprojection condition has to be
modified \cite{Asorey}.
\subsection{Wave equations for the BMS group}

Recalling the discussion on representations of the previous Section 
we focus on the Hilbert topology case and derive the covariant wave equations for 
the little groups $SU(2),\Gamma,\Theta$ and finite groups. \newpage
\begin{center}
{\large The group SU(2)}
\end{center}
Clearly the first point of the construction has already been given 
by McCarthy (\cite{Mc1}, \cite{Mc2}) togheter with a partial classification of the
orbits. For massive particles a great difference arises from the Poincar\'e
group where the only orbit with $m>0$ is the one of the $SU(2)$ little group
whereas for the BMS group, apart from $SU(2)$, we need to consider more groups
as it can be seen from the tables in the previous Section.
 
\noindent Let us now address point (2). Remember that in this case the orbit is
isomorphic to $\Re^3$ and that a fixed point on the orbit is given by a constant
function $K$ over the sphere $S^2$. Thus the orbit is uniquely characterized by
the mass $m^2=K$ and the momentum 4-vector is given by $\pi(\phi)=(m,0,0,0)$
where $\phi$ is a constant supertranslation.\\ A section for the bundle $G\left(SL(2,\mathbb{C})\ltimes
N/SU(2)\ltimes N,\pi,SU(2)\ltimes N\right)$ has been given in the previous
discussion on the Poincar\'e case through the
polar decomposition $g=\rho u$ with $u\in SU(2)$ and $\rho$ a positive
definite hermitian matrix. Thus a section $\eta$ can be written as:
\begin{equation}\label{sectionSU}
\eta[p]=\eta[\rho SU(2)]=\left(\rho,0\right).  
\end{equation} 

Point (3) is also easy to implement starting from the well known
representations of $SU(2)$ since (\ref{psi2}) becomes:
\begin{equation}\label{inducedSU}
U^{m,+,j}\left(g\right)u(gp)=e^{i<\phi,a>}D^j[\rho^{-1}_{\Lambda_g\Lambda_p}\Lambda_g\rho_{\Lambda_p}]u(p),
\end{equation}
where $g=(a,\Lambda)\in BMS$, the exponential term is the character of the 
supertranslations $a$ expressed
via the Riesz-Fischer theorem, $p$ is a point over the orbit, $U$ is a representation of
the little group over an Hilbert space $\mathcal{H}_j$ and $u(p)$ is a function in
$L^2(SL(2,\mathbb{C})\ltimes N/SU(2)\ltimes N)\otimes\mathcal{H}_j$.

After completing point (3), we can write the
general covariant wave equation. In this case we deal with
functions $f \in L^2(N)\otimes\mathcal{H}_\lambda$ ($\lambda$ is index for an
$SL(2,\mathbb{C})$ representation) i.e.
\begin{equation}\label{covariant}
\left(U^\lambda(g)f \right)(\phi)=U_\lambda(\Lambda)f (\Lambda^{-1}(\phi-\phi^\prime)),
\end{equation}
where $g=(\phi^\prime,\Lambda)$ and $U_\lambda$ is a representation of
$SL(2,\mathbb{C})$. We go to the dual space of supertranslations 
using the well defined character $\chi(\alpha)$ and 
functional integration restricting to the $SU(2)$ orbit to get 
for $\hat{f}(a)\in L^2(\hat{N})\otimes\mathcal{H}_\lambda$.
\begin{equation}\label{covariant2}
\left(U^\lambda(g)\hat{f}\right)(a)=e^{i<\phi^\prime,a>}U_\lambda(\Lambda)\hat{f}(\Lambda^{-1}a),
\end{equation}
Thus in the end we deal with functions in 
$\mathcal{\hat{H}}^{m,+,\lambda}=L^2(N,d\mu^\prime)\otimes\mathcal{H}_\lambda.$
The linear operator relating the induced and the covariant wave equations is
given for a generic point in the orbit $p=(\phi,\Lambda)$ by
$$V=U_\lambda(\rho^{-1}_p),$$
where $\rho^{-1}_p=\eta(p)^{-1}$. This acts on functions as
$$(V\hat{f})(p)=U_\lambda\left(\rho^{-1}_p\right)\hat{f}(p).$$ 
Thus if we define the space $\mathcal{H}_\eta^{m,+,\lambda}$ to coincide as a
vector space with $\mathcal{H}^{m,+,\lambda}$ but provided with the inner
product 
$$(f,f^\prime)_\eta=\int_{orbit}
d\mu^\prime(p)\left(U_\lambda(\eta(p))^{-1}f(p),U_\lambda(\eta(p)^{-1})f^\prime 
(p)\right)_{\mathcal{H}_\lambda},$$
we see that the map $V$ is indeed unitary and substituting it in
(\ref{covariant2}) through $V^{-1}U^{m,+,\lambda}V=U_\eta^{m,+,\lambda}$ we
get:
\begin{equation}\label{covariant3}
\left(U^{m,+,\lambda}(g)f\right)(p)=e^{i<\phi,a>}
U_\lambda(\rho^{-1}_p\Lambda\rho_{\Lambda^{-1}p})f(\Lambda^{-1}p),
\end{equation}
which is coincident with the canonical wave equation though 
$U_\lambda$ is still a representation of $SL(2,\mathbb{C})$.

The last point of our construction comes into play since we need to 
impose further constraints on the wave equation in order to mod out the unwanted
components arising from the fact that $U_\lambda$ is a representation of
$SL(2,\mathbb{C})$ that has to be restricted to an $SU(2)$ one. This reduction can be expressed
using a matrix $\pi$ and, for the BMS group, the discussion for $SU(2)$ is
exactly the same as in Poincar\'e; so if we choose a fixed point $\bar{\phi}$ in the orbit, the constraints we have to impose becomes:
\begin{equation}
\pi[SU(2)]f(\bar{\phi})=f(\bar{\phi}).
\end{equation}
Remembering that a representation $D^{(j_1,j_2)}\in SL(2,\mathbb{C})$ can be
decomposed into SU(2) representations through
$D^{(j_1,j_2)}=\bigoplus\limits_{j=\mid j_1-j_2\mid}^{j_1+j_2}D^j$, the matrix
$\pi$ simply selects the desired value of $j$ from the above decomposition. If,
as an example, we consider the value $j=1$ in the representation
$D^{(\frac{1}{2},\frac{1}{2})}=\bigoplus\limits_{j=0}^1 D^j$, the matrix $\pi$ is:
$$\pi[SU(2)]=diag[0,1,1,1].$$

\begin{center}
{\large The group $\Gamma$}
\end{center}

Consider now $\Gamma$, namely the double cover of $SO(2)$. An element of this group is given by a
diagonal complex matrix:
$$g=\left[\begin{array}{cc}
e^{\frac{i}{2}t} &0\\
0& e^{-\frac{i}{2}t}
\end{array}\right].$$ Again in \cite{Mc2} the orbits for this little group were 
studied and it was shown that both the squared
mass and the sign of the energy are good labels. A fixed point on this orbit is
given by a supertranslation depending only on the modulus of the complex
coordinate over $S^2$ i.e. $\psi=\psi(\mid z\mid)$ (or equivalently in real
coordinates this means that $\psi(\theta,\varphi)$ does not depends on $\phi$).Thus
in this case the projection over the four-momentum for the fixed point is
$$\pi(\psi(\theta,\varphi))=(p_0,0,0,p_3).$$

Notice also that we choose in the orbit of $\Gamma$ those functions not in
the orbit of $SU(2)$-i.e. they cannot be transformed into a constant function;
this means that any point in the orbit has the form $\psi=\Gamma\psi(\mid
z\mid)$. In \cite{Mc1} also a section for the bundle
$G(SL(2,\mathbb{C})/\Gamma,\pi,\Gamma)$ has been given: any element $g\in SL(2,\mathbb{C})$ can be decomposed as explained
before as $g=\rho u$ with $u\in SU(2)$. The element $u$ can further be
decomposed as $u=\gamma_\phi\sigma_\theta\gamma_\psi$ which implies
$g=\tau\gamma_\psi$ where now $\gamma_\psi\in\Gamma$. Thus a section can be
written as:
$$\eta(\tau[\Gamma])=(0,\tau)\in BMS.$$
This completes the first and the second point of the construction. The canonical
wave equation can be written directly from the one-dimensional representation
for $\Gamma$ acting on an element $\gamma$ as a multiplication in one complex
dimension:
$$D^s(\gamma)=e^{ist}.$$
This leads to
\begin{equation}\label{inducedg}
U^{m,+,s,p_0}(g)f(gp)=e^{i<\phi,a>}D^s(\tau^{-1}_{\Lambda_g\Lambda_p}\Lambda_g\tau_{\Lambda_p})f(p),
\end{equation}
where $g=(a,\Lambda)\in BMS$, $f(p)$ is a square integrable function over the orbit
$SL(2,\mathbb{C})/\Gamma\sim \Re^3\times S^2$ and the indices on the 
left hand side label the orbit. 

Notice that the difference from $SU(2)$ is that these
indices are not uniquely determining the orbit since in that case the fixed
point was given by the constant supertranslation $K=m$ and for this reason the
value of the mass, and the sign of $p_0$ were fixing in a unique way the point over the
orbit. On the other hand here the labels $m,+,p_0$ are fixing only a class of
points with the same 4-momentum i.e. $p^\mu=(p_0,0,0,\sqrt{m^2+p^2_3})$ whereas
the supertranslation associated is not unique since in real coordinates it has
the form:
$$\phi(\theta)=p_0+p_3cos\theta+h(\theta),$$
for all possible $h(\theta)$ $\Gamma$-invariants.

It also worth stressing that if we choose $p_0=m$ and  $p_3=0$ (i.e. the
equivalent of the rest frame), the associated supertranslation is not constant
but it has the form $\phi(\theta)=K+h^\prime(\theta)$ with $h^\prime\neq 0$
since otherwise the point would belong to the $SU(2)$ orbit which is
impossible. This can be seen as the
impossibility to define ordinary angular momentum . More generally this is related to
the statement \cite{Mc8} that the representations of BMS different from $SU(2)$
cannot be uniquely reduced to Poincar\'e ones but each of them decomposes into
many differents Poincar\'e spins as recalled before.

Point (4) of our construction goes along similar lines 
just by reducing the functional integration to the corresponding orbit. 
Thus we only need to define the linear map for a generic point over
the orbit $p=(a,\Lambda)$ 
$$V=U(\rho_p^{-1}),$$
where $\rho^{-1}(p)$ is the inverse of the section $\eta(p)$.
As in the case of $SU(2)$ this linear operator switches from the space
$\mathcal{H}^{m,+,p_0,s}=L^2(N,d\mu^\prime)\otimes\mathcal{H}_s$ to the space 
$\mathcal{H}^{m,+,p_0,s}_\eta$ which coincides as a vector space but the internal product is:
$$(\psi,\phi)_\eta=\int\limits_{\Re^3\times S^2}
d\mu^\prime(p)\left(U_\lambda(\eta(p)^{-1})\psi(p),U_\lambda(\eta(p)^{-1})\phi(p)\right)_{\mathcal{H}_s}.$$
As before substituting the map $V$ into (\ref{covariant2}), we get:
\begin{equation}\label{covariant5}
\left(U^{m,+,p_0,\lambda}(g)f \right)(p)=e^{i<\phi,a>}
U_\lambda(\tau^{-1}_p\Lambda\tau_{\Lambda^{-1}p})f(\Lambda^{-1}p),
\end{equation}
which is coincident with (\ref{inducedg}) except that the representation needs to
be reduced from $SL(2,\mathbb{C})$ to $\Gamma$. Since
 all little groups in the Hilbert topology are compact, we know that $U$ is always
completely reducible and that the desired irreducible component can be 
picked out.

As before we start from a representation $D^{(0,j)}\oplus D^{(j,0)}$ to make 
sense of parity and we impose restriction. Since the decomposition
of $D^j(SU(2))$ into $\Gamma$ is well known and has the form:
$$D^l(\Gamma)=\bigoplus\limits_{m=-l}^l U^m(\Gamma),$$
we can first project from $SL(2,\mathbb{C})$ into a representation of $SU(2)$
and using the above decomposition, select a particular value of s. The
procedure is basically:
$$D^{(j_1,j_2)}(g)=\bigoplus\limits_{j=\mid
j_1-j_2\mid}^{j_1+j_2}D^j(g)=\bigoplus\limits_{j=\mid
j_1-j_2\mid}^{j_1+j_2}\bigoplus\limits_{s=-j}^jD^s(g).$$
The subsidiary condition is:
$$\pi(\Gamma)f(\bar{\phi})=f(\bar{\phi}),$$
where $\bar{\phi}$ is a fixed point over the orbit (i.e.
$\bar{\phi}=p_0+p_3\cos\theta+h(\theta)$) and where $\pi(\Gamma)$ extracts 
from the above decomposition the desired $s$
component. As an example in the case of $s=1$ from the representation
$D^{(\frac{1}{2},\frac{1}{2})}$ we know that:
$$D^{(\frac{1}{2},\frac{1}{2})}=\bigoplus\limits_{j=0}^1D^j(g)=\bigoplus\limits_{j=0}^1\bigoplus\limits_{s=-j}^jD^s(g),$$
so that 
$$\pi= diag[0,0,0,1] $$
\begin{center}
{\large The group $\Theta$}
\end{center}

The third group we examine is the compact, non-connected
little group $\Theta=\Gamma R_2$ where $R_2$ is the set (not the group) given by
the matrix $I$ and $J=\left[\begin{array}{cc}
0 & 1\\
-1 & 0\end{array}\right]$. The orbit of this group is given by the points in the
orbit of $\Gamma$ which are also invariant under $R_2$. This last condition is
equivalent to require for the supertranslation $\psi(\theta)=\psi(-\theta)$.
Upon projection over the dual space this implies that
$\pi(\psi)=p_0+p_3cos\theta=p_0$. Thus a fixed point under the action of $\Theta$ is
an element of the orbit of $\Gamma$ with the form
$$\psi(\theta)=p_0+h(\theta),$$
with clearly $h(\theta)\neq 0$ since otherwise the supertranslation would be in
the $SU(2)$ orbit.

Another important remark is that the orbit
$SL(2,\mathbb{C})/\Theta$ is $\Re^3\times P^2$ which is the same orbit of the
group $\Gamma$ plus the antipodal identification of the points over the sphere
due to $R_2$. In order to choose a section we remember that every $g\in
SL(2,\mathbb{C})$ can be decomposed as $g=\tau\gamma$ with $\gamma\in\Gamma$; a
point in $SL(2,\mathbb{C})/\Gamma$ is thus identified with the value of $\tau$.
In our case a global section (see \cite{Mc2}) can be given noticing that
every matrix $\sigma_\theta=\left[\begin{array}{cc}
\cos\frac{\theta}{2} & i\sin\frac{\theta}{2}\\
i\sin\frac{\theta}{2} & \cos\frac{\theta}{2}\end{array}\right]$ can be decomposed as
$\sigma_\theta=\sigma_{\theta^\prime}q$ with $0<\theta^\prime<\frac{\pi}{2}$,
$q=\gamma_{\frac{\pi}{2}}r\gamma_{-\frac{\pi}{2}}$ and $r\in R_2$. Since any
element $g$ of $SL(2,\mathbb{C})$ can be written as $g=\rho
u=\rho\gamma_\Phi\sigma_\theta\gamma_\psi$, we can plug in the above
decomposition 
$$g=\rho\gamma_\Phi\sigma_{\theta^\prime}q\gamma_\psi=\beta q^\prime.$$

Thus we can see that a section $\omega:G/\Theta\to G$ is given by
$\omega(g\Theta)=\omega(\lambda q^\prime\Theta)=\omega(\lambda\Theta)=\lambda$.
From this we can easily write a section 
$SL(2,\mathbb{C})\ltimes N/\Theta\ltimes N\to SL(2,\mathbb{C})\ltimes N$ as:
$$\eta(\beta)=(\beta,0).$$
This concludes point (1) and (2) of the construction. Consider now $\Theta$ 
representations: the 1-dimensional one when $s=0$ (s is the index for the
representation of $\Gamma$) which is $U(\gamma)=1$ and $U(J)=-1$ whereas for
integer $s$ we have:
$$U(\gamma_\phi)=\left[\begin{array}{cc}
e^{\frac{i}{2}s\phi} & 0\\
0 & e^{-\frac{i}{2}s\phi} 
\end{array}\right],\;\;\; U(J)=\left[\begin{array}{cc}
0 & (-)^s\\
1 & 0\end{array}\right].$$
Recalling that the for the orbit of $\Theta$ we can apply the same
considerations and the same labels as for $\Gamma$ except that in our case
$p_3=0$ (and $p_0=m$), we can write 
\begin{equation}\label{inducedt}
U^{m,+,s}(\gamma_\theta)(g)f(gp)=e^{i<\phi,a>}D^s[\beta^{-1}_{\Lambda_g\Lambda_p}\Lambda_g\beta_{\Lambda_p}]f(p),
\end{equation} 
where here we denote with $D^s$ the representation of $\Theta$ over an Hilbert
space $\mathcal{H}_s$ and thus the function $f(p)$ is in
$\mathcal{H}^{m,+,s}=L^2(SL(2,\mathbb{C})\ltimes N/\Theta\ltimes
N)\otimes\mathcal{H}_s$. We proceed than as in the previous cases. We introduce the operator
$V=U_s(\eta(\beta(p)^{-1}))$ that sends the Hilbert space $\mathcal{H}^{m,+,s}$ to
$\mathcal{H}^{m,+,s}_\eta$ which is coincident as a vector space to the first
but it is endowed with the internal product:
$$(f,f^\prime)_\eta=\int\limits_{\Re^3\times P^2}
d\mu^\prime(p)\left(U_\lambda(\eta(p))^{-1}f(p),U_\lambda(\eta(p)^{-1})f^\prime(p)\right)_{\mathcal{H}}.$$
As usual substituting the map $V$ into (\ref{covariant2}) we get:
\begin{equation}
\left(U^{m,+,\lambda}(g)f\right)(p)=e^{i<\phi,a>}
U_\lambda(\beta^{-1}_p\Lambda\beta_{\Lambda^{-1}p})f(\Lambda^{-1}_p),
\end{equation}
where $f(p)=(V\hat{f})(p)$ and $U_\lambda$ is the restriction of
$U_\lambda$ from $SL(2,\mathbb{C})$ to $\Theta$. We now reduce
$U_\lambda$ and this is indeed possible since $\Theta$ is compact. 
This can be achieved as for the $\Gamma$
case using the character formula (see \cite{Mc8}) and every single $s$ appears
exactly once in the decomposition of $SU(2)$. We proceed then as in 
the $\Gamma$ situation with the exception that now the
projection equation $\pi[\Theta]f(\bar{\phi})=f(\bar{\phi})$ is applied to the
supertranslations which are a fixed point over the orbit of the $\Theta$ group. The form of
the matrix will indeed be the same.
\begin{center}
{\large Finite groups}
\end{center}

Only finite dimensional groups have to be considered in the massive
case. We shall address only the (double cover of the) cyclic group $C_n$ as an example since all the
others are similar cases. 

The group
$\tilde{C}_n$ is given by the diagonal matrices:
$$c_n=\left[\begin{array}{cc}
e^{\frac{\pi ik}{n}} & 0\\
0 & e^{\frac{-\pi ik}{n}}\end{array}\right],$$
where $1\leq k\leq 2n$. The orbit for this group is constructed from a fix point
which is given by those supertranslations satisfying the periodic condition
$\psi(\theta,\varphi)=\psi(\theta,\varphi+\frac{2\pi}{n})$. Thus, a part from the case
$n=1$ which is trivial, the above condition tells us that the function
$\psi(\theta,\varphi)=p_0+p_3\cos\theta+k(\theta,\varphi)$ where $k(\theta,\varphi)$ is a
pure supertranslation. Furthermore we can assign a global section to this orbit
noticing that any element of $\Gamma$ can be decomposed as:
$$\left[\begin{array}{cc}
e^{i\frac{t}{2}} &0\\
0& e^{-i\frac{t}{2}}\end{array}\right]=\left[\begin{array}{cc}
e^{i\frac{t^\prime}{2}} &0\\
0& e^{-i\frac{t^\prime}{2}}\end{array}\right]\left[\begin{array}{cc}
e^{i\frac{\pi k}{n}} &0\\
0& e^{-i\frac{\pi k}{n}}\end{array}\right],$$
where $0<t^\prime<\frac{\pi}{n}$. Since any element $g$ of
$SL(2,\mathbb{C})$ can uniquely be written as $g=\tau\gamma_\theta$, we can plug
in the above decomposition writing $g=\tau\gamma_{\theta^\prime}c_k$. Thus
$\omega(gC_n)=\omega(\tau\gamma_{\theta^\prime}c_kC_n)=\omega(\alpha C_n)=\alpha$.
This concludes both points (1) and (2) of our construction. The representation is
simply one dimensional and gives $D^k(g)=e^{\frac{ik\pi}{n}}$ with $1\leq k\leq
2n$. Thus the canonical wave equation is simply:
\begin{equation}\label{inducedc}
U^{m,+,p_0,k}(g)f(gp)=e^{i<\phi,a>}D^k[\alpha^{-1}_{\Lambda_g\Lambda_p}\Lambda_g\alpha_{\Lambda_p}]f(p),
\end{equation}
where as usual $g=(a,\Lambda)$, $p$ is a point over the orbit and $f(p)\in
L^2(SL(2,\mathbb{C}\ltimes N/C_n\ltimes N)\otimes\mathcal{H}_k$. 
The operator relating the two description is as usual $V=U_\lambda(\eta_p^{-1})$ that
acting on functions over $\mathcal{H}^{m,+,p_0,\lambda}=L^2(N,
d\mu^\prime)\otimes\mathcal{H}_\lambda$ as:
$$(V\hat{f})(p)=U_\lambda\left(\eta^{-1}(p)\right)\hat{f}(p).$$
Thus V sends the Hilbert space $\mathcal{H}^{m,+,p_0,s}$ into
$\mathcal{H}^{m,+,p_0,s}_\eta$ which are coincident as vector spaces but the
latter is endowed with the inner product:
$$(f,f^\prime)_\eta=\int\limits_{\Re^3\times\frac{P^3}{C_n}}
d\mu^\prime\left(U_\lambda(\eta(p)^{-1})f(p),U_\lambda(\eta(p)^{-1})f^\prime(p)\right)_{\mathcal{H}_\lambda}.$$
Substituting the map $V$ into (\ref{covariant3}), we get 
\begin{equation}\label{covariant10}
\left(U^{m,+,p_0,\lambda}f\right)(p)=e^{i<\phi,a>}
U_\lambda(\alpha^{-1}_p\Lambda\alpha_{\Lambda^{-1}_p})f(\Lambda^{-1}p),
\end{equation} 
which is as usual coincident with (\ref{inducedc}) except for the fact that we
need some further constraints to select the rep we like. In this case we can start from a representation of
$SL(2,\mathbb{C})$ and progressively reduce it first to $SU(2)$ then to $\Gamma$
and in the end to $C_n$. Following \cite{Mc8} we only need to consider the
reduction of a representation $D^s$ of $\Gamma$. Since on an element $c_n\in
C_n$ $D^s$ acts giving $e^{\frac{i\pi s}{n}}$ whereas the representation of
$D^k$ of $C_n$ gives $e^{\frac{i\pi k}{n}}$; thus the representation of $C_n$
appears only one or no times in $D^s$. This condition is expressed by the
equation $s=\frac{k}{2}(mod\; n)$. From this we can see that
$$D^{(j_1,j_2)}(g)=\bigoplus\limits_{j=\mid j_1 - j_2\mid}^{j_1
+j_2}\bigoplus\limits_{s=-j}^j\delta_{s,\frac{k}{2}(mod\; n)}D^k(g).$$
We can easily  now extract the orthoprojection matrix $\pi$ and write
the additional conditions
$$\pi[C_n]f(\bar{\phi})=f(\bar{\phi}),$$
where $\bar{\phi}$ is a fixed point over the orbit of the $C_n$ group.
As an example let us consider the case $k=0$ for $C_2$ in the representation
$D^{(\frac{1}{2},\frac{1}{2})}$. We find from the decomposition that the
$U^0(C_2)$ can appear only when the equation $0=\frac{p}{2}(mod 2)$ holds; since $p$
ranges only from $-1$ to $1$, this implies that the desired representation appears
only once in each $D^0(\Gamma)$.Thus 
$$\pi[C_2]=diag[1,0,1,0].$$

This concludes our analysis for the massless case and also for the massive
case, since all other discrete groups in the Hilbert topology are acting like the
cyclic. 
\subsection{Generalizations of the BMS group and possible extensions of the results}

A natural question arising both from the representation theory and wave
equations concerns the origin and meaning
 of discrete little groups. A suggestion comes from \cite{Mc7} where McCarthy studies all
possible generalizations  (42 in the end) of the BMS group. 
Among them one finds $N(S)\ltimes L_+$ where $L_+$ is the usual (connected component of the
homogeneous) Lorentz group and $N(S)$ is
the set of $C^\infty$ scalar functions (``supertranslations'' this time 
defined on a hyperboloyd and depending on three angular coordinates) from
$S=\left\{x\in\mathbb{R}^{3,1}\;\mid\;x\cdot x<0\right\}$ to $\mathbb{R}$. This
group is isomorphic to the {\it Spi group} identified by Ashtekar and Hansen
\cite{hansen} as the asymptotic symmetry group of spatial infinity $i_0$ in
asymptotically flat space-times. The study of the representations for this group
as well as for all others BMS-type groups can be carried on exactly along the 
lines of the
original BMS. Thus we have still a freedom on the choice of topology for the
``supertranslation'' subgroup and wave equations can be in principle derived
exactly in the same way as we did in the previous sections. Furthermore this
suggests that a candidate field theory living on $i_0$ with fields carrying
representations of $Spi$, should display, as well as the
theory on $\Im^\pm$, a high degree of non locality this time with even 
more degeneracy since three instead of two angular coordinates define 
supertranslations.

Finally McCarthy also identified the euclidean BMS and the
complexification of the BMS group\footnote{The euclidean BMS is the semidirect product
of $N(\mathbb{R}^4-\left\{0\right\})$ with $SO(4)$ whereas the complexification
of BMS is given by the semidirect product of the complexified Lorentz group
$\mathbb{C}SO(3,1)$ with the space of scalar functions from
$\mathcal{N}=\left\{x\in\mathbb{C}^4\;\mid\;x\cdot x=0,\;\;x\neq 0\right\}$ to
$\mathbb{C}$.}; the study of representations for these groups endowed
with Hilbert topology gives rise to discrete groups and it has been suggested to
relate them to the parametrization of the gravitational instantons 
moduli space. 

\section{Implications for the holographic mapping}

\subsection{Identifying boundary degrees of freedom}

As we have seen in the previous analysis wave functions appearing in covariant and 
canonical wave equations are functions of the supertranslations or in dual terms of 
supermomenta. We have therefore a huge degree on non locality, the fields
 depending 
on an enormous (actually infinite) number of parameters entering
 into the expansions of supertranslations (supermomenta)
 on the 2-spheres.  
 
In addition their fluctuations are supposed to spread out on
 a degenerate manifold at null infinity. Recall also that 
supertranslations act along the $u$ direction and are punctual transformations, with $u$ playing 
the role of an affine parameter. Because of the difficulties in defining a theory 
on a degenerate manifold we find therefore more natural to place these fields and the 
putative boundary dynamics on the so 
called cone space \cite{bramson}, the space of smooth cross sections on $\Im$. The BMS group is 
thus interpreted as the group of mappings of the cone space onto itself.

We can give a pictorial description as follows: fix a two sphere
 section on $\Im$ and associate 
with this a point in the cone space calling this the origin.
 Any other point in the cone space will correspond to another 
two sphere section on $\Im$ and can be obtained by moving an affine distance $u=\alpha
(\theta,\phi)$ along the original $\Im$. In this way points on cone space are mapped one to 
one to cross sections on $\Im$. 

Therefore the holographic data ought to be encoded\footnote{See 
\cite{Jan} for a similar considerations despite differences in the choice 
of screens.} on the set of 2-spheres and these in 
turn are mapped to points in cone space; we think then 
the candidate holographic description living in an 
abstract space, implying in the end of the day that 
holography should be simply an equivalence of bulk amplitudes with those derived 
from the boundary theory.\footnote{The situation can be compared to the 
BMN \cite{BMN} limit of AdS/CFT, where the boundary of a pp-wave background
 is a null one 
dimensional line and geometrical interpretation seems difficult (and may be lacking). Again 
holography seems to be thought as an equivalence between bulk/boundary amplitudes, the latter 
may be living in some smaller CFT.}

An interesting consequence of working in the cone space is that one has in principle a 
way to define a length and therefore separate, in some sense, the spheres $S^2$ along null infinity. 
One can actually choose coordinates on the cone space: they will be 
the coefficients entering in the 
expansion of supertranslations in spherical harmonics. One can then show that
there exsists 
an affine structure (infinite dimensional) on the cone space and eventually define a 
length for vectors in the cone space \cite{bramson}
\begin{equation} \label{lunghezza}
L^2 = [\int d\Omega (\alpha(\theta,\phi))^{-2}]^{-1}
\end{equation}
This should allow to define \cite{noi} a sort of ``cutoff'', a concept otherwise absent on the 
original degenerate $\Im$. In the case of AdS/CFT the dual theory is a CFT with 
no fundamental scale. There, however, one uses the fact that AdS is basically 
a ``cylinder'' to end up with the correct counting from both sides \cite{susskind}.

This goes along the direction suggested
 by Bousso in \cite{changing}. 
 Actually apart from the special AdS case where
 one can show that, moving the boundary-screen 
to infinity, the boundary theory is indeed dual since it contains no more than one degree 
of freedom per Planck area \cite{susskind}, the ``dual theory''
 approach should not work in our case and one 
should expect theories with a {\it changing} number of degrees of freedom in the case of 
null boundaries. Degrees of freedom 
should appear and disappear continuously\footnote{We thank G.'t Hooft for
stressing this point also in cosmological context}. The
 dependence of (\ref{lunghezza}) on the coefficients 
tells us that a possible cut off length can change according to the 
number of coefficients we switch on-off. In turn we have seen that
 again $\alpha(\theta ,
\phi)$ enters in the changing of asymptotic shears and we have related the possibility 
of more/less bulk production according to the vanishing of asymptotic shear. 

Interestingly, more/less bulk entropy will have in the end of the day effect on
the way one defines lengths on the 
cone space.

\subsection{Similarities with 't Hooft S-matrix Ansatz for black holes}
The final picture one gets is quite similar to the scenario proposed by 't Hooft
\cite{'thooft} in the context of black holes. In this case we have sort of holographic fields 
living on the horizon of the black hole. Time reversal symmetry is 
required and therefore one has operators living on the future and past horizons. 

The description is given in first quantized set up and the degree of non 
locality is eventually expressed in the operator algebra at the horizon
\begin{equation}
[u(\Omega),v(\Omega')]=i f(\Omega - \Omega')
\end{equation}
\noindent where $u(\Omega),v(\Omega)$ are the holographic fields living on the 
future/past horizon depending on the angular coordinates $\Omega$ 
and $f(\Omega,\Omega')$ is the Green function of the Laplacian operator 
acting on the angular horizon coordinates. Clearly the 
algebra is non local.

In this case too the angular coordinates are at the end of the day responsible 
for the counting of the degrees of freedom, even if 't Hooft S-matrix Ansatz
is derived in a sort of eikonal limit assuming therefore a resolution bigger than 
Planck scale in the angular coordinates.

The Green function $f(\Omega,\Omega')$ tells 
us how to move on the granular structure living on the horizon and it is similar 
to our supertranslations generated indeed by $P_{lm}=Y_{lm}(\Omega) \partial_u$. 
The holographic information is therefore spread out in both cases on angular 
coordinates.

Going then to a second quantized description of 't Hooft formalism, fields are 
expected to be functional of $u(\Omega)$ and $v(\Omega)$, pretty much in the same 
way of our case. A proposal of 't Hooft (for the 2+1 dimensional case) preserving 
covariance is indeed
\begin{equation}
\phi \sim \sum_{orderings} \int d\Omega \left(
\delta(x^1(\Omega)-x) \delta(x^2(\Omega)-y) \delta(x^0(\Omega)-t) \right)
\end{equation}
where fields are functionals of coordinates in turn depending on the angles. 
Of course the horizon itself is a very special null surface 
and the set up is different, since the whole 't Hooft picture is dynamically generated in a holographic reduction 
taking place in a sort of WKB limit. At the end of the day, however, angular 
coordinates and their resolutions are the basis for the book-keeping of states.

The horizon itself is a sort of computer storing-transmitting information. One looses 
in a sense the notion of time evolution. This suggests that also in our case (recall 
that $u$ acts via point transformations) the fields we have constructed 
are not quite 
required to evolve but are independent data living on the
2-spheres. What generates the dynamics should be a S-matrix in the spirit 
of 't Hooft. In this sense, the states are indeed holographic, since contain all 
bulk information. 
\subsection{Remarks on the construction of a S-matrix}
Therefore, in analogy with 't Hooft scenario, we 
could imagine \cite{noi} a S-matrix with in and out states
living on the respective cone spaces corresponding to  $\Im^+$ and
$\Im^-$ with fields carrying BMS labels. Of course the big task is to 
explicitly construct such a mapping.\footnote{One should also may be take into account 
the role of spatial infinity in gluing the past and the future null 
infinities.}

However, the motivation for a S-matrix is also due to the fact that in the asymptotically 
flat case we have problems with massive states which can change the 
geometry at infinity. In the case of AdS/CFT correspondence, on the other 
hand, it is true that one has a sort of box with walls at infinity so that
quantization of modes is similar to fields in a cavity. But via the Kaluza-Klein 
mechanism one generates a confining potential for massive modes. Note also 
that already for massless fields the scattering problem in the physical spacetime
 can be translated in a 
characteristic initial value problem \cite{frolov} at null infinity in the unphysical 
spacetime.\footnote{\cite{frolov} contains a derivation of the Hawking 
effect and an interesting discussion on  the BMS group. The main point is that
one can have in any case an unambiguous definition of positive/negative frequencies and this 
is what matters for the Hawking particle production. Recall however that there ones refers 
to fields  and their asymptotics in the bulk, not to fields carrying 
BMS representations as the ones we have derived.}

The way in which one should proceed in quantizing has to be
different from the usual one, since one does not need choices of
polarizations
to kill unwanted phase space ``volume''. This has already been pointed out
in \cite{Mc9} discussing BMS representations and therefore is
automatically
induced to fields carrying BMS representations we have constructed.
                                                                                
The choice of Hilbert topology, as already remarked, should be associated
with
bounded sources in the bulk, while the nuclear one should correspond to
unbounded
systems. In order to accommodate the unbounded systems which ought to
correspond to scattering states, one should however define a proper notion
of
conformal infinity for unbounded states and this seems a difficult
problem. Note however that to have a unitary S-matrix in a candidate
holographic
theory one {\it must} include unbounded
states into the Hilbert space, otherwise
asymptotic completeness \cite{simon} is violated.
                                                                                
In \cite{Mc4} it was also suggested to take a finer topology than the
nuclear one because of the freedom in the
topology choice for the supertranslations. It was proposed to use real
analytitc functions enlarging therefore supermomenta to real
hyperfunctions on the sphere. Interestingly quantum field theories in
which
fields are smeared by hyperfunctions show a non local behaviour and the
density of states can have a non polynomial growth.
This
might in principle allow to recover bulk locality\footnote{See \cite{Giddings}
for similar considerations even if from a different point of view.} although one
should consider hyperfunctional
solutions to the Einstein's equations. Moreover, if one assumes that the high
energy behaviour of the density of states in the bulk is dominated by black
holes, the exponential growth of states which suggests an intrinsic degree of
non locality might be explained by working with hyperfunctions. This is again in
sharp contrast with the asymptotically AdS case where the black hole density of
states grows essentially like the entropy of a CFT \cite{tom}.

\chapter{Conclusion}
In this thesis we have studied several aspects of two different systems in which
we wanted to implement the holographic principle.\\
In the previous chapter we focused our attention on asymptotically flat
space-times; we followed a road similar to the one settled by Maldacena in the
AdS/CFT scenario where the bulk datas are essentially encoded in a field
theory invariant under the action of the conformal group which is the asymptotic
symmetry group of an AdS manifold. In the asymptotically flat framework, the role
of the conformal group is taken over by the so called Bondi-Metzner-Sachs group
which is the semidirect product of the homogeneous Lorentz group with the
abelian set of $C^\infty$ maps from the 2-sphere into the real numbers. \par
We outlined the construction of the BMS group following the approach proposed by
Penrose who considered the boundary of an asymptotically flat space-time as a
separate manifold from the bulk. This line of research considers
the BMS group as the set of isometries preserving the inner structure
of the null infinity $\Im^\pm$; this suggests to look at the BMS group as a
universal feature of any asymptotically flat space-time and not only as the
set of diffeomorphism preserving some suitable asymptotic form of the metric.
Starting from these remarks, we focused our attention on the implementation of
the holographic principle in such framework. The first step in this direction
was to recognize that already at a geometrical level the asymptotically flat 
scenario has a completely different
beahviour compared to an AdS manifold. In particular we applied Bousso
covariant entropy conjecture to show from one side that the null boundaries $\Im^\pm$ are
preferred screens suitable to encode holographic datas; on the other side a key concept
for the AdS/CFT scenario such as the holographic renormalization process, is not
available in an asymptotically flat manifold due to the high focusing of light
rays which leads to the formation of caustics; thus a purely
geometrical reconstruction of the bulk is impossible. \par
In order to fully implement the holographic principle in the asymptotically flat
scenario, we switched our attention to a pure field theoretical point of view
hoping to share some light on the bulk/boundary correspondence from the
correlator of a BMS field theory. In order to complete this task, we need to
know the full spectrum of particles for the boundary theory; in this thesis we
reached this goal constructing the field equations for all the BMS invariant 
particles following a road similar to Wigner construction for the Poincar\'e
scenario \cite{Wigner}. In this approach we started from the theory of
representation for the BMS group and we introduced the covariant BMS wave
equations as maps from the set of supertranslations $N=L^2(S^2)$ into a suitable
Hilbert space\footnote{This is in no
contrast with the usual notion of the fields as maps from space-time into an
Hilbert space since the space of supertranslations is the quotient of the BMS
with the Lorentz group in the same way as Minkowski is the quotient of the
Poincar\'e with the Lorentz group.}. Moreover, from these equations, we can
conclude that, whatever is the holographic dual theory, it has to show an high
degree of non locality since the particles evolve on $N$ which is an infinite
dimensional Hilbert (and affine) space where there is no proper notion of
metric. The concept of distance itself, i.e.
$L^2=\int\frac{d\Omega}{\alpha^2(\theta,\phi)}$, is strictly dependant on the
choice of the supertranslation and it has no direct correspondence with points
on $\Im^\pm$; this is a clear key sign of non locality. \par
In the spirit of Wigner program, we also introduced the induced wave equations which are
maps from the orbit of the little groups into a suitable Hilbert space. From
these datas we achieved a twofold result: from one side we have shown that the
covariant wave equation reduces to the induced equation if we impose that each BMS
field satisfies the relation
$$\pi(\alpha)\psi(\alpha)=\psi(\alpha),$$
where $\psi$ is the covariant wave and $\pi$ is a suitable orthoprojector. Although
the interpretation of the above relation is not immediate, we have shown that in the
Poincar\'e scenario, it is totally equivalent to the usual field equations such as the
Klein-Gordon or the Dirac equations. On the other side the induced wave equations
allowed us to discover the full spectrum of the BMS field theory; comparing it with
the Poincar\'e spectrum, we can outline some important remarks:
\begin{itemize}
\item the particles coming from the $SU(2)$ little group can be easily put in
correspondence with Poincar\'e massive particles whereas the fields
associated to other little groups, most notably, the groups $\Gamma$ and $\Theta$
cannot be directly mathced with any bulk datas. 
\item in the BMS frameork, massless particles have no difference from their massive counterpart and a massless
wave equation can be derived as the limit of $m\to 0$ of the massive one. This is a
striking difference from the Poincar\'e group since in that case massless fields are
associated to unfaithful representations of the $E(2)$ group and they have no direct
correspondence with their massive counterpart. This is clearly a pure
supertranslational effect since it is possible to show that in the limit $m\to 0$, the
Poincar\'e four momentum vanishes whereas in the BMS scenario we have still the
freedom to choose a non zero supermomentum.
\end{itemize}
Another interesting point that rises from our analysis is related to the discrete
little groups. Their structure is identical to the common ADE series and we
advocate the hypotesis proposed in \cite{Mc7} to relate them to instanton
configurations of the gravitational field. This idea is extremely suggestive since it
implies that already at a classical level, we have a residual sign on the boundary of
a pure quantum gravity effect; on the holographic side this remark tells us that we
are facing a scenario which is completely different from the usual paradigm
where the boundary dual theory loses notion of the gravitational field.\par
Eventually we conclude that the analysis in this thesis strongly encourages the
research of an holographic correspondence in an asymptotically flat background
starting from a field theory invariant under the asymptotic symmetry group.
Thus, at this stage, we are now facing two possible roads in order to
continue the analysis along the lines presented in this thesis: from one side it would
be interesting to find an action functional for the BMS fields on the underlying
Hilbert space $N$. Since there is no proper notion of metric, the usual techniques are
doomed to failure; for this reason we suggest to follow a pure group theoretical
approach starting from the caodjoint orbits for the BMS group. This line of research
was first introduced by Witten for the Virasoro algebra \cite{coadjoint} and
generalized to any infinite dimensional Lie group in \cite{delius}. The underlying
idea is based upon the existence on any coadjoint orbit $\mathcal{O}$ of a symplectic 
form $\Omega$ which allow us to interpret the pair $(\mathcal{O},\Omega)$ as a
symplectic manifold and as the phase space for the correspondent BMS particle. For a
system with vanishing hamiltonian such as a conformal field theory, the above datas
are sufficient in order to introduce the action for the system, whereas in our
framework we need to find a proper hamiltonian for each particle in order to fully
understand the underlying dynamic. A possible evolution in this direction comes both
from the analysis in \cite{Symplectic} and \cite{Chrusciel} where a global BMS invariant hamiltonian is
introduced although it is unknown at present how to extract from it the corresponding
hamiltonian for each single particle (see \cite{Arcioni3} for a recent progress).\par
A second line of research, which we also strongly advocate, follows 't Hooft suggestion to
encode the dynamic of field evolution in an asymptotically flat space-time through a
proper S-matrix connecting the states on $\Im^+$ with those on $\Im^-$. In this
approach there is no direct reference to any hamiltonian and we refer only to
the usual techniques used to analyize Hilbert spaces. Following in particular the
lines of Ashtekar work \cite{Ashtekar-Magnon}, this idea seems very promising
but it still lacks
direct contact with the pure geometrical datas about the BMS group \cite{Symplectic} and it is
still difficult to concieve a way to explicitly write the S-matrix.\par
Although a full understanding of the holographic correspondence in asymptotically flat
space-times is still far away, another interesting point which emerges from this
thesis refers to the implementation of holography in our framework in a different
dimension. From the analysis of the asymptotic symmetry group, the situation seems
quite different from the one cexpected from past experiences in the AdS/CFT
scenario. In particular, if we consider $d>4$, the situation can be quite easier since
there are indications that it is possible to extract a unique Poincar\'e group from the
BMS group although a universally accepted demonstration of this property is still
lacking. On the opposite if we consider a three dimensional manifold, the situation is
completely different and apparently much more complicated since in this case the BMS
group is the semidirect product of the Lorentz group with the diffeomorphism of the
circle. We cannot avoid to notice a situation very similar to $AdS_3/CFT_2$ since both
$CFT_2$ and $BMS_3$ have a conformal algebra. Moreover this leads us to conjecture
that, in the same spirit as for $CFT_2$, \cite{Brown} a possible classical central charge emerge
for the BMS$_3$ algebra and preliminary works in this direction \cite{Barnich} 
could suggest that this a concrete chance.  \par
For sake of completeness let us also mention that, in the spirit of finding an
holographic correspondence, recently new lines of research have been proposed,
the most notable one by Banks who suggests to slice an
asymptotically flat space-times through a diamond $D$. The dual theory should be constructed on the 
boundaries of such a diamond and the holographic regime would be reached only
in the limit where $D$ is coincident with the boundaries. Although this approach is interesting and
avoids a lot of technical problems, we still advocate a direct study of the
asymptotic symmetry group since the above mentioned approach does not take into
account both the null nature of $\Im^\pm,$ and the origin of the BMS group. As
far as $D$ is in the bulk, the field theory living on its boundaries is
by construction Poincar\'e invariant and the BMS will not naturally
appear in an asymptotic regime. Moreover by sending $D$ to infinity we have to
consider not only the null boundaries $\Im^\pm$ but also the spatial infinity
$i_0$ where the asymptotic symmetry group, the Spi, is different from the BMS
and it is unclear to us how in a potential dual theory both the BMS
and the Spi invariance could cohabit at the same time.

\vspace{0.3cm}

On a different ground, in this thesis we studied discretized systems and in particular triangulated surfaces. Our starting point was the tentative
in \cite{Arcioni} to implement the holographic principle in the context of Ponzano-Regge calculus. Unfortunately, although a scheme to find a 
boundary partition function $Z_{bound}$ was introduced, the boundary functional itself was difficult to interpret. The leading idea was that, since 
in the continuum counterpart, there is a strict relation between a Chern-Simons theory living in the bulk of a three dimensional manifold $M$ and
a WZW model living on $\partial M$, the functional $Z_{bound}$ should be related to an SU(2) WZW model. The main result of chapter 2 and 3 has
thus been to provide the correct definition of such a model on random Regge triangulation and in particular we constructed the partition function
for the case $SU(2)$ at level $1$. Morover the results of this thesis on the ground of discretized models open a lot of interesting
line of research.

As an example, in the language we have introduced, 2D gravity can be promoted to a dynamical role by summing (\ref
{FinPart}) over all possible Regge polytopes (\emph{i.e.}, over all possible
metric ribbon graphs $\{\Gamma ,\{L(p,r)\}\}$). It is clear, from the
edge-lenght dependence in (\ref{FinPart}), that the formal Regge functional
measure $\propto \prod_{\{\rho ^{1}(p,r)\}}dL(p,r)$, involved in such a
summation, inherits an anomalous scaling related to the presence of the weighting
factor (to be summed over all isospin channels $j(r,p)$)

\begin{equation}
\prod_{\{\rho ^{1}(p,r)\}}^{N_{1}(T)}L(p,r)^{-2H_{j_{(r,p)}}},  \label{scale}
\end{equation}
where the exponents $\{H_{j_{(r,p)}}\}$ characterize the conformal dimension
of the boundary insertion operators $\{\psi _{j_{(r,p)}}^{j_{p}j_{r}}\}$. 
A dynamical triangulation prescription (\emph{i.e.}, holding fixed the $
\{l(p,r)\}$ and simply summing over all possible topological ribbon graphs $
\{\Gamma \}$) feels such a scaling more directly via the two-point function 
(\ref{twopoints}), and (\ref{critedge})(again to be summed over all possible isospin 
channels $j(r,p)$) which exhibit the same exponent dependence.\\
This remark is extremely important from the view point of the critical
string where we still lack a universally accepted explenation for the anomalous
scaling related to the conformal factor. We suggest that the definition of a WZW model on a 
triangulated Riemann surface thus open the chance to find in the language of
discretized models a way to correctly describe such scaling.

More in the spirit of the duality CS/WZW model, it could be of great interest to better understand how (\ref{FinPart}) relates with the bulk dynamics in the
double $\widetilde{V}_{M}$ of the 3-manifold $V_{M}$ associated with the
triangulated surface $M$. Since we are in a discretized setting, such a
connection manifests itself, not surprisingly, with an underyling structure
of $\ Z^{WZW}(|P_{T_{l}}|,\{\widehat{h}(S_{\theta (i)}^{(+)})\})$ which
directly calls into play, via the presence of the (quantum) $6j$-symbols,
the building blocks of the Turaev-Viro construction. This latter theory is
an example of topological, or more properly, of a cohomological model. When
there are no boundaries, it is characterized by a small (finite dimensional)
Hilbert space of states; in the presence of boundaries, however, cohomology
increases and the model provides an instance of a holographic correspondence
where the space of conformal blocks of the boundary theory (\emph{i.e.}, the
space of pre-correlators of the associated CFT) can be also understood as
the space of physical states of the bulk topological field theory. A
boundary on a Riemann surface, for instance, makes the cohomology bigger and
this is precisely the case we are dealing with since we are representing a
(random Regge) triangulated surface $|T_{l}|\rightarrow M$ \ by means of a
Riemann surface with cylindrical ends. Thus, we come to a full circle: the
boundary discretized degrees of freedom of the $SU(2)$ WZW theory coupled
with the discretized metric geometry of the supporting surface, give rise to
all the elements which characterize the discretized version of the
Chern-Simons bulk theory on $\widetilde{V}_{M}$. What is the origin of such
a Chern-Simons model? The answer lies in the observation that by considering 
$SU(2)$ valued maps on a random Regge polytope, the natural outcome is not
just a WZW model generated according to the above prescription. The
decoration of the pointed Riemann surface $((M;N_{0}),\mathcal{C})$ with the
quadratic differential $\phi $, naturally couples the model with a gauge
field $A$. In order to see explicitly how this coupling works, we observe
that on the Riemann surface with cylindrical ends $\partial M$, associated
with the Regge polytope $|P_{T_{l}}|\rightarrow {M}$, we can introduce $
\mathfrak{su}(2)$ valued flat gauge potentials $A_{(i)}$ locally defined by 
\begin{gather}
A_{(i)}\doteq \gamma _{i}\left[ \sqrt{\phi (i)}\left( \frac{\lambda (i)}{
\kappa }\mathbf{\sigma }_{3}\right) -\frac{\sqrt{-1}}{2\pi }L(i)\left( \frac{
\lambda (i)}{\kappa }\mathbf{\sigma }_{3}\right) d\ln \left| \zeta
(i)\right| \right] \gamma _{i}^{-1}=  \notag \\
\\
=\frac{\sqrt{-1}}{4\pi }L(i)\gamma _{i}\left( \frac{\lambda (i)}{\kappa }
\mathbf{\ \sigma }_{3}\right) \gamma _{i}^{-1}\left( \frac{d\zeta (i)}{\zeta
(i)}-\frac{d\overline{\zeta }(i)}{\overline{\zeta }(i)}\right) ,  \notag
\end{gather}
around each cylindrical end $\Delta _{\theta (i)}^{\ast }$ of base
circumference $L(i)$, and where $\gamma _{i}\in SU(2)$. (It is worthwhile to
note that the geometrical role of $\ $\ the connection $\{A_{(i)}\}$ is more
properly seen as the introduction, on the cohomology group $H^{1}((M,N_{0});
\mathcal{C})$ of the pointed Riemann surface $((M,N_{0});\mathcal{C})$, of
an Hodge structure analogous to the classical Hodge decomposition of $
H^{h}(M;\mathcal{C})$ generated by the spaces $\mathcal{H}^{r,h-r}$ of
harmonic $h$-forms on $(M;\mathcal{C})$ of type $(r,h-r)$. Such a
decomposition does not hold, as it stands, for punctured surfaces since $
H^{1}((M,N_{0});\mathcal{C})$ can be odd-dimensional, but it can be replaced
by the mixed Deligne-Hodge decomposition). The action $S_{|T_{l=a}|}^{WZW}(
\eta )$ gets correspondingly dressed according to a standard prescription
(see \emph{e.g.} \cite{gawedzki}) and one is rather naturally led to the
familiar correspondence between states of the bulk Chern-Simons theory
associated with the gauge field $A$, and the correlators of the boundary WZW
model.


 Let us also stress that the relation between (\ref{FinPart}) and a triangulation of the
bulk 3-manifold $\widetilde{V}_{M}$, say,  the association of tetrahedra to
the (quantum) $6j$-symbols characterized by (\ref{seigei}), is rather
natural under the doubling procedure giving rise to $\widetilde{V}_{M}$ and
to the Schottky double $M^{D}$. Under such doubling, the trivalent vertices $
\{\rho ^{0}(p,q,r)\}$ of \ $|P_{T_{l}}|\rightarrow {M}$ yield two preimages
in $\widetilde{V}_{M}$, say $\sigma _{(3)}^{0}(\alpha )$ and $\sigma
_{(3)}^{0}(\beta )$, whereas the outer boundaries $S_{\theta (p)}^{(+)}$, $
S_{\theta (q)}^{(+)}$, $S_{\theta (r)}^{(+)}$ associated with the vertices $
\sigma ^{0}(p)$, $\sigma ^{0}(q)$, and $\sigma ^{0}(r)$ in $
|T_{l}|\rightarrow M$ are left fixed under the involution $\Upsilon $
defining $M^{D}$. Fix our attention on $\sigma _{(3)}^{0}(\alpha )$, and let
us consider the tetrahedron $\sigma _{(3)}^{3}(p,q,r,\alpha )$ with base the
triangle $\sigma ^{2}(p,q,r)\in |T_{l}|\rightarrow M$ and apex $\sigma
_{(3)}^{0}(\alpha )$. According to our analysis of the insertion operators $
\{\psi _{j_{(r,p)}}^{j_{p}j_{r}}\}$, to the edges $\sigma ^{1}(p,q)$, $
\sigma ^{1}(q,r)$, and $\sigma ^{1}(r,p)$ of the triangle  
$\sigma ^{2}(p,q,r)$ we must
associate the primary labels $j(p,q)$, $j(q,r)$, and $j(r,p)$, respectively.
Similarly, it is also natural to associate with the edges $\sigma
_{(3)}^{1}(p,\alpha )$, $\sigma _{(3)}^{1}(q,\alpha )$, and $\sigma
_{(3)}^{1}(r,\alpha )$ the labels $j_{p}$, $j_{q}$, and $j_{r}$,
respectively. Thus, we have the tetrahedron labelling 
\begin{equation}
\sigma _{(3)}^{3}(p,q,r,\alpha )\longmapsto \left(
j(p,q),j(q,r),j(r,p);j_{p},j_{q},j_{r}\right) .
\end{equation}
The standard prescription for associating the $6j$-symbols to a $
SU(2)_{Q}$-labelled tetrahedron such as $\sigma _{(3)}^{3}(p,q,r,\alpha )$
provides 
\begin{equation}
\sigma _{(3)}^{3}(p,q,r,\alpha )\longmapsto \left\{ 
\begin{array}{ccc}
j_{(q,p)} & j_{p} & j_{q} \\ 
j_{r} & j_{(q,r)} & j_{(p,r)}
\end{array}
\right\} _{Q=e^{\frac{\pi }{3}\sqrt{-1}}},
\end{equation}
which (up to symmetries) can be identified with (\ref{seigei}). \ In this
connection, one can observe that the partition function (\ref{FinPart}) has
a formal structure not too dissimilar (in its general representation
theoretic features) from the boundary partition function discussed in 
\cite{Arcioni}. Such a correspondence should be analyized in details since 
a possible matching between the partition function of a discretized $SU(2)$ WZW
model and the functional $Z_{bound}$ would definitely allow us to understand the
mechanism of the holographic principle above all in a system where it is apparently realized in a way completely different from the "canoncial" one
seen in the AdS/CFT. 

\newpage     

\begin{center}
{\bf Acknowledgments}
\end{center}

\vspace{0.5cm}

\noindent First of all and above all I wish to deeply thank my supervisor Mauro Carfora for his help, teaching and ability to understand me in the past 
3 years. Without his help all my work and this thesis simply would not exist. 
I want also to deeply thank Annalisa Marzuoli for the help, the encouragements
and the useful discussions during my work in Pavia. I also deeply thank 
professor Renate Loll for useful discussions and for granting me the chance to visit the Spinoza institute
during last year where I learned really a lot. I am also grateful to professor P.J. McCarthy for fruitful correspondence on the BMS group. 
For usueful and interesting discussions I also thank all the people in Spinoza
Institute and in Pavia University and in particular professor G. 't Hooft, Dr. Henning 
Samtlebeen, Dr Omar Maj, Drs Alberto Orlandi, Dr Paoloplacido Lopresti, Drs Zoltan Kadar, Drs Laura Tamassia, 
Drs Valeria Gili, Drs Mathias Karadi and Drs Dario Benedetti. 
A special thanks goes to Dr. Mario Trigiante for the discussions, his patience and his advices during the realization 
of the paper on holography and the BMS group. Another deep special thank you goes to my friend and collaborator Dr. Giovanni Arcioni for his help during my stay in 
Utrecht and in the physics world.


\appendix
\chapter[Induced representations...]{Induced representations for semidirect product groups}
We review in the following the theory of induced representations for 
semidirect product groups. Notations and conventions are those of \cite{Simms}. 

Choose a locally compact group $G$ and a closed subgroup $K\subset G$
whose unitary representations $\sigma: K\to U(M)$ on a Hilbert space $M$ are
known. On the topological product $G\times V$, define an
equivalence relation
$$(gk,v)\sim (g,\sigma(k)v)\;\;\;\forall k\in K,$$
and the natural map
\begin{equation}\label{hilb}
\pi:G\times_K V\to G/K,
\end{equation}
assigning to equivalence classes $[g,k]$ the element $gK$ on the coset $G/K$.
The structure given in (\ref{hilb}) is clearly the one of a fiber bundle where the generic fiber over a base point 
($\pi^{-1}(gK)=\left\{[g,v]\right\}$) uniquely determines the element $v\in M$.
This bijection gives to $\pi^{-1}(gK)$ the structure of a Hilbert
space\footnote{A Hilbert bundle is defined as $\pi:X\to Y$ where both $X,Y$
are topological spaces and a structure of a Hilbert space is given to
$\pi^{-1}(p)$ for each $p\in Y$.}.

One can also introduce then a {\it Hilbert G-bundle} which is a Hilbert
bundle as before with the action of a group $G$ on both $X,Y$ such
that the map 
$$\alpha_g:x\longrightarrow gx,\;\;\;\beta_g:y\longrightarrow gy$$ is a Hilbert
bundle automorphism for each group element.\\
Choose then a unique invariant measure class on the space $G/K$ defined as
above-i.e. for any $g\in G$, for any Borel set $E$ and for a given measure $\mu$ also
$\mu_g(E)=\mu(g^{-1}E)$ is a measure in the same class. It can be
extended to any G-Hilbert bundle $\zeta=(\pi:X\to Y)$ where $G$ is a topological group; besides given
the Hilbert inner product on a fiber $\pi^{-1}(p)$ and an invariant measure
class $\mu$ we can introduce 
$$\mathcal{H}=\left\{\psi\mid \psi\; {\textrm\; a\; Borel\; section\; of\; the\; bundle,}
\int\limits_Y<\psi(p),\psi(p)>d\mu(p) < \infty\right\}.$$

A unitary G-action on $\mathcal{H}$ is given by:
$$(g\psi)(p)=\sqrt{\frac{d\mu_g}{d\mu}(p)}g(\psi(g^{-1}\cdot p)),\;\;\;\forall
p\in Y.$$
This action is also a representation of $G$ on $\mathcal{H}$ and does not
 depend on the measure $\mu$. Moreover this construction grants us that
for any G-Hilbert bundle $\zeta_\sigma=(\pi:G\times_K V\to G/K)$ defined by a
locally compact group $G$ and by a K-representation $\sigma$, it is possible
 to derive an {\it induced representation}
$T(\zeta_\sigma)$ which basically tells us that from any representation $\sigma$
of $K$ we can de facto induce a representation $T$ to $G$. 
Consider now a group $G$ which contains an
abelian normal subgroup $N$. If we choose a subgroup $H$ such that the map
$N\times H$ is bijective, we can show that there exists an isomorphism between G
and the semidirect product of $N\ltimes H$.

A character of $N$ is a continuous homomorphism
\begin{equation}
\chi:N\longrightarrow U(1).
\end{equation}
The set of all these maps forms an abelian group called the {\it dual group}:
$$\hat{N}=\left\{\chi\mid (\chi_1\chi_2)(n)=\chi_1(n)\chi_2(n)\right\}$$
Define then a G-action ($G\sim N\ltimes H$) onto the dual space
induced from $G\times N\to N$ letting $(g,n)\to g^{-1}ng$ such that
$$G\times\hat{N}\to\hat{N}$$
gives $g\chi(n)=\chi(g^{-1}ng)$. Thus for any element $\chi\in\hat{N}$, one can
define the {\it orbit} of the character as:
$$G\chi=\left\{g\chi\;\mid\; g\in G\right\},$$
and the {\it isotropy group} of a character under the G-action as:
$$G_\chi=\left\{g\; \mid\; g\in G,\;\;g\chi=\chi\right\}.$$
Clearly the set $G_\chi$ is never empty due to the fact that $N$ acts trivially
onto a character. Introduce now $L_\chi=H\cap G_\chi$, then
$$G_\chi=N\ltimes L_\chi.$$
The group $L_\chi$ is called the {\it little group} of $\chi$ and it is the
isotropy group of the character $\chi$ under the action of the subgroup $H\subset
G$.

Consider now a unitary representation $\sigma$ for the little group $L_\chi$ acting on a
vector space $V$. Then the map
$$\chi\sigma: N\times V\to U(N\times V),$$ 
such that $(n,v)\to \chi(n)\sigma(v)$, is a unitary representation of
$G_\chi$ on the vector space $V$. So one can
introduce an Hilbert G-bundle $\pi:G\times_{G_\chi} V\to G/G_\chi$ with a base
space isomorphic to the space of orbits $G\chi$ defined for every representation
$\chi\sigma$. One finally has \cite{mackey}: 
\begin{theorem}[Mackey]
Let $G=N\ltimes H$ be a semidirect group as above and suppose that $\hat{N}$
contains a Borel subset meeting each orbit in $\hat{N}$ in just one point. Then
\begin{itemize}
\item The representation $T(\zeta)$ induced by the bundle $\pi:G\times_{G_\chi}
V\to G/G_\chi$ is an irrep of $G\sim N\ltimes H$ for any $\chi$ and for any
$\sigma$.
\item each irrep of $G$ is equivalent to a representation $T(\zeta)$ as above
with the orbit $G\chi$ uniquely determined and $\sigma$ determined only up to
equivalence.
\end{itemize}  
\end{theorem}
\section{BMS representations in Hilbert topology}

We endow the supertranslation group
with an Hilbert inner product:
\begin{equation}
<\alpha,\beta>=\int\limits_{S^2}\alpha(x)\beta(x)d\Omega;
\end{equation}
where $x\in S^2$, and the supertranslations $\alpha,\beta$ are scalar maps
$S^2\to\Re$.
Therefore $N=L^2(S^2)$ is an abelian
topological group. 

Any element $\alpha$ in the supertranslation group can be decomposed as:
$$\alpha(\theta,\phi)=\sum\limits_{l,m}\alpha_{lm}Y_{lm}(\theta,\phi).$$
This decomposition is topology independent but in the case we are
 considering the complex coefficients $\alpha_{lm}$
have to satisfy
$$\bar{\alpha}_{lm}=(-)^m\alpha_{l,-m}.$$
Notice that supertranslations
admit a natural decomposition into the direct sum of two orthogonal (under the
Hilbert space internal product) subspaces (i.e. subgroups): translations and
proper supertranslations. In particular, for any $\alpha(\theta,\phi)\in
L^2(S^2)$, one can write $\alpha=\alpha_0+\alpha_1$ with:
$$\alpha_0=\sum\limits_{l=0}^1\sum\limits_{m=-l}^l\alpha_{lm}Y_{lm}(\theta,\phi),\;\;\;\alpha_1=\sum\limits_{l>1}\sum\limits_{m=-l}^l\alpha_{lm}Y_{lm}(\theta,\phi).$$
Thus $N$ can be written as:
$$N\sim A\oplus B,$$
where $A$ is the translation group and $B=N-A$. One has however to keep in mind
that this decomposition, as an isomorphism, is not preserved under the action of the
$SL(2,\mathbb{C})$ group. 

Consider the dual of the supertranslation space, the character
space $\hat{N}$ whose elements can be written exploiting the Reisz-Fischer theorem for
Hilbert spaces as: 
$$\chi\in\hat{N}\Longrightarrow \chi(\alpha)=e^{i<\phi,\alpha>},$$
where $\phi\in N$ is uniquely determined. The G-action on $\hat{N}$ is defined as
the map $G\times\hat{N}\to\hat{N}$ sending the pair $(g,\chi)$ to
$g\chi(\alpha)=\chi(g^{-1}(\alpha))$; instead from the point of view of an
element $\phi\in N$,
the action $G\times N\to N$ is:
$$g\phi(z,\bar{z})=K_g^{-3}(z,\bar{z})\phi(gz,g\bar{z}).$$

The above relation tells us that the
dual space $\hat{N}$ is isomorphic to the supertranslations space
 $N$ and there exists a decomposition of $\hat{N}$ as a direct sum of two 
 subgroups-i.e.
$\hat{N}=A_0\oplus B_0$, where $A_0$ is (isomorphic to) the space of linear functionals   
vanishing on $A$ whereas $B_0$ is the space of linear functionals vanishing on
$B_0$.
This means that $A_0$ is composed by those character
mapping all the elements of $A$ into the unit number and the same holds also for $B_0$.
As in the supertranslation case, this decomposition is only true at the level of vector
fields since it is not G-invariant. The onlyù space which is not changing under group
transformations is the subspace $A_0$.

Since one can associate a unique element of $N$, namely $\phi$ to each $\chi(\alpha)$ one 
can
also decompose this field as:
$$\phi(\theta,\phi)=\sum\limits_{l=0}^1\sum\limits_{m=-l}^l
p_{lm}Y_{lm}(\theta,\phi)+\sum\limits_{l>1}\sum\limits_{m=-l}^l
p_{lm}Y_{lm}(\theta,\phi),$$
where the first piece in the sum is in one to one correspondence with the quadruplet
$(p_0, p_1, p_2, p_3)$ which can be thought as the components of a momentum vector
related to the Poincar\'e group. For this reason one can introduce a new space
$A^\prime$ isomorphic both to $A$ and $\hat{A}$ which is given by the set of all
possible functions $\phi$ and which is often referred as the supermomentum space.

One can now find representations of the BMS group in Hilbert topology with the
help of Mackey's theorem applying it to this infinite dimensional (Hilbert)-Lie
group in the spirit of \cite{Mc1}. The first step consists in finding the orbits of $SL(2,\mathbb{C})$ in
$\hat{N}$ which are homogeneous spaces that can be classified as the elements of the set of non
conjugate subgroups of $SL(2,\mathbb{C})$. In order to find a representation for the BMS
group, after classifying the homogeneous spaces $M$, we shall find a character $\chi_0$
fixed under $M$ and then identify each $M$ with its little group associated with the
orbit $G\chi_0\sim G/L$. 

As a starting point we shall consider only connected subgroups of
$SL(2,\mathbb{C})$. The list of these groups is well known but most of
them do not admit a non trivial fixed point in $N$. This request restricts them
to

\begin{center}
\begin{tabular}{|c|c|c|c|}
\hline
Little group & Character & Fixed point & Orbit\\
\hline
$SU(2)$ & $\chi(\alpha)=e^{i<K,\alpha>}$ & $\phi(\theta,\varphi)=K$ &
$PSL(2,\mathbb{C})$\\
\hline
$\Gamma$ & $\chi(\alpha)=e^{i<\zeta(\mid
z\mid),\alpha>}$ & $\zeta(\mid z\mid)$ & $G/\Gamma$\\
\hline
$Z_2$ & $\chi(\alpha)=e^{i<\phi_0(z,\bar{z}),\alpha>}$ & $\phi_0(z,\bar{z})$ &
$G/SU(2)$\\
\hline
\end{tabular}
\end{center}

\noindent where $\Gamma$, which consists of diagonal matrices, is the double cover of
$SO(2)$ and where $Z_2$ is not formally a connected group but nonetheless it is the center of
$SL(2,\mathbb{C})$ and for this reason it acts in a trivial way.

At this point one needs to express explicitly the induced representations;
this operation consists in giving a unitary irrep $U$ of $L_\chi$ on a suitable
Hilbert space $H$ for any little group and a G-invariant measure on the orbit of
each little group. For the connected subgroups, one has \cite{Mc1}:
\begin{itemize}
\item the group $Z_2$ has only two unitary irreps, the identity $D^0$ and a second
faithful representation $D^1$ both acting on the Hilbert space of complex numbers
$\mathbb{C}$ as:
$$D^0(\pm I)=1,\;\;\;D^1(\pm I)=\pm 1.$$
\item the unitary irreps of $\Gamma(\sim \pi\frac{R}{4}Z)$ are instead indexed
by an integer or half integer number $s$ acting on the Hilbert space of complex numbers
$\mathbb{C}$ as:
$$D^s(g)=e^{ist},$$
where $g\in\Gamma$ and
$$g=\left[\begin{array}{cc}
e^{\frac{it}{2}} & 0\\
0 &  e^{\frac{-it}{2}}
\end{array}\right].$$ 
\item the unitary irreps of $SU(2)$ are the usual ones acting on a $2j+1$
dimensional complex Hilbert space with $j\in\frac{Z}{2}$.
\end{itemize}

Consider now the case of non connected little
groups; the hope is that all these groups are compact since this
grants us that their representations can be labelled only by
finite indices. For the BMS group in the Hilbert
topology this is indeed the case since it was shown in
\cite{Mc2} (see theorem 1) that all little groups are compact. 
Besides, since the homogeneous Lorentz
group admits $SO(3)$ as maximal compact subroup, we need to analyse
only subgroups of $SO(3)$. The list of these subgroups has been in \cite{Mc2}:
$$ C_n\;\;\; D_n\;\;\; T\sim A_4\;\;\; O\sim S_4\;\;\; I\sim
A_5\;\;\; \Theta=\Gamma R_2,$$
where $R_2=\left\{I, J=\left[\begin{array}{cc}
0 &1\\
-1 & 0
\end{array}\right]\right\}$. It shows that we are dealing only with groups with finite dimensional
representations which means that there are no continuous indices
labelling states invariant under BMS group with Hilbert
topology.

As in the usual approach, one can then construct on each orbit a
constant function, namely the Casimir, in order to classify them.
Instead of working on $\hat{N}$ it is easier to consider the space
of scalar functions $N^\prime$ isomorphic to $\hat{N}$ and endow
it with a bilinear application assigning to the pair $(\phi_1,\phi_2)$ the number
$B(\phi_1,\phi_2)=\pi(\phi_1)\cdot\pi(\phi_2)$ where $\pi$ is the
projection on the momentum components (i.e. $\pi:N^\prime\to
A^\prime$) and the dot denotes the usual Lorentz inner product. It
is also straightforward to see that the bilinear application is
G-invariant.

This last property implies that on each orbit in $\hat{N}$, the
function $B$ is constant and its value can be calculated since
$\pi(\phi)=(p_0,p_1,p_2,p_3)$ so that:
$$B(\phi,\phi)=\pi(\phi)\cdot \pi(\phi)=m^2. $$

One can thus label each orbit in the
character space with an invariant, the squared mass, together with
the sign of the ``temporal'' component i.e. $sgn(p_0)$. These invariants
grant only a partial classification since, for example, in the case
of unfaithful representations, $\pi(\phi)=0$, which implies
that the above invariants are trivial. For faithful
representations too, we cannot conclude that the classification is
complete since different orbits can correspond to the same value
for the mass. One can also find a constant
number to label the orbits corresponding to unfaithful
representations-i.e. an bilinear invariant application mapping at
least $A^0$ to real numbers. This has been done in
\cite{Mc2}:
\begin{equation}
Q^2=\pi^2\int\int\frac{\mid z_1-z_2\mid^2\ln\mid z_1-z_2\mid}{(1+\mid
z_1\mid^2)(1+\mid
z_2\mid^2)}\phi(z_1,\bar{z}_1)\phi(z_2,\bar{z}_2)d\mu(z_1,\bar{z}_1)d\mu(z_2,\bar{z}_2),
\end{equation}
where $\phi$ is a function of class $C^\infty(S^2)$. Thus $Q^2$ is
defined only for a subset dense in the Hilbert space $L^2(S^2)$
which from the physical 
point of view makes no difference.
\section{BMS representations in nuclear topology}
The study of BMS group to label elementary particles started with
 the hope to remove the difficulty  with 
 the Poincar\'e group concerning the continuous representations
associated to the non compact $E(2)$ subgroup. Unfortunately, continuous
representations appear if one chooses
for the supertranslations a different topology (for istance $C^k(S^2))$. This was for
the first time pointed out in \cite{Girardello} where
it was shown that also non compact little
groups appear (for instance $E(2)$).  

Nonetheless it is worth studying representations for the BMS group with $N$
endowed with the topology $C^\infty(S^2)$. In this case the action of $SL(2,\mathbb{C})$ on the space of
supertranslations is given by a representation $T$ equivalent to the irrep of $SL(2,\mathbb{C})$
on the space $D_{(2,2)}$ introduced by Gel'fand. This implies that we have to
use techniques proper of {\it rigged} Hilbert spaces. 

The main object we shall deal
with is $D_{(n,n)}$ which is the space of functions $f(z,\omega)$ of
class $C^\infty$ except at most in the origin. These functions also satisfy the relation $f(\sigma
z,\sigma w)=\mid\sigma\mid^{(2n-2)}f(z,w)$ for any $\sigma\in\mathbb{C}$. At the
end of the day, one has the following chain of isomorphisms
$$BMS=N\ltimes G\longleftrightarrow D_{(2,2)}\ltimes G\longleftrightarrow D_{2}\ltimes
G,$$
where $D_2$ is the space of $C^\infty$ functions $\zeta(z)$ depending on a
single complex variable such that any element $g(z,w)\in D_{(2,2)}$ can be
written as $g(z,w)=\;\;\mid z\mid^{2}\zeta(z_1)=\mid w\mid^2\hat{\zeta}(z_1)$
with $z_1=\frac{w}{z}$ and $\hat{\zeta}(z_1)=\mid z_1\mid^2\zeta(z_1^{-1})$.

Irreducible representation can arise (see theorem 2 in \cite{Mc4}) can arise either from a transitive G action in the
supermomentum space or from a cylinder measure $\mu$ with respect to the G
action is strictly ergodic i.e. for every measurable set $X\subset N^\prime$ $\mu(X)=0$ or
$\mu(N^\prime-X)=0$ and $\mu$ is not concentrated on a single G-orbit in
$N^\prime$. 

The first step is to classify all little groups; they can either
be discrete subgroups, non-connected non discrete Lie subgroups and
connected Lie subgroups. 

Discrete subgroups can be derived exactly as in the Hilbert
case and so the only connected little groups for the BMS group are:
$$SU(2),\;\;\Gamma,\;\;\Delta,\;\;SL(2,\mathbb{R}).$$

Here one can point out the first difference between the Hilbert and the nuclear
topology which consists in the appearance of the $SL(2,\mathbb{R})$ little group
which will contribute only to unfaithful representations.

Non connected non discrete subgroups $S$ can be derived using still theorem
5 in \cite{Mc4} since each $S$ is a subgroup of the normalizer $N(S_0)$ where
$S_0$ is the identity component of $S$. Here is the list:
\begin{itemize} 
\item $S_1$ which is the set of matrices $$\left\{\left[\begin{array}{cc}
\sigma^r & 0\\
0 & \sigma^r\end{array}\right] {\rm with}\;\; \sigma=e^{\frac{2\pi i}{n}},\;\;0\leq
r\leq(n-1)\right\},$$ 
\item $S_2$ which is the set of matrices $$\left\{\left[\begin{array}{cc}
e^{qr} & 0\\
0 & e^{-qr}\end{array}\right] \textrm{where q is a fixed non negative number and
r $\in\mathbb{Z}$}\right\},$$
\item $S_3$ which is the set of matrices $$\left\{\left[\begin{array}{cc}
z_1^{r}z_2^s & 0\\
0 & z_1^{-r}z_2^{-s}\end{array}\right] {\rm where}\; z_1, z_2\in\mathbb{C}\;
\textrm{and r, s are integers},\right\},$$
\item $S_4$ which is the set of matrices $$\left[\begin{array}{cc}
1 & r\\
0 & 1\end{array}\right].$$
\end{itemize}

To establish the faithfulness of the irreps, one needs to calculate the
projection on the supermomentum space of supertranslation:
$$\pi(\phi)(z^\prime)=\frac{i}{2}\int
dzd\bar{z}(z-z^\prime)(\bar{z}-\bar{z^\prime})\phi(z)\neq 0.$$

The only connected groups occuring as little
group for faithful representations are $\Gamma,\Delta,SU(2)$ with vanishing
square mass $0$. It is also interesting to
notice that the orbit invariant $Q^2$ defined for the Hilbert topology for
unfaithful representations is not available in the nuclear topology since it is
not defined for distributional supermomenta. Finally no information is available about discrete subgroups
since it very difficult to classify them above all for infinite discrete
subgroups. This means that the study of BMS representations in the nuclear
topology has to be completed yet.
\chapter{Wave equations in fiber bundle approach}
We are going to briefly review in the following definitions and notations 
used in the derivations of the BMS wave equations following \cite{Asorey}. 
As said we use sort of diagrams to facilitate the reader even if they are
not rigorous mathematically speaking.

Consider then
$$\xymatrix{P(H,M)\ar@{-->}[r]^{\sigma}\ar[d]^{\pi} & GL(V)\ar@{-->}[d]^{*} \\
M & E(M,V)\ar[l]^{\pi_E}}$$
In our  case $P(H,M)$ is a group $G$ and a principal bundle whose
fiber $H$ is a closed subgroup of $G$ and whose base space is the homogeneous space
given by the coset $G/H$;
each linear representation $\sigma:H\to GL(V)$ automatically defines the vector bundle
(that's the reason why we used the dotted lines) $E(M,V)=P\times_H V$ whose generic
element is the equivalence class $[u,a]$ with $u\in P$ and $a\in V$. The equivalence
relation defining this class is given by
$(u,a)\sim(uu^\prime,a)=(u,\sigma(u^\prime)a)$. 

One defines than a G-action; in particular on
$P=G$ it is the obvious one i.e.
$g(uh)=(gu)h$ whereas on $M$ the action is induced through the projection $\pi$ as
$g\pi(u)=\pi(gu)$. Finally on the E bundle the G-action is $g[u,a]=[gu,a]$. i
if we consider a generic section for the E bundle i.e. $\psi:M\to E(M,V)$, we can act
on it through a linear representation of G as:
\begin{equation}\label{psi1}
\left(U(g)\psi\right)(gm)=g\psi(m).
\end{equation}
This representation is exactly the ``induced'' representation of $G$ constructed from
the given one $\sigma$ of the subgroup $H$.
Moreover if $\sigma$ is pseudo-unitary and it exists an invariant G-measure on M we can
define an internal product $(,):\Gamma(E)\times\Gamma(E)\to\mathbb{C}$ as:
$$<\psi,\phi>=\int\limits_M d\mu h_m(\psi,\phi),$$
where $h_m$ is the induced internal product on the fibre $\pi^{-1}_E(x)$.

In the specific case of semidirect product of groups i.e. $G=N\ltimes K$ (N
abelian), one can define a G-action on the character space $\hat{N}$:
$$g\chi(gn)=\chi(n).$$

For any element $\chi_0$ one can construct its stability (little) group
$G_{\chi_0}:N\ltimes K_{\chi_0}$ and assign a representation $\sigma:K_{\chi_0}\to
GL(V)$. This induces a representation $\chi_0\sigma:G_{\chi_0}\to V$ which 
associates with the
couple (n,g) the element $\chi_0(n)\sigma(g)$. Thus in our diagram the group
$G_{\chi_0}$ is playing the role of $H$ and $M$ becomes the coset space
$G/G_{\chi_0}$. The diagram is then: 
$$\xymatrix{G(N\ltimes K_{\chi_0}, G/G_{\chi_0})\ar@{-->}[r]^{\:\:\;\;\;\;\;\;\;\;\chi_0\sigma}\ar[d]^{\pi} & GL(V)\ar@{-->}[d]^{*} \\
G/G_{\chi_0} & E(G/G_{\chi_0},V)\ar[l]^{\pi_E}}$$
Since we want to describe vector valued functions $f:G/G_{\chi_0}\to V$, a representation of  
$U$ can be made fixing a section $s:G/G_{\chi_0}\to G$ and remembering that for any
element $\psi\in\Gamma(E)$ there exists a function $\tilde{f}_{\psi}:P\to V$:
$$\psi(\pi(u))=[u,\tilde{f}_\psi(u)].$$
A vector valued function is:
$$ f_\psi=\tilde{f}_\psi\circ s.$$

Let us notice that this construction makes sense only if the section $s$ is global
otherwise $f$ is not defined everywhere; this happens only if the bundle $G$ is
trivial i.e. $G=M\times H$ which is always the case in the situation we are interested
in.

Moreover equation (\ref{psi1}) translates in:
\begin{equation}\label{psi2}
\left[U(g)f_\psi\right](gx)=\sigma(\gamma(g,x))f_\psi(x),
\end{equation}
where in our case $\gamma:(N\ltimes K)\times G/G_{\chi_0}\to G_{\chi_0}$ is defined
as:
\begin{equation}\label{gamma}
s(gx)\gamma(g,x)=gs(x)
\end{equation}
From now on we shall call (\ref{psi2}) the \emph{canonical (or induced) wave
equation}.

In physical relevant situations covariant representations are used
instead of induced ones. In this case we deal with a principal bundle
$G(X,G_{x_0})$ where $x_0\in X$ and the interesting representations are those
preserving locality on the physical relevant space $X$ and acting on a vector
valued function $f:X\to V$ as:
\begin{equation}\label{psi3}
[T(g)f](gx)=A(g,x)f(x),
\end{equation}
where $A$ is a map from $G\times X$ to $GL(V)$ satisfying the property:
$$A(g_1g_2,x)=A(g_1,g_2x)A(g_2,x).$$
Examples of these representations are those induced from the isotropy group
$G_{x_0}$ when expressed in term of sections $s:X\to G$. Let us also notice that
there exists a map $\Sigma:G_{x_0}\to GL(V)$ assigning to an element $\gamma$
the matrix $\Sigma(\gamma)=A(\gamma,x_0)$ and the induction of such a
representation from the isotropy group to the entire $G$ generates the
representation
$$A^\prime(g,x)=\Sigma(\gamma(g,x)),$$
where $\gamma$ is defined as in (\ref{gamma}) and $g=(\phi,\Lambda)$.

\end{document}